\documentclass[
reprint,
bibnotes,
amsmath,preprintnumbers,amssymb,
aps,
prl
]{revtex4-2}

\usepackage{amssymb,amsmath,mathtools}
\usepackage{amsthm}
\usepackage[normalem]{ulem}
\usepackage{titletoc}
\usepackage{graphicx}
\usepackage{dcolumn}
\usepackage{enumitem}
\usepackage{bm}
\usepackage{float}
\usepackage[many]{tcolorbox}
\usepackage{shuffle}
\usepackage{xcolor}
\usepackage{hyperref}
\hypersetup{hypertexnames=false}
\usepackage{physics}
\usepackage{tabularx}
\usepackage{diagbox}
\usepackage{multirow}
\usepackage{tikz}
\usepackage{tikz-feynman}
\usepackage{dsfont}

\usetikzlibrary{decorations.pathmorphing}
\usetikzlibrary{arrows}
\usepackage{circuitikz}
\usetikzlibrary{arrows.meta}
\usetikzlibrary{shapes.misc}
\usetikzlibrary{positioning}
\usetikzlibrary{decorations.markings}
\usetikzlibrary{intersections}
\usetikzlibrary{calc}
\usepackage{verbatim}
\usepackage{xcolor}

\definecolor{Maroon}{RGB}{128, 0, 0}
\definecolor{Green}{rgb}{0.0, 0.5, 0.0}
\definecolor{Purple}{rgb}{0.5, 0.0, 0.5}
\definecolor{Blue}{rgb}{0.0, 0.0, 0.6}

\newcommand{\head}[1]{\par\medskip\noindent\textbf{\textit{#1.}}---}

\begin{document}

\let\oldaddcontentsline\addcontentsline
\renewcommand{\addcontentsline}[3]{}

\title{A Low Scale for Quark--Lepton Unification}

\author{Lawrence J. Hall$^{1,2}$}
\email{ljh@berkeley.edu}
\author{Claudio Andrea Manzari$^{3}$}
\email{manzari@ias.edu}
\author{Clara Murgui$^{4}$}
\email{clara.murgui@cern.ch}

\affiliation{%
\begin{tabular}{@{}c@{}}
\resizebox{\linewidth}{!}{%
$^{1}$Leinweber Institute for Theoretical Physics, Department of Physics,
University of California, Berkeley, CA 94720, USA%
}\\[3pt]
$^{2}$Theoretical Physics Group, Lawrence Berkeley National Laboratory,
Berkeley, CA 94720, USA\\[3pt]
$^{3}$School of Natural Sciences, Institute for Advanced Study,
Princeton, NJ 08540, USA\\[3pt]
$^{4}$Theoretical Physics Department, CERN,
CH-1211 Geneva 23, Switzerland
\end{tabular}%
}

\preprint{CERN-TH-2026-256}

\begin{abstract}
The measured fermion masses may be pointing to the scale of quark--lepton unification. Studying the renormalization-group evolution of the Standard Model Yukawa couplings, we find that $y_\tau=y_b$ and $y_\mu=3y_s$ occur at nearly the same scale, around $10^6$~GeV, while the first generation reaches the Georgi--Jarlskog relation $y_e\simeq y_d/3$ at order $10^8$ GeV and can arise from the same Clebsch structure through a simple hierarchical flavor texture. This pattern naturally suggests low-scale Pati--Salam symmetry. We show that flavor mixing and finite threshold corrections preserve this picture. If the electron mass has an independent origin, the two heavier-generation relations can be realized in theories with a unification scale as low as $(5$--$7)\times10^5$~GeV. We identify two simple theories that accommodate all three relations and in which the measured masses fix the flavor structure. The first unifies all three generations at $4.6^{+1.2}_{-0.9}\times10^6$~GeV; the second, with symmetric Yukawa matrices, at $1.5^{+0.7}_{-0.5}\times10^7$~GeV, and predicts $|V_{cb}|=0.042\pm0.002$. Their shared light-family texture fixes the relevant vector-leptoquark couplings, making $\mu\to e$ conversion in nuclei a direct test of unification: upcoming searches will probe part of the region selected by the two theories, and proposed facilities could cover it completely.
\end{abstract}

\maketitle

\begin{figure}[b]
\raggedright
\rule{0.3\columnwidth}{0.4pt}\\[2pt]
{\small\textit{We dedicate this work to the memory of Mark B. Wise.}}
\end{figure}

Quark--lepton unification is one of the most economical ideas for reducing
the otherwise arbitrary flavor structure of the Standard Model (SM). When
quarks and leptons belong to common gauge multiplets, Yukawa interactions
that are independent in the SM become related. In canonical renormalizable
realizations, including Pati--Salam (PS)~\cite{Pati:1973uk,Pati:1974yy}, $SU(5)$~\cite{Georgi:1974sy} and $SO(10)$~\cite{Georgi:1974my,Fritzsch:1974nn}, individual
down-quark and charged-lepton Yukawa structures can satisfy schematically
\cite{Georgi:1979df}
\begin{equation}
	Y_e^{(a)}=\mathcal C_aY_d^{(a)},\qquad
	\mathcal C_a=1\ \hbox{or}\ -3,
	\label{eq:clebsch}
\end{equation}
where $a$ labels the underlying unified Yukawa structure. In PS the origin of these coefficients is especially transparent. Since quarks and leptons belong to $F\sim\mathbf 4$ and $F^c\sim\bar{\mathbf 4}$ of $SU(4)_C$, and $\mathbf 4\otimes\bar{\mathbf 4}=\mathbf 1\oplus\mathbf{15}$, renormalizable Yukawa interactions involve only $SU(4)_C$ singlet and adjoint scalars. The singlet couples quarks and leptons identically, giving $\mathcal C=1$, whereas the electroweak-doublet component of the adjoint lies along the $B-L$ direction ${\rm diag}(1,1,1,-3)$, which singles out the fourth color and gives $\mathcal C=-3$.

Equation~\eqref{eq:clebsch} is fundamentally a statement about Yukawa
\emph{structures}, not directly about physical masses. With several
contributions, the singular values of the complete matrices need not inherit
an exact Clebsch factor. Precisely for this reason, simple relations among
physical masses are informative: they may indicate a flavor structure in which unified Yukawa structures with definite Clebsch coefficients leave a recognizable imprint after diagonalization.

Historically, quark--lepton unification has been largely studied in the context of theories of grand unification. In particular, imposing bottom-tau Yukawa unification, $y_b=y_\tau$, at a high scale was already shown in early studies of $SU(5)$ to reproduce the observed bottom--tau mass splitting through renormalization-group (RG) evolution alone, within the experimental precision available at the time~\cite{Chanowitz:1977ye,Buras:1977yy}. Together with the approximate meeting of the gauge couplings~\cite{Georgi:1974yf}, this provided one of the early phenomenological successes of grand unified theories (GUTs). Subsequent studies tested whether several fermion mass relations could hold at a common scale ~\cite{Harvey:1980je,Dimopoulos:1991yz, Dimopoulos:1991za,Arason:1992eb,Barger:1992ac,Anderson:1993fe}.
Today the fermion masses are known with sufficient precision to test quark--lepton relations far more sharply than was possible in the early studies: while the charged-lepton masses are measured far below the per-mille level and the bottom mass at the few per-mille level, recent lattice determinations fix the strange- and down-quark masses at the percent level~\cite{ParticleDataGroup:2024cfk,FlavourLatticeAveragingGroupFLAG:2024oxs}. Can the observed spectrum reveal a scale at which a simple quark--lepton structure emerges?
Departures from the elementary relations in Eq.~\eqref{eq:clebsch} are expected once a realistic flavor and symmetry-breaking sector is included~\cite{Ross:2007az}. Therefore, their occurrence would point toward a flavor sector in which only a small number of unified Yukawa structures dominate the relevant masses, allowing simple group-theoretical relations to remain visible in the physical spectrum.  Although more general constructions remain possible, this raises the tantalizing possibility that precision fermion data can reveal the scale of quark--lepton unification as well as something about the underlying structure of flavor.\\
In this work we show that the measured fermion spectrum does point to such a scale. Under SM running, the relations $y_\tau=y_b$ and $y_\mu=3y_s$ are realized at a common scale near $10^6$~GeV, far below conventional grand unification, while the first generation can follow from the same Clebsch structure through a simple hierarchical texture. Such low scales single out PS quark--lepton unification, which can be realized consistently at low scales~\cite{Smirnov:1995jq,FileviezPerez:2013zmv,Calibbi:2017qbu,Dolan:2020doe,Iguro:2021kdw,FileviezPerez:2021lkq,FileviezPerez:2022rbk,FileviezPerez:2023rxn,DiLuzio:2017vat,Bordone:2017bld,Fuentes-Martin:2020pww,Davighi:2022bqf}, and has been considered in connection with the bottom--tau mass relation~\cite{Volkas:1995yn}. Here we show that all three generations point to a common low scale, and that realistic flavor structures and finite threshold corrections preserve this picture, with the preferred scales within reach of rare flavor-violating processes.\\

\textbf{\textit{Running of Quark and Lepton Yukawas.}}---
We start by studying the running of down-quark and charged lepton Yukawa couplings.
We evolve them together with the gauge couplings and Higgs quartic with the two-loop SM RG equations, following Refs.~\cite{Luo:2002ti,Buttazzo:2013uya}.  The $\overline{\rm MS}$ inputs at the $Z$ pole-scale, $\mu=m_Z$, and their uncertainties are taken from Ref.~\cite{Antusch:2025fpm}. We solve the coupled system numerically and propagate the input uncertainties through a Monte Carlo ensemble. The results are shown in Fig.~\ref{fig:running} while further details are given in Appendix~\ref{app:rge}.  It is convenient to define 
\begin{equation}
R_3\equiv\frac{y_\tau}{y_b},
\qquad
R_2\equiv\frac{y_\mu}{3y_s},
\qquad
R_1\equiv\frac{3y_e}{y_d}.
\label{eq:Rdefs}
\end{equation}

\begin{figure}[t]
\centering
\includegraphics[width=\columnwidth]{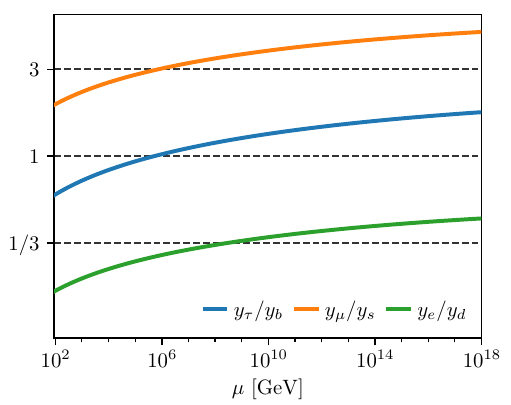}
\caption{Two-loop SM running of the quark--lepton Yukawa ratios.  The third generation reaches $y_\tau/y_b=1$ and the second reaches $y_\mu/y_s=3$ near $10^6$~GeV, whereas the first reaches $y_e/y_d=1/3$ near $3\times10^8$~GeV.  The $1\sigma$ bands, barely visible, arise from the $m_Z$-scale input uncertainties.}
\label{fig:running}
\end{figure}

As shown in Fig.~\ref{fig:running}, over the range from $m_Z$ to $10^{18}$ GeV, the physical ratios evolve in the range
\begin{align}
\begin{split}
&\frac{y_\tau}{y_b} \in [0.61,1.74]\,,\qquad
\frac{y_\mu}{y_s} \in [1.91,4.82]\,, \\ & 
\frac{y_e}{y_d} \in[0.18,0.45].
\end{split}
\label{eq:ratios_range}
\end{align}

Hence, if the physical eigenvalues directly inherit one of the elementary Clebsch relations discussed above, the running already selects which one is attainable: the third-generation ratio crosses $1$ but never reaches $3$, the second-generation ratio crosses $3$ but never reaches $1$, while $y_e/y_d$ never crosses $1$ nor $3$.

The first interesting observation is that for the two heavier generations we find, at 68\% C.L.,
\begin{align}
R_3(\mu_3)&=1,
&
\mu_3&=5.20^{+0.90}_{-0.76}\times10^5~{\rm GeV}\,,
\nonumber\\
R_2(\mu_2)&=1,
&
\mu_2&=7.20^{+3.69}_{-2.33}\times10^5~{\rm GeV}\,,
\label{eq:mu_cross}
\end{align}
with
\begin{equation}
\frac{\mu_2}{\mu_3}\in[0.92,2.12]\,.
\label{eq:ratio}
\end{equation}
This is a nontrivial coincidence: four independently measured Yukawa couplings realize two different Clebsch relations at a common scale within uncertainties. The scale itself is equally striking. Both relations are realized near $10^6$~GeV, far below the scale conventionally associated with grand unification. SM running does not cause the two Yukawa couplings to reconverge at higher energies. Already at one loop, below the Planck scale, the dominant contribution comes from QCD, $\beta_{y_b}\propto -8g_3^2 y_b$, and since it has no counterpart in $\beta_{y_\tau}$, it drives $y_\tau/y_b$ away from unity above $\mu_3$. Higher-order corrections do not alter this conclusion; as discussed in Appendix~\ref{app:rge}, the perturbative expansion is well converged and the residual truncation uncertainty is subdominant to the experimental one. Within SM evolution, the measured second and third generation Yukawa couplings therefore select a common Clebsch-crossing scale of order $10^6$~GeV.

One may ask how robust this crossing scale is to physics beyond the SM. New states without direct Yukawa couplings to the SM fermions affect the Yukawa ratios through the running of the gauge couplings. At one loop,
\begin{equation}
	16\pi^2
	\frac{d\ln(y_\tau/y_b)}{d\ln\mu}\bigg|_{\rm gauge}
	=
	-2g_1^2+8g_3^2,
\end{equation}
so that only hypercharged states can raise the crossing scale, but only weakly before $g_1$ becomes nonperturbative.
Appendix~\ref{app:betaext} makes this statement quantitative studying the effect of several scalar and fermion gauge representations.
As we will see, however, substantial shifts can arise from a non-trivial flavor structure and threshold corrections.\\

Turning to the first generation, $y_e/y_d\simeq0.29$ at $10^6$ GeV. While Clebsch ratios of $1$ or $3$ are not attained, the ratio satisfies the Georgi--Jarlskog relation~\cite{Georgi:1979df} within the range of scales considered,
\begin{equation}
R_1(\mu_1)=1
\qquad 
\mu_1=2.97^{+2.44}_{-1.34}\times10^8~{\rm GeV}.
\label{eq:mu1}
\end{equation}
This is particularly suggestive, since such a relation can arise naturally through a simple seesaw mechanism~\cite{Georgi:1979df}. Consider the following texture for the first two generations
\begin{equation}
	Y_d \simeq
	\begin{pmatrix}
		0 & a\\
		a^{\prime} & e
	\end{pmatrix},
	\qquad
	Y_e \simeq
	\begin{pmatrix}
		0 & a\\
		a^{\prime} & -3e
	\end{pmatrix},
\end{equation}
with $|a|,|a^{\prime}|\ll |e|\,.$ After diagonalization
\begin{equation}
	\frac{y_\mu}{y_s}\simeq 3,
	\qquad
	\frac{y_e}{y_d}\simeq \frac{1}{3}. 
\end{equation}
Thus the first-generation inverse factor need not correspond to a new Clebsch coefficient, but can emerge from the same factor responsible of the second generation ratio of $3$ and a simple flavor structure.

For comparison, at the conventional grand unification scale ($\Lambda_{\rm GUT} \sim 2\times10^{16}$~GeV), $R_3\simeq1.7$, $R_2\simeq1.6$ and $R_1\simeq1.3$: a flavor sector in which correction of $30-70\%$ are required. 

The data therefore single out a relatively small window between roughly $10^5$ and $10^9$~GeV, where quark--lepton unification may be realized. In the following, we study concrete ultraviolet realizations and ask whether, once realistic flavor structures and threshold corrections are included, the three relations can hold at a common scale, and at what scale.\\

\textbf{\textit{UV completion.}}---
To unify quarks and leptons in common gauge multiplets, the QCD gauge group must itself be embedded in a larger group. Indeed, quarks transform as triplets of $SU(3)_C$, whereas leptons are color singlets, so they cannot belong to the same irreducible multiplet as long as $SU(3)_C$ remains an independent gauge factor. An elegant approach to quark-lepton unification is to identify leptons as the fourth color, enlarging  $SU(3)_C$ to $SU(4)_C$~\cite{Pati:1973uk,Pati:1974yy}. At the low scales suggested by the Yukawa running, conventional simple-group unification is obstructed by proton decay from its baryon-number-violating gauge bosons. PS avoids that gauge-mediated obstruction and is the minimal framework for relating quarks and leptons. Low-scale bottom--tau unification and its sensitivity to scalar thresholds in PS theories were previously explored in Ref.~\cite{Volkas:1995yn}. The simultaneous first- and second-generation relations supply an additional motivation here.

We focus on the completion
\begin{equation}
G_{422}=SU(4)_C\times SU(2)_L\times SU(2)_R,
\end{equation}
as in the original PS construction~\cite{Pati:1973uk,Pati:1974yy}. Each SM family together with a right-handed neutrino is embedded as
\begin{equation}
F_i\sim(4,2,1), \qquad F_i^c\sim(\bar4,1,2),
\end{equation}
The Yukawa sector contains an $SU(4)_C$ singlet and an adjoint bidoublets,
\begin{equation}
\Phi_1\sim(1,2,2),\qquad \Phi_{15}\sim(15,2,2).
\label{eq:PSbidoublets}
\end{equation}
 A neutral fourth-color vev of $\chi_R\sim(4,1,2)$ can break $G_{422}$ directly to the SM. We denote the breaking scale as $\Lambda_{\rm PS}$, with $M_X=g_4\Lambda_{\rm PS}/\sqrt2$ for the vector leptoquark in this convention. 
Each complex bidoublet admits two independent Yukawa invariants, with matrices $\lambda_r$ and $\widetilde\lambda_r$ for $r=1,15$. Their component interactions and vacuum projections are given in Appendix~\ref{app:PStheory}. A closely related realization is provided by the $G_{421}$ model, in which $SU(2)_R$ is replaced by an abelian $U(1)_R$. The discussion below applies to both gauge structures, with the corresponding differences in the right-handed charges.

In minimal PS realizations, the neutrino Dirac masses are tied to those of the up-type quarks, making a purely Dirac neutrino spectrum phenomenologically unacceptable. A low quark-lepton unification scale can instead be reconciled with light neutrino masses through an inverse seesaw mechanism~\cite{Mohapatra:1986aw,Mohapatra:1986bd}, as proposed in Refs.~\cite{Volkas:1995yn,FileviezPerez:2013zmv}. This requires the introduction of at least two gauge-singlet fermions $N_i$, permitting the coupling $F^c\chi_R N$. In the lepton-number-conserving limit, one-family neutral masses have the structure
\begin{equation}
\mathcal M_\nu=\begin{pmatrix}
0&m_D&0\\m_D&0&M_N\\0&M_N&0
\end{pmatrix},
\label{eq:neutral_fermion_mass}
\end{equation}
in the basis $(\nu, \nu_c, N)$.
There is a heavy Dirac pair of mass $m_S=\sqrt{M_N^2+m_D^2}$ and an exactly massless state. A small lepton-number-violating entry $\mu_N N N$ gives the inverse-seesaw scaling $m_\nu\simeq\mu_N(m_D/M_N)^2$~\cite{Mohapatra:1986aw}. Realistic neutrino flavor requires a choice of the neutral-sector matrices. The splitting between the heavy neutral partner and the corresponding up quark is important in the discussion of threshold corrections below (see also Appendix~\ref{app:thresholds}). For concreteness, the renormalizable Yukawa interactions are
\begin{align}
\begin{split}
{\cal L}^{422}_Y={}&
F^T\left[\lambda_1\Phi_1^*-\widetilde\lambda_1\,\epsilon_L\Phi_1\epsilon_R\right]F^c\\
&+F^T\left[\lambda_{15}\Phi_{15}^*-\widetilde\lambda_{15}\,\epsilon_L\Phi_{15}\epsilon_R\right]F^c\\
&+\lambda_N\,F^c\chi_R N+\tfrac12\mu_N NN+{\rm h.c.},
\end{split}
\label{eq:L422_main}
\end{align}
where $\epsilon_{L,R}=i\sigma_2$ act on the $SU(2)_{L,R}$ indices.

The $SU(4)$ singlet treats quarks and leptons identically, while the color-singlet direction in the adjoint is proportional to ${\rm diag}(1,1,1,-3)$. Writing the singlet and adjoint contributions to the vacuum-aligned Yukawa matrices as $Y_f^{1}$ and $Y_f^{15}$ gives
\begin{align}
Y_d&=Y_d^{1}+Y_d^{15},& Y_e&=Y_d^{1}-3Y_d^{15},\nonumber\\
Y_u&=Y_u^{1}+Y_u^{15},& Y_\nu^D&=Y_u^{1}-3Y_u^{15}.
\label{eq:mass_projections}
\end{align}
When each entry is generated by either the singlet contribution $Y_f^{1}$ or the adjoint contribution $Y_f^{15}$, but not both, the corresponding quark and lepton Yukawa entries are related by a factor of $1$ or $-3$, respectively. The imprint of these entrywise relations on the physical mass ratios then depends on the flavor structure and its diagonalization.

The observed mass pattern points to a particularly economical structure, in which $\Phi_1$ provides the dominant contribution to the $12$, $21$, and $33$ entries and $\Phi_{15}$ to the $22$ entry. We may summarize this assignment by the single schematic texture
\begin{equation}
	{\cal Y}^{(0)}
	\sim
	\begin{pmatrix}
		0 & \Phi_1 & 0\\
		\Phi_1 & \Phi_{15} & 0\\
		0 & 0 & \Phi_1
	\end{pmatrix}.
	\label{eq:leading_PS_texture}
\end{equation}
This leads to:
\begin{equation}
Y_d^{(0)}=\begin{pmatrix}0&a&0\\a'&e&0\\0&0&d\end{pmatrix},\qquad
Y_e^{(0)}=\begin{pmatrix}0&a&0\\a'&-3e&0\\0&0&d\end{pmatrix}.
\end{equation}
For $|a|,|a'|\ll|e|\ll|d|$, the heavy entries give $y_\tau\simeq y_b$ and $y_\mu\simeq3y_s$. The first-family masses arise from the hierarchical light block: $y_d\simeq|aa'/e|$ and $y_e\simeq|aa'/(3e)|$. The same Clebsch factor $-3$ coming from the coupling with $\Phi_{15}$ that gives the second-generation factor therefore generates the inverse factor $1/3$ for the first generation through diagonalization~\cite{Georgi:1979df}. Consequently,
\begin{equation}
R_1\simeq R_2\simeq R_3\simeq1.
\label{eq:leading_three_relations}
\end{equation}
All three characteristic quark--lepton relations can consequently arise from a remarkably sparse assignment of only two PS Yukawa structures.\\

A realistic completion also contains additional states associated with the Pati--Salam multiplets. The bidoublets contain extra electroweak doublets, while the adjoint contains color-octet doublets and scalar leptoquarks. We assume that only the SM Higgs combination remains light, with Pati--Salam-breaking interactions splitting the remaining components. For the minimal scalar content considered here in the $G_{421}$ theory, baryon number is conserved at the renormalizable level~\cite{Dorsner:2016wpm,Assad:2017iib}, while potentially dangerous higher-dimensional baryon-number-violating operators are sufficiently suppressed for a high cutoff scale~\cite{Murgui:2021bdy}. In the $G_{422}$ realization, however, dimension-five operators involving the Pati--Salam-breaking scalar $\chi_R\sim(4,1,2)$ can induce diquark couplings of the surviving color-triplet scalar. Assuming Planck-suppressed operators with order-one Wilson coefficients, nucleon-decay constraints can impose lower bounds on the Pati--Salam breaking scale of several $10^6$ GeV~\cite{Dutka:2022lug}. These proton-decay bounds therefore rely on assumptions about the Wilson coefficients of Planck-suppressed operators and cannot be regarded as unavoidable lower limits on the Pati--Salam breaking scale.

These ingredients generically perturb the simple relations above. Flavor mixing modifies the connection between entrywise Clebsch factors and physical masses, while split Pati--Salam multiplets generate threshold corrections at the symmetry-breaking scale. We therefore turn next to the size of these effects and to the extent to which the common low scale suggested by the leading relations survives in realistic realizations.

\head{Corrections in realistic theories}

The relations in Eq.~\eqref{eq:leading_three_relations} follow from the leading hierarchical texture, including the diagonalization of its $1$--$2$ block. In a realistic realization they are modified in two stages. First, the full three-generation flavor structure shifts the tree-level singular-value ratios away from $R_i=1$. Second, finite matching contributions from the split PS spectrum further modify these ratios. At the matching scale $M$, the ratios obtained from SM running satisfy
\begin{equation}
R_i(M)=\left(1+\Delta_i(M)\right)\left(1+\epsilon_i\right),
\label{eq:matching_condition}
\end{equation}
where $1+\Delta_i(M)$ are the tree-level ratios after flavor diagonalization at the scale M and $\epsilon_i$ are the threshold corrections, which depend on the masses and couplings of the heavy states. The corrections need not be independent: the same flavor entries and scalar interactions can affect several ratios.

\begin{center}
\textit{Flavor diagonalization}
\end{center}

The light block of Eq.~\eqref{eq:leading_PS_texture} reproduces $R_1 = 1$ at leading order, with subleading corrections arising from the mixing between the first two generations. Denoting by $\theta_{12}^{dL}$ and $\theta_{12}^{dR}$ the left- and right-handed $1$-$2$ rotation angles that diagonalize the down-type Yukawa matrix, one finds at leading order
\begin{equation}
\left|\theta_{12}^{dL}\theta_{12}^{dR}\right|\simeq\frac{y_d}{y_s}.
\end{equation}
If the Cabibbo angle is predominantly generated in the down sector and the two rotations have comparable magnitudes, this yields the familiar leading-order relation $|V_{us}|\simeq\sqrt{y_d/y_s}$~\cite{Gatto:1968ss}. With the quark masses now determined to the percent level of accuracy \cite{ParticleDataGroup:2024cfk,FlavourLatticeAveragingGroupFLAG:2024oxs}, a diagonalization beyond leading order is required, giving the exact $|V_{us}| = \sqrt{y_d/(y_d+y_s)} = 0.219 \pm 0.002$. The deviation of this from the experimental value of $0.2243 \pm 0.0008$ can easily originate from several sources. For instance, if $Y_u$ has the same light-block structure, its exact $1$--$2$ rotation is $\sin\theta_{12}^u=\sqrt{y_u/(y_u+y_c)}\simeq0.044$, and
\begin{align}
|V_{us}|^2=\frac{y_dy_c+y_uy_s-2\sqrt{y_dy_sy_uy_c}\,\cos\varphi}{(y_d+y_s)(y_u+y_c)}\,,
\end{align}
where $\varphi$ is the relative phase between the up and down light blocks; the measured value is reproduced for $\varphi\simeq(92\pm3)^\circ$, very suggestive of a spontaneous origin for CP violation~\cite{Arkani-Hamed:2026uos}~\footnote{Since a left--right $C$ symmetry enforces symmetric Yukawa matrices, and $P$ can be imposed as well, implying $CP$, $G_{422}$ offers a suggestive framework for flavor and spontaneous CP violation that deserves further study. Imposing $C$ or $P$ requires the left-handed partner of $\chi_R$, $\chi_L\sim(4,2,1)$, which we assume to be split from $\chi_R$ by $D$-parity breaking at a high scale~\cite{Chang:1983fu}, close to the cutoff of the theory.}. We find the prediction to be sufficiently significant that in this paper we restrict our attention to $Y_{d,e}$ matrices that are symmetric in the light $2 \times 2$ block and have a sufficiently small 11 entry. This gives
\begin{align}
\delta_{12}&=\frac49\left(|\theta_{12}^{dL}|^2+|\theta_{12}^{dR}|^2\right)
\simeq0.045,\nonumber\\
R_1&\simeq1+\delta_{12},\qquad
R_2\simeq1-\delta_{12}.
\label{eq:delta12_main}
\end{align}
The corrections are a few percent and have opposite signs. They cannot by themselves account for the first-generation deficit near $\mu_{2,3}$, but they are part of the correction pattern that a realistic theory must include.

Mixing with the third generation produces further deformations. Let the $23$ and $32$ charged-lepton entries differ from their down-quark partners by $C_{23},C_{32}\in\{1,-3\}$, and define
\begin{equation}
	\begin{split}
	\delta_{23} &\equiv	
		\left(1+\frac{C_{23}C_{32}}{3}\right)\operatorname{Re}\!\left[\frac{(Y_d)_{23}(Y_d)_{32}}
		{(Y_d)_{22}(Y_d)_{33}}
		\right]\,,\\
	\delta_3 &\equiv
	\frac{C_{23}^2-1}{2}\left|\theta_{23}^{dL}\right|^2
	+
	\frac{C_{32}^2-1}{2}\left|\theta_{23}^{dR}\right|^2\,.
	\end{split}
	\label{eq:main_23_delta3}
\end{equation}
For real entries, assuming that $y_s$ is dominated by a positive $(Y_d)_{22}$ entry, the hierarchical $2$-$3$ block gives 
\begin{equation}
    \frac{(Y_d)_{23}(Y_d)_{32}}
{(Y_d)_{22}(Y_d)_{33}}\simeq \frac{y_b}{y_s} \, \theta_{23}^{dL}\theta_{23}^{dR}.
\end{equation} 
The hierarchy $y_b/y_s\sim50$ allows CKM-sized rotations to generate ten-percent light-family corrections, whose sign depends on the interference. By comparison, $\delta_3$ has no hierarchy enhancement and is nonnegative: it vanishes for the $(1,1)$ assignment and equals $4(|\theta_{23}^{dL}|^2+|\theta_{23}^{dR}|^2)$ for $(-3,-3)$. The right-handed rotations and their phases are not fixed by the CKM matrix. Therefore, to leading order in the small mixing angles, the corrections are
\begin{align}
	\begin{split}
	\Delta_1&\simeq
	\delta_{12}-\delta_{23}\,,\\
	\Delta_2&\simeq-\delta_{12}
	+\delta_{23}-\delta_3\,,\quad
	\Delta_3\simeq\delta_3.
	\end{split}
	\label{eq:main_23_interference}
\end{align}

Determinants provide useful checks on these correlations. The isolated $12$ block obeys $R_1R_2=1$ exactly. With common singlet $12$, $21$, and $33$ entries and $11=13=31=0$, the three-family texture instead obeys $R_1R_2R_3=1$ at tree level, including nonzero $23$ and $32$ entries. 

Nonzero $13$ and $31$ entries can modify the determinant relation and generate additional corrections predominantly to $R_1$. For an approximately symmetric $1$--$2$ block, the fractional first-generation corrections involving one $1$--$3$ and one $2$--$3$ rotation scale as
\begin{equation}
\Delta_1
\sim
\frac{y_b}{\sqrt{y_dy_s}}
\left(
\theta_{13}^{dL}\theta_{23}^{dR}
+
\theta_{23}^{dL}\theta_{13}^{dR}
\right),
\end{equation}
while the direct $13$--$31$ contribution scales as
$y_b/y_d\,
\theta_{13}^{dL}\theta_{13}^{dR}$.
The large ratios $y_b/\sqrt{y_dy_s}\sim200$ and $y_b/y_d\sim10^3$ can therefore compensate for small mixing angles, allowing even CKM-sized rotations to induce percent-level shifts in $R_1$.
The corresponding corrections to $R_2$ remain sub-percent for CKM-sized rotations, and those to $R_3$ are smaller still. These entries therefore provide additional freedom in the first-generation relation, although their contributions to the light masses must remain subleading if the $1$--$2$ seesaw is to retain its leading role. Explicit expressions are given in Appendix~\ref{app:Flavor}, together with a discussion of an alternative texture in which the strange-quark mass is also generated by a seesaw mechanism.

\begin{center}
\textit{Threshold corrections}
\end{center}

We evaluate finite threshold corrections in the tree-level fermion mass basis. Writing
$U_f^\dagger Y_fV_f=\operatorname{diag}(y_{f_1},y_{f_2},y_{f_3})$
with positive singular values, the strictly one-loop shifts are
\begin{equation}
    \delta y_{f_i}
    =
    \operatorname{Re}\!\left[
    \left(U_f^\dagger\delta Y_fV_f\right)_{ii}
    \right] + {\cal O}(\delta Y_f^2).
    \label{eq:threshold_mass_shift_main}
\end{equation}
Thus the tree-level matrices are diagonalized first, and only the diagonal elements of the rotated correction enter the masses at this order. Off-diagonal corrections affect the singular values only at quadratic order in $\delta Y_f$. The threshold correction in Eq.~\eqref{eq:matching_condition} is then, at one loop,
\begin{equation}
\epsilon_i=\frac{\delta y_{e_i}}{y_{e_i}}-\frac{\delta y_{d_i}}{y_{d_i}}.
\label{eq:epsilon_definition}
\end{equation}
A model independent contribution is given by self-energy diagrams with the exchange of a vector-leptoquark. Matching at $\mu\simeq M_X$, the leading generation-independent vector-leptoquark contribution is
\begin{equation}
\epsilon_i^X=\frac{5g_4^2}{32\pi^2}\simeq0.01,
\label{eq:vector_threshold_main}
\end{equation}
where $g_4 = \sqrt{4\pi \alpha_s}$.
The neutral vector $Z'$ gives a smaller additional contribution. Larger effects can arise from scalar self-energy diagrams with a chirality flip on the internal fermion line. A charged scalar and an up-type quark correct the down-quark mass, while the corresponding charged-lepton diagram contains a neutral fermion. In the exact PS limit, the two contributions are identical and cancel in $R_i$. However, as discussed above, a realistic neutrino sector separates the light-neutrino masses from the up-quark masses and supplies heavy neutral partners $S_j$. Their mixing satisfies $m_{S_j}\sin\theta_j=m_{D_j}$: the chirality-changing numerator therefore remains proportional to the PS-related Dirac mass, while the heavy-neutral-fermion mass enters the loop function.

For the texture discussed in Eq.~\eqref{eq:leading_PS_texture}, the $33$ entries arise from the singlet bidoublet $\Phi_1$, implying $m_{D_3}=m_t$. The dominant charged-scalar contribution to the third-generation ratio is then
\begin{align}
\begin{split}
	&\epsilon_3^H
	\simeq
	-\frac{y_t\,(\widehat B_{\Phi_1})_{33}(\widehat A_{\Phi_1})_{33}}
	{16\pi^2y_b} 
	\left[
	F\!\left(\frac{m_{S_3}^2}{m_H^2}\right)
	-
	F\!\left(\frac{m_t^2}{m_H^2}\right)
	\right]\,,
\end{split}
	\label{eq:threshold_top_main}
\end{align}
where $(\widehat B_{\Phi_1})_{33}(\widehat A_{\Phi_1})_{33}$ are the couplings at the two charged-scalar vertices, defined in Appendix~\ref{app:PStheory}, $F(r)=r\ln r/(r-1)$, and $F(1)=1$. The two functions arise from the charged-lepton and down-quark diagrams, respectively, and cancel when the top and its neutral partner are degenerate. With a split spectrum, the enhancement $y_t/y_b$ permits $\mathcal{O}(10\%)$ corrections with perturbative couplings~\cite{Volkas:1995yn}. The two independent bidoublet Yukawa invariants of the $422$ theory allow either sign of $(\widehat B_{\Phi_1})_{33}(\widehat A_{\Phi_1})_{33}$, including the positive correction needed for matching above the heavier-generation crossings. In the corresponding $421$ realization, the analogous dominant contribution is instead strictly negative. Details are provided in Appendix~\ref{app:thresholds}.

The adjoint sector provides additional charged-scalar and scalar-leptoquark diagrams. Among the leading chirality-enhanced contributions, a charge-$5/3$ leptoquark and an up-type quark correct the charged-lepton mass, while a leptoquark with charge $-1/3$ and a neutral fermion correct the down-quark mass. A small electroweak vev of $\Phi_{15}$ enhances its underlying Yukawa couplings, allowing sizable charm-mediated corrections to the second generation. Suitable flavor-changing couplings of both chiralities can instead produce the larger enhancement $y_t/y_s$. These corrections also feed into the first generation through the $1$--$2$ mixing. The charged-scalar and scalar-leptoquark corrections share the parametric structure of Eq.~\eqref{eq:threshold_top_main}, with
the appropriate flavor indices, order-one Clebsch and color factors,
and spectrum-dependent loop functions. For the leading texture, a deformation confined to the $22$ entry preserves the determinant of the light-family block, giving opposite fractional shifts of its two singular values and hence $\epsilon_1\simeq-\epsilon_2$. This correlation need not hold for more general loop-induced deformations.

The scalar couplings and threshold expressions are collected in Appendices~\ref{app:PStheory} and~\ref{app:thresholds}. Throughout the remainder of this work, we assume that the new particles lie near the PS scale and include their finite threshold corrections in a common-scale matching. Large mass splittings would instead require renormalization-group evolution through the appropriate intermediate effective theories.\\

\head{Phenomenological Implications}

\begin{figure*}[t]
    \centering
    \includegraphics[width=1.0\linewidth]{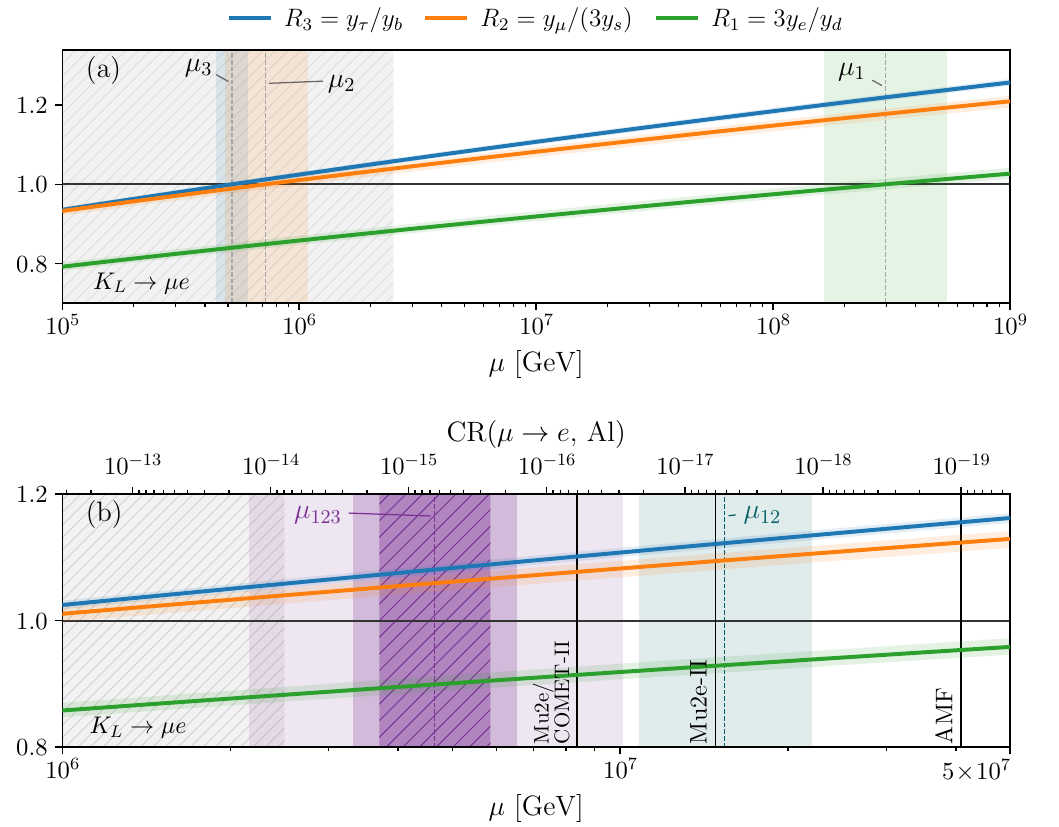}
    \caption{Two-loop SM running of the normalized Yukawa ratios $R_3=y_\tau/y_b$, $R_2=y_\mu/(3y_s)$, and $R_1=3y_e/y_d$, with $68\%$ C.L.\ bands from the input uncertainties at $m_Z$. {\it Upper panel (a):}~Running between $10^5$ and $10^9$~GeV; dashed vertical lines and bands mark the unit crossings $\mu_3$, $\mu_2$, and $\mu_1$ and their $68\%$ C.L.\ intervals.  {\it Lower panel (b):}~Enlargement of the region $10^6$--$5\times10^7$~GeV of panel (a), showing the scales $\mu_{123}$ and $\mu_{12}$, defined by $R_1R_2R_3=(1+\epsilon_g^X)^3$ and $R_1R_2=(1+\epsilon_g^X)^2$, respectively. The teal band is the $68\%$ C.L.\ interval of $\mu_{12}$. The darker violet band is the $68\%$ C.L.\ interval of $\mu_{123}$ without a third-generation threshold, $\epsilon_3^H=0$, reflecting only the input uncertainties propagated through the RG evolution; the other two violet bands are the envelopes of the $68\%$ C.L.\ intervals of $\mu_{123}$ for $|\epsilon_3^H|\le1\%$ and $|\epsilon_3^H|\le5\%$, respectively. In both panels the hatched region is excluded by the experimental bound on $K_L\to\mu^\pm e^\mp$~\cite{BNL:1998apv}. The upper axis of (b) gives the corresponding $\mu\to e$ conversion rate in aluminum, $\mathrm{CR}(\mu\to e,\,\mathrm{Al})$, for a vector-leptoquark mass $M_X=\mu$. The solid lines in (b) mark the projected $\mu\to e$ conversion reaches in aluminum of Mu2e and COMET Phase~II~\cite{COMET:2018auw,Mu2e:2022ggl,COMET:2025sdw}, Mu2e-II~\cite{Mu2e-II:2022blh}, and the Advanced Muon Facility (AMF)~\cite{CGroup:2022tli}. The bounds and the conversion-rate axis are evaluated with the light-family vector-leptoquark couplings common to the $123$ and $12$ theories, fixed by the symmetric $1$--$2$ block, with $M_X$ identified with the PS matching scale.}
    \label{fig:phen_running}
\end{figure*}

The running Yukawa couplings suggest a striking possibility: quarks and leptons may unify at scales closer to $10^6$~GeV. Such low scales are incompatible with simple unified embeddings in which gauge interactions induce rapid proton decay. PS theories are qualitatively different. The embedding $SU(3)_C\subset SU(4)_C$ introduces vector leptoquarks without the gauge-mediated baryon number violation characteristic of simple GUTs. As we discussed, the scalar spectrum and flavor structure naturally generate corrections to the quark--lepton Yukawa relations. These observations raise precise questions: how low can the scale of quark--lepton unification actually be, what do realistic theories look like, and how can these possibilities be tested?

Even in PS theories, scales this low are directly probed by rare flavor-violating processes~\cite{Gedeonova:2022,Valencia:1994cj,Kuznetsov:1994,Smirnov:2018,Dolan:2020doe,Iguro:2021kdw,Iguro:2022ozl,Debnath:2026zow,Babu:2025,CMS:2024LQ,CMS:2025LQ}. In particular, tree-level vector-leptoquark exchange induces $K_L\to\mu^\pm e^\mp$ even in the absence of quark--lepton flavor mixing, yielding a lower bound of a few $\times10^6$~GeV for unsuppressed light-family couplings, as discussed in Appendix~\ref{app:Kmue}. This already lies above the nearly coincident crossing scales $\mu_2 \sim \mu_3$ suggested by the second- and third-generation Yukawa relations.

One possibility for reconciling a PS scale near $\mu_{2,3}$ with the kaon bound is to give the first-generation charged-lepton mass an independent origin. Mixing with additional vector-like fermions has been shown to substantially relax the rare-meson constraints on Pati--Salam unification~\cite{Calibbi:2017qbu,Dolan:2020doe,Iguro:2021kdw}. As illustrated in Appendix~\ref{app:chiralKmue}, this can suppress both chiral couplings to the vector leptoquark while preserving, at leading order, the muon--strange and tau--bottom relations. The kaon constraint can then be satisfied near $\mu_{2,3}$, retaining the intriguing heavier-generation coincidence at the price of relinquishing the first-generation Yukawa prediction. This remains a very interesting possibility. Since $K_L\to\mu^\pm e^\mp$ and $\mu\to e$ conversion are then suppressed, the leading probes become processes involving only muons, such as $K_{L,S}\to\mu^+\mu^-$, including the $K_S$--$K_L$ interference in the time-dependent $K \to \mu^+\mu^-$ rate~\cite{DAmbrosio:2017klp,Dery:2021mct,Dery:2021vql,Dery:2022yqc}.

More strikingly, we now show that it is possible to retain all three Yukawa relations. \hyperref[fig:phen_running]{Fig.~\ref*{fig:phen_running}(a)} shows that the three ratios remain close to unity over a remarkably low range of scales. If flavor and threshold corrections are restricted to the $\mathcal O(20\%)$ level, all three relations can be accommodated between roughly $10^5$ and $10^9$~GeV, many orders of magnitude below conventional grand-unification scales. The upper end of this window and certainly even higher scales are however difficult to reach with the simple hierarchical texture discussed above. Above $\mu_1$ all three ratios exceed unity, so that $R_1 R_2$ departs appreciably from one. Since the first-generation masses arise from a seesaw in the light block, the corrections that preserve its determinant shift $R_1$ and $R_2$ in opposite directions and cannot move both toward unity. Nonetheless, the window should not be interpreted as a model-independent upper bound on the PS scale: larger corrections or more elaborate textures can extend the viable region. Rather, it identifies the range in which quark--lepton unification can emerge with particularly modest deformations of the zeroth-order Yukawa relations. Within this range, two simple theories stand out. In each, the flavor structure and the choice of Clebsch coefficients leave no free flavor parameters, so that the measured masses fix the unification scale.\\

In the ``123 theory", the deformations arise entirely from flavor diagonalization. For the texture of Eq.~\eqref{eq:leading_PS_texture} supplemented only by $23$ and $32$ entries, $\det Y_e=\det Y_d$ implies the exact tree-level relation $R_1R_2R_3=1$. Including the unavoidable vector threshold of Eq.~\eqref{eq:vector_threshold_main}, the matching condition becomes $R_1R_2R_3=(1+\epsilon_g^X)^3$. The previous identity selects, at $68\%$ C.L.,
\begin{equation}
\mu_{123}=4.6^{+1.2}_{-0.9}\times10^6~{\rm GeV}.
\label{eq:mu123}
\end{equation}
The $1$--$2$ block is symmetric, as suggested by $|V_{us}|\simeq\sqrt{m_d/m_s}$, so the light-family correction is fixed, $\delta_{12}\simeq\frac89\,m_d/m_s\simeq0.045$. For $(C_{23},C_{32})=(-3,-3)$, the remaining two ratios then determine both $2$--$3$ rotations, so that the scale and all flavor parameters of the texture are fixed: exact diagonalization reproduces the corresponding tree-level ratios $(R_1,R_2,R_3)/(1+\epsilon_g^X)\simeq(0.891,1.049,1.070)$, evaluated at $\mu_{123}$, for $\theta_{23}^{dR}\simeq0.135$ and $\theta_{23}^{dL}\simeq0.009$. Only this assignment is viable: for $(1,1)$, $\delta_3=0$ leaves $R_3=1$, which cannot match the required $R_3\simeq1.07$, while the mixed assignments, $(1,3)$ and $(3,1)$, give no hierarchy-enhanced $\delta_{23}$, leaving $R_1\simeq1+\delta_{12}>1$, opposite to the required $R_1<1$. The small left-handed rotation is consistent with $V_{cb}$ arising predominantly from the up-quark sector, which disfavors the solution with the two rotations exchanged. A residual charged-scalar top loop, Eq.~\eqref{eq:threshold_top_main}, modifies only $R_3$ at leading order. For $|\epsilon_3^H|\le1\%$ and $5\%$, as motivated by $m_{S_3}/m_H\sim\mathcal{O}(1)$ and small $\sin\beta$, the envelopes of the $68\%$ C.L.\ intervals span $\mu_{123}\simeq(3.3$--$6.5)\times10^6$~GeV and $(2.2$--$10)\times10^6$~GeV, respectively, shown by the dark and light violet bands in \hyperref[fig:phen_running]{Fig.~\ref*{fig:phen_running}(b)}. The ``123 theory'' can therefore be realized in both $G_{422}$ and $G_{421}$ through flavor diagonalization alone, without requiring additional scalar threshold corrections. Such corrections can nevertheless shift the preferred matching scale, with $\epsilon_3^H\leq0$ in $G_{421}$.

In the ``12 theory", all Yukawa matrices are symmetric, as can be enforced by a left--right $C$ symmetry in $G_{422}$. This accounts for $|V_{us}|\simeq\sqrt{m_d/m_s}$ and, with small $23$ and $32$ entries, leaves the third-generation relation to be corrected by the charged-scalar threshold. In this case $\delta_3$ is negligible and the light block is effectively isolated, so that the tree-level relation $R_1R_2=1$ holds. As discussed, threshold corrections to the $22$ entry shift $R_1$ and $R_2$ in opposite directions, leaving the light-block determinant unchanged. Including the vector threshold, we are therefore left with the matching condition $R_1R_2=(1+\epsilon_g^X)^2$. Parametrizing the net deviation of the individual ratios by $x_{12}$, which incorporates both flavor diagonalization and potential scalar threshold corrections, we get
\begin{align}
\begin{split}
R_1(\mu_{12})&=(1+\epsilon_g^X)(1-x_{12})\,,\\
R_2(\mu_{12})&=\frac{1+\epsilon_g^X}{1-x_{12}}.
\end{split}
\label{eq:12_shift}    
\end{align}
Since both ratios increase monotonically with the renormalization scale, this condition selects, at $68\%$ C.L.,
\begin{equation}
\mu_{12}=1.5^{+0.7}_{-0.5}\times10^7~{\rm GeV},
\qquad
x_{12}\simeq0.08.
\label{eq:mu12}
\end{equation}
With the symmetric light block, $\delta_{12}\simeq0.045$. Neglecting potential scalar threshold corrections to the light block, the required light-family shift must be supplied by the interference term of Eq.~\eqref{eq:main_23_interference}, $\delta_{23}=\delta_{12}+x_{12}\simeq0.12$. For $(C_{23},C_{32})=(1,1)$, $\delta_3$ vanishes and this condition fixes only the product of the two $2$--$3$ rotations, $\theta_{23}^{dL}\theta_{23}^{dR}\simeq\frac34(y_s/y_b)\,\delta_{23}$, which is of order $|V_{cb}|^2$. Since $Y_d$ is symmetric, the two rotations are equal and exact diagonalization gives $\theta_{23}^{dL}=\theta_{23}^{dR}=0.045$; the leading-order relation underestimates the angle by about 5\%. This predicts, after running to low energies,
\begin{equation}
|V_{cb}|=0.042\pm0.002,
\label{eq:Vcb_prediction}
\end{equation}
in agreement with the measured value $0.0411\pm0.0012$~\cite{ParticleDataGroup:2024cfk}, provided the up sector contributes negligibly to $2$--$3$ mixing. The uncertainty is dominated by the light quark masses, through $m_d/m_s$, and would shrink correspondingly with improved lattice determinations. At this scale $R_3/(1+\epsilon_g^X)\simeq1.11$, which in the $G_{422}$ theory the charged-scalar top loop of Eq.~\eqref{eq:threshold_top_main} can provide for $(174\,{\rm GeV}/u)^2\sin\beta\,F(m_{S_3}^2/m_H^2)\simeq0.35$, where $u$ is the electroweak vev of $\Phi_1$ and $ \beta$ parametrizes its two neutral vev components. For $u\simeq v$ and $m_{S_3}/m_H$ between $2$ and $10$, this requires $\sin\beta\simeq0.07$--$0.17$. The same values of $\sin\beta$ make the up-sector condition for the $|V_{cb}|$ prediction mild. In $G_{422}$ the $\widetilde\lambda_1$ contribution to $(Y_u)_{23}$ is suppressed by $\sin\beta$, while the $\lambda_1$ contribution to the up rotation is suppressed relative to its down-sector counterpart by $y_b/(y_t\tan\beta)$, a factor of a few to ten. The prediction \eqref{eq:Vcb_prediction} then holds to the accuracy quoted provided $|(\lambda_1)_{23}|\lesssim3\,|(\widetilde\lambda_1)_{23}|$.\\

Both theories unify all three generations between a few $\times10^6$ and order $10^7$~GeV, with only two Yukawa structures, symmetric light blocks, and corrections of about ten percent to the leading Yukawa relations. The lowest scales allowed for $\mu_{123}$ reach the present $K_L\to\mu^\pm e^\mp$ bound~\cite{BNL:1998apv}, and both theories lie within reach of upcoming charged-lepton-flavor-violation searches. Since the same flavor structure fixes the leading vector-leptoquark couplings, these searches directly test the scale inferred from the Yukawa relations. The symmetric $1$--$2$ block, motivated by $|V_{us}|\simeq\sqrt{m_d/m_s}$, determines the rotations between the flavor and mass bases of both the down quarks and the charged leptons, and hence the vector-leptoquark couplings of the light families. Because of the Clebsch factor $-3$ in the $22$ entry, the two rotations proceed in opposite directions and differ by $\alpha \simeq\frac{4}{3}\sqrt{m_d/m_s}\simeq0.3$, with approximately equal magnitudes in both chiralities. Interestingly, despite the nontrivial light-family mixing, the leading chirality-enhanced $K_L\to\mu^\pm e^\mp$ amplitude becomes independent of $\alpha$ and therefore unsuppressed, as shown in Appendix~\ref{app:Kmue}, up to $\mathcal O(10\%)$ corrections from $2$--$3$ mixing.  The same couplings induce coherent $\mu\to e$ conversion in nuclei at tree level. Retaining only the down-quark contribution, we find~\cite{Kitano:2002mt} (see Appendices~\ref{app:Kmue} and~\ref{app:future_constraints})
\begin{equation}
    \begin{split}
        &\mathrm{CR}(\mu\,{\rm Al}\to e\,{\rm Al})\simeq2.8\times10^{-17}\,(10^7\,{\rm GeV}/M_X)^4\\
        &\mathrm{Br}(K_L\to\mu^\pm e^\mp)\simeq1.7\times10^{-14}\,(10^7\,{\rm GeV}/M_X)^4
    \end{split}
\end{equation}
The present kaon bound, $4.7\times10^{-12}$~\cite{BNL:1998apv}, implies $M_X\gtrsim2.5\times10^6$~GeV, whereas the SINDRUM~II limit on gold, ${\rm CR}_{\rm Au}<7\times10^{-13}$~\cite{SINDRUMII:2006dvw}, implies only $M_X\gtrsim9\times10^5$~GeV. The next generation of experiments changes this picture. Mu2e and COMET Phase~II, aiming at ${\rm CR}_{\rm Al}\simeq6\times10^{-17}$ at $90\%$ C.L.~\cite{COMET:2018auw,Mu2e-II:2022blh}, reach $M_X\simeq8\times10^6$~GeV and cover most of the $\mu_{123}$ range; the proposed Mu2e-II upgrade, with ${\rm CR}_{\rm Al}\simeq6\times10^{-18}$~\cite{Mu2e-II:2022blh}, reaches $1.45\times10^7$~GeV, covering $\mu_{123}$ entirely and reaching the central value of $\mu_{12}$; and an aluminum-equivalent sensitivity ${\rm CR}_{\rm Al}\sim10^{-19}$, as envisaged for the Advanced Muon Facility~\cite{CGroup:2022tli}, would reach about $4\times10^7$~GeV and cover both theories (\hyperref[fig:phen_running]{Fig.~\ref*{fig:phen_running}(b)}). Additional scalar-exchange contributions, potentially enhanced for a small adjoint vev, are not included.

\head{Outlook}
The measured fermion masses may be revealing the scale of quark--lepton unification. In canonical unified theories, individual renormalizable Yukawa structures relate down quarks and charged leptons only through the Clebsch coefficients $1$ and $-3$, which in PS arise from $SU(4)_C$ singlet and adjoint scalars, respectively. In a simple hierarchical texture these coefficients survive in the physical masses as the leading relations $y_\tau=y_b$, $y_\mu=3y_s$ and, through a seesaw in the light block, $y_e=y_d/3$. Two-loop SM running realizes the first two near $10^6$~GeV, at $\mu_3\simeq5\times10^5$~GeV and $\mu_2\simeq7\times10^5$~GeV, and the third at $\mu_1\simeq3\times10^8$~GeV. Such low scales single out PS as a natural UV completion, which unifies quarks and leptons without gauge-mediated proton decay.

If the electron mass has an independent origin, the second- and third-generation relations can be realized in theories with a unification scale as low as $\mu_{2,3}\simeq(5$--$7)\times10^5$~GeV, with processes such as $K_{L,S}\to\mu^+\mu^-$ as leading probes.

A common scale for all three relations requires departures of only about ten percent in the Yukawa ratios, the natural size of the flavor-mixing and threshold corrections that the same theories generate. With corrections of $\mathcal{O}(20\%)$, the relations can be accommodated over roughly $10^5$--$10^9$~GeV, a region that larger corrections could extend.

Two simple theories make this picture concrete, with the flavor structure fixed by the measured masses. In the first, the ``$123$ theory", flavor diagonalization corrects all three relations and selects the unification scale $\mu_{123}\simeq5\times10^6$~GeV. In the second, the ``$12$ theory", symmetric Yukawa matrices select $\mu_{12}\simeq1.5\times10^7$~GeV, with a charged-scalar threshold of about $11\%$ completing the third-generation relation. This theory predicts $|V_{cb}|=0.042\pm0.002$ when the up sector contributes little to $2$--$3$ mixing.

The precision of these predictions is limited mainly by the light down-type quark masses. The unification scales depend on the product $m_dm_s$, through $R_1R_2\propto y_\mu y_e/(y_dy_s)$, and their uncertainties come almost entirely from the percent-level errors on $m_d$ and $m_s$, which account for about $90\%$ of the variance of $\mu_{12}$ and $75\%$ of that of $\mu_{123}$, the remainder coming mostly from $\alpha_s$. The $|V_{cb}|$ prediction depends instead on the ratio $m_d/m_s$. Reducing the uncertainties on $m_d$ and $m_s$ would sharpen the region to be probed by $\mu\to e$ conversion, and would turn the $|V_{cb}|$ prediction into a more stringent test of the second theory. Improved lattice determinations of the light-quark masses are therefore directly relevant to testing low-scale quark--lepton unification.

The simplicity of these theories is part of their appeal. The Yukawa couplings are among the most precisely measured and yet least understood parameters of the Standard Model, and attempts to explain them often require navigating a vast parameter space and many competing mechanisms. Even if these theories are not unique, there is real value in finding simple constructions: it is fascinating that the apparently disordered pattern of fermion masses may be the low-energy imprint of a simple order in the UV.

Most excitingly, this possibility is not experimentally remote. The vector leptoquarks of PS symmetry mediate rare flavor-violating processes at tree level, and in both theories their light-family couplings are fixed at leading order by the same flavor structure that accounts for the Yukawa relations. The present $K_L\to\mu^\pm e^\mp$ bound already reaches the lower edge of the $\mu_{123}$ region. Mu2e and COMET-II are expected to probe most of this region, the proposed Mu2e-II upgrade could cover it entirely and reach into the $\mu_{12}$ region, and the sensitivity envisaged for the Advanced Muon Facility would cover both theories. Electric dipole moments offer a complementary probe of additional CP-violating flavor structure. That the scale of quark--lepton unification may lie within present or foreseeable experimental reach makes the possibility suggested by the measured fermion masses even more intriguing and motivates a dedicated and exciting experimental program.\\

\noindent {\bf Acknowledgements.}---We thank Nima Arkani-Hamed, David Brown, Yuval Grossman, Alberto Ramos, and Mark Wise for helpful discussions. The work of L.J.H. was supported by Department of Energy grant DE-AC02-05CH11231 and NSF grant PHY-2515115. The work of C.A.M. is supported by the Friends of the Institute for Advanced Study. Our numerical analyses were carried with the help of OpenAI Codex and Claude Code.

\bibliographystyle{apsrev4-2}
\bibliography{refs}



\clearpage

\let\addcontentsline\oldaddcontentsline
\onecolumngrid
\begin{center}
  \textbf{\large Supplementary Material for Low Scale Quark--Lepton Unification}\\[.2cm]
  \vspace{0.05in}
  {Lawrence John Hall, Claudio Andrea Manzari and Clara Murgui}
\end{center}

\setcounter{equation}{0}
\setcounter{figure}{0}
\setcounter{table}{0}
\setcounter{section}{0}
\setcounter{page}{1}

\makeatletter
\renewcommand{\theequation}{S\arabic{equation}}
\renewcommand{\thefigure}{S\arabic{figure}}
\renewcommand{\thetable}{S\arabic{table}}
\makeatother 

\setcounter{secnumdepth}{2}
\renewcommand{\thesection}{\Roman{section}}
\renewcommand{\thesubsection}{\thesection.\alph{subsection}}

\begingroup
\renewcommand{\tocname}{Contents of the Supplemental Material}
\makeatletter
\let\l@subsection\@gobbletwo
\let\l@subsubsection\@gobbletwo
\makeatother
\tableofcontents
\endgroup
\medskip

\section{Two-loop Standard Model renormalization group equations}
\label{app:rge}

We collect here the two-loop SM renormalization-group equations used
throughout this work, for the gauge couplings $g_i$ ($i=1,2,3$, with
$g_1$ in $SU(5)$/GUT normalization,
$g_1^2=\tfrac53 g^{\prime\,2}$), the Yukawa matrices
$Y_f$ ($f=u,d,l$), and the Higgs quartic $\lambda$, with
$t=\ln\mu$. We follow the two-loop results of Refs.~\cite{Luo:2002ti,Luo:2002ey,Buttazzo:2013uya}, with $n_g=3$ generations. We also define
\begin{align}
\begin{split}
Y_2 &= \mathrm{Tr}\!\left(
3Y_u^\dagger Y_u+3Y_d^\dagger Y_d+Y_e^\dagger Y_e
\right),\\
H &= \mathrm{Tr}\!\left(
3(Y_u^\dagger Y_u)^2+3(Y_d^\dagger Y_d)^2
+(Y_e^\dagger Y_e)^2
\right),\\
Y_4 &=
\left(\tfrac{17}{20}g_1^2+\tfrac94g_2^2+8g_3^2\right)
\mathrm{Tr}(Y_u^\dagger Y_u)\quad+
\left(\tfrac14g_1^2+\tfrac94g_2^2+8g_3^2\right)
\mathrm{Tr}(Y_d^\dagger Y_d)\quad+
\left(\tfrac34g_1^2+\tfrac34g_2^2\right)
\mathrm{Tr}(Y_e^\dagger Y_e),\\
\chi_4 &= \tfrac94\mathrm{Tr}\Big[
3(Y_u^\dagger Y_u)^2+3(Y_d^\dagger Y_d)^2
+(Y_e^\dagger Y_e)^2\quad
-\tfrac13\{Y_u^\dagger Y_u,\,Y_d^\dagger Y_d\}
\Big].
\end{split}
\end{align}

\textbf{\textit{Yukawa couplings.}}---
At one loop,
\begin{align}
\begin{split}
16\pi^2\frac{dY_u^{(1)}}{dt}
&=Y_u\Big[
\tfrac32Y_u^\dagger Y_u-\tfrac32Y_d^\dagger Y_d+Y_2\mathbf 1 
-\left(\tfrac{17}{20}g_1^2+\tfrac94g_2^2+8g_3^2\right)\mathbf{1}
\Big],\\
16\pi^2\frac{dY_d^{(1)}}{dt}
&=Y_d\Big[
\tfrac32Y_d^\dagger Y_d-\tfrac32Y_u^\dagger Y_u+Y_2\mathbf 1 -\left(\tfrac14g_1^2+\tfrac94g_2^2+8g_3^2\right)\mathbf{1}
\Big],\\
16\pi^2\frac{dY_e^{(1)}}{dt}
&=Y_e\Big[
\tfrac32Y_e^\dagger Y_e+Y_2\mathbf 1
-\left(\tfrac94g_1^2+\tfrac94g_2^2\right)\mathbf{1}
\Big].
\end{split}
\end{align}

At two loops,
\begin{align}
\begin{split}
(16\pi^2)^2\frac{dY_u^{(2)}}{dt}
&=Y_u\bigg[
\tfrac32(Y_u^\dagger Y_u)^2
-(Y_u^\dagger Y_u)(Y_d^\dagger Y_d)
-\tfrac14(Y_d^\dagger Y_d)(Y_u^\dagger Y_u)
+\tfrac{11}4(Y_d^\dagger Y_d)^2
+Y_2\!\left(\tfrac54Y_d^\dagger Y_d-\tfrac94Y_u^\dagger Y_u\right)
 \\
&\quad
-\chi_4\mathbf{1}
+6\lambda^2\mathbf{1}
-12\lambda Y_u^\dagger Y_u
+\tfrac52Y_4\mathbf{1}
+\left(\tfrac{223}{80}g_1^2+\tfrac{135}{16}g_2^2+16g_3^2\right)
Y_u^\dagger Y_u
 \\
&\quad
-\left(\tfrac{43}{80}g_1^2-\tfrac9{16}g_2^2+16g_3^2\right)
Y_d^\dagger Y_d
+\left(\tfrac9{200}+\tfrac{29}{45}n_g\right)g_1^4\mathbf{1}
-\tfrac9{20}g_1^2g_2^2\mathbf{1}
+\tfrac{19}{15}g_1^2g_3^2\mathbf{1}
 \\
&\quad
-\left(\tfrac{35}4-n_g\right)g_2^4\mathbf{1}
+9g_2^2g_3^2\mathbf{1}
-\left(\tfrac{404}3-\tfrac{80}9n_g\right)g_3^4\mathbf{1}
\bigg],\\[4pt]
(16\pi^2)^2\frac{dY_d^{(2)}}{dt}
&=Y_d\bigg[
\tfrac32(Y_d^\dagger Y_d)^2
-(Y_d^\dagger Y_d)(Y_u^\dagger Y_u)
-\tfrac14(Y_u^\dagger Y_u)(Y_d^\dagger Y_d)
+\tfrac{11}4(Y_u^\dagger Y_u)^2
+Y_2\!\left(\tfrac54Y_u^\dagger Y_u-\tfrac94Y_d^\dagger Y_d\right)
 \\
&\quad
-\chi_4\mathbf{1}
+6\lambda^2\mathbf{1}
-12\lambda Y_d^\dagger Y_d
+\tfrac52Y_4\mathbf{1}
+\left(\tfrac{187}{80}g_1^2+\tfrac{135}{16}g_2^2+16g_3^2\right)
Y_d^\dagger Y_d
 \\
&\quad
-\left(\tfrac{79}{80}g_1^2-\tfrac9{16}g_2^2+16g_3^2\right)
Y_u^\dagger Y_u
-\left(\tfrac{29}{200}+\tfrac1{45}n_g\right)g_1^4\mathbf{1}
-\tfrac{27}{20}g_1^2g_2^2\mathbf{1}
+\tfrac{31}{15}g_1^2g_3^2\mathbf{1}
 \\
&\quad
-\left(\tfrac{35}4-n_g\right)g_2^4\mathbf{1}
+9g_2^2g_3^2\mathbf{1}
-\left(\tfrac{404}3-\tfrac{80}9n_g\right)g_3^4\mathbf{1}
\bigg],\\[4pt]
(16\pi^2)^2\frac{dY_e^{(2)}}{dt}
&=Y_e\bigg[
\tfrac32(Y_e^\dagger Y_e)^2
-\tfrac94Y_2Y_e^\dagger Y_e
-\chi_4\mathbf{1}
+6\lambda^2\mathbf{1}
-12\lambda Y_e^\dagger Y_e
+\tfrac52Y_4\mathbf{1}
 \\
&\quad
+\left(\tfrac{387}{80}g_1^2+\tfrac{135}{16}g_2^2\right)
Y_e^\dagger Y_e
+\left(\tfrac{51}{200}+\tfrac{11}{5}n_g\right)g_1^4\mathbf{1}
+\tfrac{27}{20}g_1^2g_2^2\mathbf{1}
-\left(\tfrac{35}4-n_g\right)g_2^4\mathbf{1}
\bigg].
\end{split}
\end{align}

\textbf{\textit{Higgs quartic.}}---
With the Higgs quartic normalized as $V(\Phi)\supset\lambda\,(\Phi^\dagger\Phi)^2$,
so that $m_h^2=2\lambda v_{\rm SM}^2$ at tree level, and keeping only the diagonal third-generation Yukawas $y_t,y_b,y_\tau$, at one loop
\begin{align}
16\pi^2\frac{d\lambda}{dt}
&=24\lambda^2
+12\lambda\left(y_t^2+y_b^2\right)+4\lambda y_\tau^2
-9\lambda g_2^2-\tfrac95\lambda g_1^2
-6(y_t^4 + y_b^4)-2y_\tau^4
+\tfrac98g_2^4+\tfrac9{20}g_2^2g_1^2+\tfrac{27}{200}g_1^4,
\end{align}
and at two loops
\begin{align}
\begin{split}
(16\pi^2)^2\frac{d\lambda}{dt} &= -312\lambda^3-144\lambda^2\left(y_t^2+y_b^2\right)-48\lambda^2y_\tau^2+108\lambda^2g_2^2+\tfrac{108}5\lambda^2g_1^2 \\
&\quad+\lambda y_t^2\!\left(-3y_t^2-42y_b^2+80g_3^2+\tfrac{45}2g_2^2+\tfrac{17}2g_1^2\right)
+\lambda y_b^2\!\left(-3y_b^2+80g_3^2+\tfrac{45}2g_2^2+\tfrac52g_1^2\right) \\
&\quad+\lambda y_\tau^2\!\left(-y_\tau^2+\tfrac{15}2g_2^2+\tfrac{15}2g_1^2\right)
+\lambda\!\left(-\tfrac{73}{8}g_2^4+\tfrac{1887}{200}g_1^4+\tfrac{117}{20}g_2^2g_1^2\right) \\
&\quad+y_t^4\!\left(30y_t^2-6y_b^2-32g_3^2-\tfrac85g_1^2\right)
+y_b^4\!\left(-6y_t^2+30y_b^2-32g_3^2+\tfrac45g_1^2\right) \\
&\quad+y_\tau^4\!\left(10y_\tau^2-\tfrac{12}5g_1^2\right)
+y_t^2\!\left(-\tfrac94g_2^4-\tfrac{171}{100}g_1^4+\tfrac{63}{10}g_2^2g_1^2\right)
+y_b^2\!\left(-\tfrac94g_2^4+\tfrac9{20}g_1^4+\tfrac{27}{10}g_2^2g_1^2\right) \\
&\quad+y_\tau^2\!\left(-\tfrac34g_2^4-\tfrac94g_1^4+\tfrac{33}{10}g_2^2g_1^2\right)
+\tfrac{305}{16}g_2^6-\tfrac{3411}{2000}g_1^6-\tfrac{289}{80}g_2^4g_1^2-\tfrac{1677}{400}g_2^2g_1^4.
\end{split}
\end{align}

\textbf{\textit{Gauge couplings.}}---
At one loop,
\begin{align}
16\pi^2\frac{dg_1}{dt}&=\tfrac{41}{10}g_1^3,\qquad
16\pi^2\frac{dg_2}{dt}=-\tfrac{19}{6}g_2^3,\qquad
16\pi^2\frac{dg_3}{dt}=-7g_3^3,
\end{align}
while at two loops, keeping only the third-generation Yukawa
contributions,
\begin{align}
\begin{split}
(16\pi^2)^2\frac{dg_1}{dt}
&=g_1^3\Big[
\tfrac{199}{50}g_1^2+\tfrac{27}{10}g_2^2+\tfrac{44}5g_3^2
-\left(\tfrac{17}{10}y_t^2+\tfrac12y_b^2+\tfrac32y_\tau^2\right)
\Big],\\
(16\pi^2)^2\frac{dg_2}{dt}
&=g_2^3\Big[
\tfrac9{10}g_1^2+\tfrac{35}6g_2^2+12g_3^2
-\left(\tfrac32(y_t^2+y_b^2)+\tfrac12y_\tau^2\right)
\Big],\\
(16\pi^2)^2\frac{dg_3}{dt}
&=g_3^3\Big[
\tfrac{11}{10}g_1^2+\tfrac92g_2^2-26g_3^2
-2(y_t^2+y_b^2)
\Big].
\end{split}
\end{align}
Together with the $m_Z$-scale inputs of Ref.~\cite{Antusch:2025fpm}, these equations are integrated numerically to obtain Fig.~\ref{fig:running}, Fig.~\ref{fig:phen_running}, and
Eqs.~\eqref{eq:ratios_range},~\eqref{eq:mu_cross},~\eqref{eq:mu1},~\eqref{eq:mu123},~\eqref{eq:mu12}.\\

\textbf{\textit{Perturbative uncertainty.}}---
Because the relevant Yukawa ratios run rather slowly, the crossing scales are more sensitive to higher-order corrections than the ratios themselves. Near $\mu_3$ we find
\begin{equation}
\frac{d\ln(y_\tau/y_b)}{d\ln\mu}\simeq0.038,
\label{eq:slope}
\end{equation}
corresponding to a change of only $8.8\%$ per decade. Consequently, a
small perturbative correction to $y_\tau/y_b$ can induce a noticeably larger relative shift in the scale at which the ratio crosses unity.

Going from one to two loops changes $y_\tau/y_b$ in the vicinity of the crossing by $+3.4\%$, while shifting the crossing from $\mu_3\simeq1.3\times10^6$~GeV to the value quoted in Eq.~\eqref{eq:mu_cross}. A naive estimate based on the square of this correction suggests a three-loop effect on the ratio of order $0.1\%$, corresponding through Eq.~\eqref{eq:slope} to only a few percent uncertainty in $\mu_3$.

We verify this explicitly using the full three-loop SM beta functions~\cite{Bednyakov:2012rb,Chetyrkin:2012rz}, as implemented in \textsc{smdr}~\cite{Martin:2019lqd}. The three-loop evolution shifts $\mu_3$ by $-4.0\%$ relative to our two-loop result, in agreement with the estimate above. The known four- and five-loop corrections in the gauge and QCD sectors move it by a further $<0.05\%$, showing that the perturbative expansion has effectively converged by three loops. As an independent cross-check, the two-loop \textsc{smdr} evolution reproduces our values of $\mu_3$ and $\mu_2$ at the $2\%$ and $1\%$ levels, respectively; the residual differences are dominated by the choice of low-energy boundary conditions. Most importantly, the agreement between the two heavier generations is stable under the inclusion of higher orders:
\begin{equation}
\frac{\mu_2}{\mu_3}=1.39
\quad\text{at two loops},\qquad
\frac{\mu_2}{\mu_3}=1.40
\quad\text{at three loops}.
\end{equation}
The common-scale behavior in Eq.~\eqref{eq:ratio} is therefore not an artifact of the perturbative truncation.

Although the slow evolution makes the crossing scale locally sensitive to small corrections, displacing it from the intermediate scale to a conventional GUT scale requires a genuinely large modification of the running. The approximately $9$ decades between $\mu_3$ and $M_{\rm GUT}$ correspond, using Eq.~\eqref{eq:slope}, to an ${\cal O}(1)$ change in $y_\tau/y_b$. Higher-order SM corrections are far too small to generate such an effect. The uncertainty associated with perturbative truncation is therefore subdominant to the experimental uncertainty quoted in Eq.~\eqref{eq:mu_cross} and does not affect the central conclusion of the analysis.

\section{Beta-function shifts from Standard Model extensions}
\label{app:betaext}

We ask whether the scale at which $y_\tau=y_b$ can be raised by adding new gauge-charged matter below the PS scale.  We restrict here to states with no direct Yukawa couplings to the third generation, so that their leading effect is through the running of the SM gauge couplings.

At one loop,
\begin{equation}
\frac{d}{dt}\ln\frac{y_\tau}{y_b} =
\frac{1}{16\pi^2} \left[ 8g_3^2-2g_1^2 + \frac32\left(y_\tau^2- y_b^2+y_t^2\right) \right]\,,
\label{eq:dlnratio}
\end{equation}
where $g_1$ is in $SU(5)$ normalization.  The $SU(2)_L$ contribution
cancels exactly.  The gauge-mediated evolution of the ratio is therefore
controlled at leading order by $g_3$ and $g_1$, with the QCD contribution
both parametrically larger and of opposite sign to the hypercharge one.

A matter multiplet $R=(d_3,d_2,Y)$ shifts the one-loop coefficients
defined by
\begin{equation}
\frac{dg_i}{dt}=\frac{b_i}{16\pi^2}g_i^3
\end{equation}
according to $\Delta b_i=c_R\,T_i(R)\,d_{\rm other}$ with $c_R= \frac{4}{3},\frac{2}{3},\frac{1}{3},\frac{1}{6}$ for a Dirac fermion, Weyl fermion, complex scalar, and real scalar,
respectively.  Here
$T_1=(3/5)Y^2$, while $T_2$ and $T_3$ are the usual Dynkin indices, and
$d_{\rm other}$ denotes the product of the dimensions under the
remaining gauge factors.

The sign of the QCD contribution already implies a general obstruction.
Any additional colored matter satisfies
\begin{equation}
\Delta b_3>0 ,
\end{equation}
and therefore weakens QCD asymptotic freedom.  Consequently $g_3$ is
larger above the new threshold and, from Eq.~\eqref{eq:dlnratio},
$y_\tau/y_b$ grows more rapidly toward the ultraviolet.  Colored matter
thus shifts the $y_\tau=y_b$ crossing scale downward, opposite to the
direction required here.  A negative $\Delta b_3$ would instead require
additional gauge dynamics rather than matter.

For color-singlet states, $\Delta b_3=0$ and the leading effect arises
through hypercharge.  Its smallness can be seen from the leading-log
expansion.  For a common threshold $\mu_{\rm low}$ and
$T\equiv\ln(\mu_{\rm high}/\mu_{\rm low})$, one has
\begin{equation}
\Delta g_i^2(t)
\simeq
\frac{\Delta b_i}{8\pi^2}\,\bar g_i^{4}\,t \,,
\end{equation}
where $\bar g_i$ denotes a representative average of the SM gauge coupling over the interval. The resulting fractional shift in the bottom--tau Yukawa ratio is
\begin{equation}
\delta
\equiv \frac{(y_\tau/y_b)_{\rm SM+R}}
{(y_\tau/y_b)_{\rm SM}}-1
\simeq
\frac{T^2}{256\pi^4}\left( 8\,\bar g_3^{4}\Delta b_3 -2\,\bar g_1^{4}\Delta b_1
\right).
\label{eq:deltab}
\end{equation}
For
$\mu_{\rm low}=10^3~{\rm GeV}$ and $\mu_{\rm high}=10^6~{\rm GeV}$ this becomes numerically
\begin{equation}
\delta[\%]
\simeq 1.02\,\Delta b_3 -0.020\,\Delta b_1 ,
\label{eq:delta_num}
\end{equation}
up to subleading higher-order effects.  The response to colored matter
is therefore roughly fifty times larger than that to hypercharge, but
has the wrong sign for raising the bottom--tau crossing scale.  The $SU(2)_L$ contribution is absent at this order,  because $g_2$ cancels identically from
Eq.~\eqref{eq:dlnratio} at one loop.

As a numerical illustration, we evolve the two-loop SM RGEs and, above $\mu_{\rm low}$, include the one-loop shifts $\Delta b_i$ of Eq.~\eqref{eq:deltab}. Representative examples are summarized in Table~\ref{tab:smext}. The table is not intended as an exhaustive classification, but rather to illustrate the general behavior implied by Eq.~\eqref{eq:dlnratio}.

For instance, from the SM running, we find
\begin{equation}
\left.\frac{y_\tau}{y_b}\right|_{\mu=10^6\,{\rm GeV}}
\simeq 1.024 ,
\end{equation}
so restoring $y_\tau=y_b$ at this scale requires $\delta_{\rm req} \simeq -2.4\%$.
Two features are apparent from Table~\ref{tab:smext}. As anticipated from Eq.~\eqref{eq:deltab}, individual color-singlet multiplets fall well short of this correction, while colored matter increases the ratio and moves the crossing scale in the opposite direction."

One may nevertheless ask whether a sufficiently large multiplicity of color-singlet states could overcome the small hypercharge coefficient. In this regime the linearized expression~\eqref{eq:delta_num} is no longer adequate, since the $U(1)_Y$ coupling itself runs rapidly. Keeping its one-loop evolution,
\begin{equation}
g_1^2(\mu) =
\frac{g_1^2(\mu_{\rm low})}
{1-\dfrac{(b_1+\Delta b_1)g_1^2(\mu_{\rm low})}{8\pi^2}
\ln(\mu/\mu_{\rm low})},
\label{eq:g1exact}
\end{equation}
shows that generating a correction of order $\delta\simeq-2.4\%$ between $10^3$ and $10^6$~GeV requires $\Delta b_1\simeq {\cal O}(40)$. For example, this corresponds roughly to ${\cal O}(10)$ copies of a charged vector-like singlet such as $E''\sim(1,1,-2)$. Such a large coefficient simultaneously drives hypercharge rapidly toward strong coupling.  The corresponding one-loop Landau pole,
\begin{equation}
\Lambda_{1} = 
\mu_{\rm low}
\exp\,\left[
\frac{8\pi^2}
{(b_1+\Delta b_1)g_1^2(\mu_{\rm low})}
\right],
\label{eq:landau}
\end{equation}
lies at only a few $10^6$~GeV for $\Delta b_1\simeq {\cal O}(40)$.  Thus hypercharge running can generate the needed effect only by driving $U(1)_Y$ close to strong coupling at a scale parametrically coincident with the desired PS scale.

The obstruction becomes stronger for a higher matching scale. We therefore conclude that perturbative gauge-only matter extensions do not provide a viable mechanism for substantially raising the $y_\tau=y_b$ crossing scale.

\begin{table}[t]
\caption{\label{tab:smext}
Representative SM matter extensions. The $\Delta b_i$ denote their one-loop contributions to the gauge beta functions, with $g_1$ in $SU(5)$ normalization. The fractional change $\delta$ in $y_\tau/y_b$ is evaluated at $10^6$~GeV using a common threshold $\mu_{\rm low}=10^3$~GeV, while $\delta\mu_3$ denotes the shift of the $y_\tau=y_b$ crossing relative to its SM value, $\mu_3^{\rm SM}=5.20\times10^5$~GeV. The new states are assumed to have no direct couplings to third-generation fermions.}
\begin{ruledtabular}
\begin{tabular}{llcccrr}
Field
& $(SU(3)_C,SU(2)_L,U(1)_Y)$
& $\Delta b_1$
& $\Delta b_2$
& $\Delta b_3$
& $\delta\,[\%]$
& $\delta\mu_3\,[\%]$\\
\hline

\multicolumn{7}{l}{\textit{Color-singlet fermions
(vector-like Dirac, unless noted)}}\\
$N$
& $(1,1,0)$
& $0$
& $0$
& $0$
& $\phantom{-}0.000$
& $\phantom{+}0.00$
\\
$E$
& $(1,1,-1)$
& $\frac45$
& $0$
& $0$
& $-0.016$
& $+0.35$
\\
$E''$
& $(1,1,-2)$
& $\frac{16}{5}$
& $0$
& $0$
& $-0.068$
& $+1.47$
\\
$L$
& $(1,2,-\frac12)$
& $\frac25$
& $\frac23$
& $0$
& $-0.010$
& $+0.21$
\\
$L'$
& $(1,2,-\frac32)$
& $\frac{18}{5}$
& $\frac23$
& $0$
& $-0.079$
& $+1.70$
\\
$\Sigma$
& $(1,3,0)$, Majorana
& $0$
& $\frac43$
& $0$
& $-0.003$
& $+0.07$
\\
$\Sigma_1$
& $(1,3,-1)$
& $\frac{12}{5}$
& $\frac83$
& $0$
& $-0.058$
& $+1.24$
\\
\hline
\multicolumn{7}{l}{\textit{Color-singlet scalars
(complex, unless noted)}}\\
$S$
& $(1,1,0)$, real
& $0$
& $0$
& $0$
& $\phantom{-}0.000$
& $\phantom{+}0.00$
\\
$S^+$
& $(1,1,1)$
& $\frac15$
& $0$
& $0$
& $-0.004$
& $+0.09$
\\
$S^{++}$
& $(1,1,2)$
& $\frac45$
& $0$
& $0$
& $-0.016$
& $+0.35$
\\
$H_2$
& $(1,2,\frac12)$
& $\frac1{10}$
& $\frac16$
& $0$
& $-0.002$
& $+0.05$
\\
$\Delta_0$
& $(1,3,0)$, real
& $0$
& $\frac13$
& $0$
& $-0.001$
& $+0.02$
\\
$\Delta$
& $(1,3,1)$
& $\frac35$
& $\frac23$
& $0$
& $-0.014$
& $+0.30$
\\

\hline
\multicolumn{7}{l}{\textit{Colored fermions
(vector-like Dirac, unless noted)}}\\
$D$
& $(3,1,-\frac13)$
& $\frac4{15}$
& $0$
& $\frac23$
& $+0.696$
& $-14.1$
\\
$U$
& $(3,1,\frac23)$
& $\frac{16}{15}$
& $0$
& $\frac23$
& $+0.679$
& $-13.8$
\\
$Q$
& $(3,2,\frac16)$
& $\frac2{15}$
& $2$
& $\frac43$
& $+1.442$
& $-26.0$
\\
$\lambda_8$
& $(8,1,0)$, Majorana
& $0$
& $0$
& $2$
& $+2.252$
& $-36.0$
\\

\hline
\multicolumn{7}{l}{\textit{Colored scalars
(complex, unless noted)}}\\
$S_1$
& $(\bar3,1,\frac13)$
& $\frac1{15}$
& $0$
& $\frac16$
& $+0.170$
& $-3.8$
\\
$\tilde R_2$& $(3,2,\frac16)$
& $\frac1{30}$
& $\frac12$
& $\frac13$
& $+0.343$
& $-7.4$
\\
$R_2$
& $(3,2,\frac76)$
& $\frac{49}{30}$
& $\frac12$
& $\frac13$
& $+0.310$
& $-6.8$
\\
$S_3$
& $(\bar3,3,\frac13)$
& $\frac15$
& $2$
& $\frac12$
& $+0.512$
& $-10.7$
\\
$O$
& $(8,1,0)$, real
& $0$
& $0$
& $\frac12$
& $+0.522$
& $-10.9$
\\
$\Phi_8$
& $(8,2,\frac12)$
& $\frac45$
& $\frac43$
& $2$
& $+2.232$
& $-35.8$
\\
\end{tabular}
\end{ruledtabular}
\end{table}

\section{Flavor corrections}
\label{app:Flavor}

In this Appendix we collect the flavor corrections used in the main text. We take all entries real in order to display the signs explicitly.  All unsquared rotation angles below are therefore signed.  The complex generalization is obtained with unitary singular-value decompositions and the corresponding rephasing-invariant interference products.

\subsection{Nonzero $Y_{22}$}

Before starting the discussion, it is useful to recall the well-known empirical relation
\begin{equation}
 |V_{us}|\simeq \sqrt{\frac{m_d}{m_s}},
\end{equation}
which holds at the percent level and has for decades provided valuable guidance in the search for organizing principles behind flavor \cite{Gatto:1968ss,Weinberg:1977hb,Fritzsch:1977za,Wilczek:1977uh,Wilczek:1978xi,Georgi:1979df}. In the hierarchical regime, if the first-family mass is generated predominantly by the $12$--$21$ block,
\begin{equation}
 \frac{y_d}{y_s}\simeq
 \left|\theta_{12}^{dL}\theta_{12}^{dR}\right|,
 \qquad
 \theta_{12}^{dL}\simeq\frac{(Y_d)_{12}}{y_s},
 \qquad
 \theta_{12}^{dR}\simeq\frac{(Y_d)_{21}}{y_s}.
\end{equation}
If, in addition, the Cabibbo angle originates predominantly in the down sector,
\begin{equation}
 |\theta_{12}^{dL}|\simeq |V_{us}|
 \simeq\sqrt{\frac{y_d}{y_s}},
\end{equation}
the observed values of $y_d/y_s$ and $V_{us}$ imply
\begin{equation}
 |\theta_{12}^{dR}|
 \simeq|\theta_{12}^{dL}|
 \simeq|V_{us}|,
 \qquad
 |(Y_d)_{12}|\simeq |(Y_d)_{21}|.
\end{equation}
Thus the approximate symmetry of the $1$--$2$ block is not imposed arbitrarily, but is suggested by the observed quark masses and mixing. Motivated by this intriguing possibility, in the following we adopt a symmetric $1$--$2$ block.  This assumption can be relaxed without altering the qualitative conclusions of the analysis. We consider
\begin{equation}
 Y_d=
 \begin{pmatrix}
 0&a&f\\
 a&e&b\\
 g&c&d
 \end{pmatrix},
 \qquad
 Y_e=
 \begin{pmatrix}
 0&a&C_{13}f\\
 a&-3e&C_{23}b\\
 C_{31} g&C_{32}c&d
 \end{pmatrix},
 \label{eq:app_nonzero22_texture}
\end{equation}
with $C_{ij}=1,-3$ and $|a|\ll |e|$, and $|e|,|b|,|c|,|f|,|g|\ll |d|$. To leading order,
\begin{equation}
 \theta_{12}^{dL}\simeq\theta_{12}^{dR}
 \simeq\frac{a}{e},
 \qquad
 \theta_{23}^{dL}\simeq\frac{b}{d},
 \qquad
 \theta_{23}^{dR}\simeq\frac{c}{d},
 \qquad
 \theta_{13}^{dL}\simeq\frac{f}{d},
 \qquad
 \theta_{13}^{dR}\simeq\frac{g}{d}.
 \label{eq:app_nonzero22_angles}
\end{equation}
It is useful to define
\begin{equation}
 \delta_{12}\equiv
 \frac49\left[
 (\theta_{12}^{dL})^2+
 (\theta_{12}^{dR})^2
 \right],
 \qquad
 \eta\equiv\frac{bc}{ed}
 \simeq
 \frac{y_b}{y_s}\,
 \theta_{23}^{dL}\theta_{23}^{dR},
 \label{eq:app_nonzero22_defs}
\end{equation}
and
\begin{equation}
 \delta_3\equiv
 \frac{C_{23}^2-1}{2}(\theta_{23}^{dL})^2+
 \frac{C_{32}^2-1}{2}(\theta_{23}^{dR})^2,
 \qquad
 \delta_{23}\equiv
 \left(1+\frac{C_{23}C_{32}}{3}\right)\eta .
 \label{eq:app_nonzero22_23defs}
\end{equation}
In the absence of explicit $13$ and $31$ entries, at leading nonvanishing order in the small mixing angles,
\begin{equation}
 R_1\simeq1+\delta_{12}-\delta_{23},
 \qquad
 R_2\simeq1-\delta_{12}+\delta_{23}-\delta_3,
 \qquad
 R_3\simeq1+\delta_3\,.
 \label{eq:app_nonzero22_no13}
\end{equation}
The isolated $2$--$3$ block also obeys the exact relation
\begin{equation}
 R_2R_3\big|_{a=f=g=0}
 =
 \frac{\left|1+C_{23}C_{32}\eta/3\right|}
      {|1-\eta|},
 \label{eq:app_nonzero22_exact23}
\end{equation}
which should be used instead of expanding in $\eta$ when the determinant
correction is not small.

Restoring the $13$ and $31$ entries, the complete determinants are
\begin{align}
 \det Y_d={}&
 -a^2d+abg+afc-efg,
 \nonumber\\
 \det Y_e={}&
 -a^2d+C_{23}C_{31}abg
 +C_{13}C_{32}afc
 +3C_{13}C_{31}efg ,
 \label{eq:app_nonzero22_dets}
\end{align}
and therefore
\begin{align}
 R_1R_2R_3=
 \left|\frac{\det Y_e}{\det Y_d}\right|\,.
 \label{eq:app_det_identity}
\end{align}
Equation~\eqref{eq:app_det_identity} is useful for obtaining $R_1$ once $R_2$ and $R_3$ have been computed and automatically retains the
correlations among the three ratios.

The explicit $13$--$31$ correction to the second ratio is, to quadratic
order in the small $13$ rotations,  $R_2\simeq
 1-\delta_{12}+\delta_{23}-\delta_3
 +\delta_{13}^{(2)}$, with
\begin{align}
\begin{split}
 \delta_{13}^{(2)}\simeq{}&
 \left(1-\frac{C_{13}C_{32}}9\right)
 \frac{y_b}{y_s}\,
 \theta_{12}^{dL}\theta_{13}^{dL}\theta_{23}^{dR} +
 \left(1-\frac{C_{31}C_{23}}9\right)
 \frac{y_b}{y_s}\,
 \theta_{12}^{dR}\theta_{13}^{dR}\theta_{23}^{dL}
 \\
 &+
 \frac12\left(\frac{C_{13}^2C_{32}^2}{9}-1\right)
 \left(
 \frac{y_b}{y_s}\theta_{13}^{dL}\theta_{23}^{dR}
 \right)^2
 +
 \frac12\left(\frac{C_{31}^2C_{23}^2}{9}-1\right)
 \left(
 \frac{y_b}{y_s}\theta_{13}^{dR}\theta_{23}^{dL}
 \right)^2 \,.
 \end{split}
 \label{eq:app_nonzero22_R2_13}
\end{align}

For the third ratio one finds $R_3\simeq1+\delta_3+\delta_{13}^{3}$, with
\begin{align}
\begin{split}
 \delta_{13}^{(3)}\simeq{}&
 (C_{13}C_{32}-1)
 \frac{y_s}{y_b}\,
 \theta_{12}^{dL}\theta_{13}^{dL}\theta_{23}^{dR}
 +
 (C_{31}C_{23}-1)
 \frac{y_s}{y_b}\,
 \theta_{12}^{dR}\theta_{13}^{dR}\theta_{23}^{dL}
 +
 \frac{C_{13}^2-1}{2}(\theta_{13}^{dL})^2
 +
 \frac{C_{31}^2-1}{2}(\theta_{13}^{dR})^2 .
 \end{split}
 \label{eq:app_nonzero22_R3_13}
\end{align}

Expanding Eq.~\eqref{eq:app_det_identity} gives the leading hierarchy-enhanced first-family correction
\begin{align}
\begin{split}
 \delta_{13}^{(1)}\simeq{}&
 -(C_{23}C_{31}-1)
 \frac{y_b}{\sqrt{y_dy_s}}\,
 \theta_{23}^{dL}\theta_{13}^{dR}
 -(C_{13}C_{32}-1)
 \frac{y_b}{\sqrt{y_dy_s}}\,
 \theta_{13}^{dL}\theta_{23}^{dR}
 -(3C_{13}C_{31}+1)
 \frac{y_b}{y_d}\,
 \theta_{13}^{dL}\theta_{13}^{dR}\,.
\end{split}
 \label{eq:app_nonzero22_R1_13}
\end{align}
For
\begin{equation}
 \theta_{12}^{dL,R}\sim |V_{us}|,\qquad
 \theta_{23}^{dL,R}\sim |V_{cb}|,\qquad
 \theta_{13}^{dL,R}\sim |V_{ub}|,
 \label{eq:app_nonzero22_CKM}
\end{equation}
the explicit $13$--$31$ corrections to $R_2$ are sub-percent and those to $R_3$ are smaller still, whereas Eq.~\eqref{eq:app_nonzero22_R1_13} can remain sizeable because of the first-family hierarchy.

\subsection{Vanishing $Y_{22}$}

We now consider the case
\begin{equation}
 (Y_d)_{22}=(Y_e)_{22}=0,
 \qquad
 C_{23}=-3,
 \qquad
 C_{32}=1 .
 \label{eq:app_zero22_texture}
\end{equation}
In this case the strange Yukawa is generated by $2$--$3$ mixing.  Keeping the right-handed rotation exact and expanding only in the small left-handed angle gives
\begin{equation}
 \left|\frac{y_s}{y_b}\right|
 \simeq
 \left|
 \theta_{23}^{dL}\tan\theta_{23}^{dR}
 \right|.
 \label{eq:app_zero22_mass}
\end{equation}
We define
\begin{equation}
 t_R\equiv\tan\theta_{23}^{dR}.
 \label{eq:app_zero22_tR}
\end{equation}
Thus $\theta_{23}^{dL}\sim |V_{cb}|\simeq0.04$ and $y_s/y_b\simeq0.02$ imply $|t_R|\sim0.5$.

It is useful to define the angles sourced by the explicit first-family entries after the $2$--$3$ rotation,
\begin{equation}
 \theta_{12}^{dL}\equiv
 \frac{(Y_d)_{12}\cos\theta_{23}^{dR}}{y_s},
 \qquad
 \theta_{12}^{dR}\equiv
 \frac{(Y_d)_{21}\cos\theta_{23}^{dL}}{y_s},
 \qquad
 \theta_{13}^{dL}\equiv
 \frac{(Y_d)_{13}\cos\theta_{23}^{dR}}{y_b},
 \qquad
 \theta_{13}^{dR}\equiv
 \frac{(Y_d)_{31}\cos\theta_{23}^{dL}}{y_b}.
 \label{eq:app_zero22_thetas}
\end{equation}
The lightest eigenvalue is still generated by the 1--2 rotation
\begin{equation}
 \frac{y_d}{y_s}
 \simeq
 \left|
 \theta_{12}^{dL}\theta_{12}^{dR}
 \right|,
 \label{eq:app_zero22_yd}
\end{equation}
up to corrections of order $(y_s/y_b)^2$. On the other hand, if the Cabibbo angle originates predominantly in the down sector, the observed relation $|V_{us}|^2\simeq y_d/y_s$ implies $|\theta_{12}^{dL}| \simeq |\theta_{12}^{dR}| \simeq |V_{us}|$, which does not require a symmetric 1--2 block in the interaction-basis. Rather,
\begin{equation}
 \left|
 \frac{(Y_d)_{12}}{(Y_d)_{21}}
 \right|
 \simeq
 \frac{\cos\theta_{23}^{dL}}
      {\cos\theta_{23}^{dR}}
 \simeq
 \sqrt{1+t_R^2}.
 \label{eq:app_zero22_12ratio}
\end{equation}
For $t_R\simeq0.5$ the asymmetry is about $10\%$.

We define
\begin{equation}
 \delta_{12}^{(0)}
 \equiv
 \frac49\left[
 (\theta_{12}^{dL})^2+
 (\theta_{12}^{dR})^2
 \right].
 \label{eq:app_zero22_delta12}
\end{equation}
Allowing the explicit $13$ and $31$ entries to carry Clebsch coefficients
$C_{13},C_{31}\in\{1,-3\}$, and keeping $t_R$ arbitrary, the complete
quadratic expressions for the two heavier ratios are
\begin{align}
\begin{split}
 R_2\simeq{}&
 1-\delta_{12}^{(0)}
 -4(\theta_{23}^{dL})^2
 +
 \left(1-\frac{C_{13}}9\right)
 \theta_{12}^{dL}
 \frac{\theta_{13}^{dL}}{\theta_{23}^{dL}}
 +
 \left(1+\frac{C_{31}}3\right)
 \theta_{12}^{dR}
 \frac{\theta_{13}^{dR}}{t_R}
 +
 \frac{C_{13}^2-9}{18}
 \left(
 \frac{\theta_{13}^{dL}}{\theta_{23}^{dL}}
 \right)^2
 +
 \frac{C_{31}^2-1}{2}
 \left(
 \frac{\theta_{13}^{dR}}{t_R}
 \right)^2 ,
 \end{split}
 \label{eq:app_zero22_R2}
\end{align}
and
\begin{align}
 R_3\simeq{}&
 1+4(\theta_{23}^{dL})^2
 +
 (C_{13}-1)
 \theta_{12}^{dL}\theta_{23}^{dL}
 t_R^2\theta_{13}^{dL}
 -
 (3C_{31}+1)
 \theta_{12}^{dR}(\theta_{23}^{dL})^2
 t_R\theta_{13}^{dR}
 +
 \frac{C_{13}^2-1}{2}
 (\theta_{13}^{dL})^2
 +
 \frac{C_{31}^2-1}{2}
 (\theta_{13}^{dR})^2 .
 \label{eq:app_zero22_R3}
\end{align}

Using the determinant identiy in Eq.~\eqref{eq:app_det_identity}, and introducing 
\begin{equation}
 {\cal B}\equiv
 (1+t_R^2)
 \frac{\theta_{13}^{dR}}
      {\theta_{12}^{dR}t_R},
 \qquad
 {\cal C}\equiv
 \frac{\theta_{13}^{dL}}
      {\theta_{12}^{dL}\theta_{23}^{dL}},
 \label{eq:app_zero22_BC}
\end{equation}
the first family relation is
\begin{equation}
 R_1=
 \frac{1}{R_2R_3}
 \left|
 \frac{1+3C_{31}{\cal B}-C_{13}{\cal C}}
      {1-{\cal B}-{\cal C}}
 \right|\,.
 \label{eq:app_zero22_R1}
\end{equation}

Taking
\begin{equation}
 \theta_{12}^{dL,R}\sim |V_{us}|,
 \qquad
 \theta_{23}^{dL}\sim |V_{cb}|,
 \qquad
 t_R\sim0.5,
 \qquad
 \theta_{13}^{dL,R}\sim |V_{ub}|,
 \label{eq:app_zero22_CKM}
\end{equation}
Eq.~\eqref{eq:app_zero22_R3} reduces to very good accuracy to
\begin{equation}
 R_3\simeq
 1+4(\theta_{23}^{dL})^2 ,
 \label{eq:app_zero22_R3_reduced}
\end{equation}
while in $R_2$ and $R_1$ the additional terms that can naturally reach the percent level are
\begin{align}
\begin{split}
 R_2\simeq &
 1-\delta_{12}^{(0)}
 -4(\theta_{23}^{dL})^2
 + \left(1-\frac{C_{13}}9\right)
 \theta_{12}^{dL}
 \frac{\theta_{13}^{dL}}{\theta_{23}^{dL}}\,,\\
 R_1\simeq &
 1+\delta_{12}^{(0)}
 +(1-C_{13})
 \frac{\theta_{13}^{dL}}
 {\theta_{12}^{dL}\theta_{23}^{dL}}
 +
 (3C_{31}+1)(1+t_R^2)
 \frac{\theta_{13}^{dR}}
 {\theta_{12}^{dR}t_R} \,.
 \end{split}
 \label{eq:app_zero22_R2R1_reduced}
\end{align}
The enhancement by $1/\theta_{23}^{dL}$ follows directly from the fact that the strange mass is generated by the $23$--$32$ entries. The complete expressions in Eqs.~\eqref{eq:app_zero22_R2}--\eqref{eq:app_zero22_R1} should be used for
percent-level numerical work.

\section{Pati--Salam scalar and Yukawa interactions}
\label{app:PStheory}

In this Appendix we collect the scalar Yukawa interactions used in the phenomenological discussion. We consider both $SU(4)_C\times SU(2)_L\times U(1)_R$ and $SU(4)_C\times SU(2)_L\times SU(2)_R$.

We denote by $\lambda$ the Yukawa matrices appearing in the
Pati--Salam theory. By contrast, $Y_f$ denotes the coupling of the
fermion $f$ to the electroweak doublet $H_v$ aligned with the vacuum. In
the convention used throughout this work,
\begin{equation}
	\langle H_v^0\rangle=v,
	\qquad
	v\simeq174~{\rm GeV},
	\qquad
	M_f=vY_f.
	\label{eq:appPS_Ydefinition}
\end{equation}
Thus the eigenvalues of $Y_f$ are the running Yukawa couplings discussed
in the main text, whereas the underlying $\lambda$ matrices also control the interactions of the additional Higgs states, scalar leptoquarks, and color-octet scalars.

We use left-handed Weyl fermions and denote the antisymmetric contraction of $SU(2)_L$ indices by
\begin{equation}
	A\cdot B\equiv A^T\epsilon B,
	\qquad
	\epsilon=i\sigma_2,
	\label{eq:appPS_dot}
\end{equation}
where the transpose and $\epsilon$ act only in $SU(2)_L$ space.
For
\begin{equation}
	Q=\begin{pmatrix}u\\ d\end{pmatrix},
	\qquad
	L=\begin{pmatrix}\nu\\ e\end{pmatrix},
	\qquad
	H=\begin{pmatrix}H^+\\ H^0\end{pmatrix},
\end{equation}
one has
\begin{equation}
	Q\cdot H=uH^0-dH^+,
	\qquad
	L\cdot H=\nu H^0-eH^+.
	\label{eq:appPS_dot_expanded}
\end{equation}
Family and color indices are displayed when useful and otherwise left
implicit.

\subsection{$SU(4)_C\times SU(2)_L\times U(1)_R$}
\label{app:PS421}

We first consider
\begin{equation}
	G_{421}=SU(4)_C\times SU(2)_L\times U(1)_R,
\end{equation}
with
\begin{align}
	F_i&=(Q_i,L_i)^T\sim(4,2,0)_{\rm PS}\,,\qquad
	F^c_{u\,i}=(u_i^c,\nu_i^c)^T\sim(\bar 4,1,-1/2)_{\rm PS}\,,\qquad
	F^c_{d\,i}=(d_i^c,e_i^c)^T\sim(\bar 4,1,+1/2)_{\rm PS}.
	\label{eq:app421_fermions}
\end{align}
Here and in the following, the subscript ${\rm PS}$ denotes quantum numbers under the quark-lepton unified theory, while the subscript ${\rm SM}$ will be used for quantum numbers under the Standard Model gauge group.

Hypercharge is
\begin{equation}
	Y=R+\frac{B-L}{2}.
	\label{eq:app421_hypercharge}
\end{equation}

We introduce two scalar multiplets,
\begin{equation}
	H_1\sim(1,2,1/2)_{\rm PS},
	\qquad
	\Phi_{15}\sim(15,2,1/2)_{\rm PS},
	\label{eq:app421_scalars}
\end{equation}
with independent up- and down-type Yukawa matrices
\begin{equation}
	\lambda_1^u,\quad \lambda_1^d,
	\qquad
	\lambda_{15}^u,\quad \lambda_{15}^d.
\end{equation}
With explicit $SU(4)_C$ indices $A,B$ and $SU(2)_L$ indices $a,b$, the renormalizable interactions can be written as
\begin{align}
	\begin{split}
		{\cal L}^{421}_Y={}&
		\epsilon_{ab}F_i^{Aa}
		\left[
		(\lambda_1^u)_{ij}(H_1)^b\delta_A^{\ B}
		+
		(\lambda_{15}^u)_{ij}(\Phi_{15}^{\,b})^B{}_{A}
		\right]
		(F^c_{u\,j})_B \\
		&+ F_i^{Aa}
		\left[
		(\lambda_1^d)_{ij}(H_1^\dagger)_a\delta_A^{\ B}
		+
		(\lambda_{15}^d)_{ij}(\Phi_{15\,a}^\dagger)^B{}_{A}
		\right]
		(F^c_{d\,j})_B
		+{\rm h.c.}
	\end{split}
	\label{eq:app421_PS_yukawa}
\end{align}

Under $SU(4)_C\rightarrow SU(3)_C\times U(1)_{B-L}$, the adjoint
doublet decomposes as
\begin{align}
	\Phi_{15}=& 
	\underbrace{(1,2,1/2)_{\rm SM}}_{H_{15}}
	 \ \oplus \
	\underbrace{(8,2,1/2)_{\rm SM}}_{\Phi_8}
	\ \oplus \
	\underbrace{(3,2,7/6)_{\rm SM}}_{R_2} \
	\oplus \
	\underbrace{(\bar 3,2,-1/6)_{\rm SM}}_{\widetilde R_2^c}.
	\label{eq:app421_15decomp}
\end{align}
We have defined $\widetilde R_2^{\,c}\equiv -\,\epsilon\,\widetilde R_2^*$, so as to make contact with the scalar-leptoquark notation commonly used in the literature.
A convenient matrix parametrization is
\begin{equation}
	\Phi_{15}=
	\begin{pmatrix}
		\displaystyle
		\Phi_8+c_{15} H_{15}\,\mathbf 1_3
		&
		\displaystyle
		r_{15} R_2
		\\[3mm]
		\displaystyle
		r_{15}\widetilde R_2^c
		&
		\displaystyle
		-3 c_{15}H_{15}
	\end{pmatrix}.
	\label{eq:app421_matrix_decomp}
\end{equation}
The $3+1$ block structure reflects the decomposition of the $SU(4)_C$ fundamental index into quark and lepton directions. The color-singlet diagonal direction is proportional to
\begin{equation}
T_{15} = \frac{1}{2\sqrt6}\,\mathrm{diag}(1,1,1,-3),
\end{equation}
while the off-diagonal blocks correspond to the leptoquark components. The relative coefficients are fixed by tracelessness and the normalization of the $SU(4)_C$ generators:
\begin{equation}
	c_{15}\equiv\frac{1}{2\sqrt6},
	\qquad
	r_{15}\equiv\frac{1}{\sqrt2}.
	\label{eq:appPS_c15r15}
\end{equation}
Above the color indices of $R_2$ and $\widetilde R_2$ are understood,
and
\begin{equation}
	\Phi_8=\Phi_8^aT^a,
	\qquad
	{\rm Tr}(T^aT^b)=\frac12\delta^{ab}.
\end{equation}

Using Eq.~\eqref{eq:app421_matrix_decomp}, the color-singlet interactions
become
\begin{align}
	{\cal L}^{421}_{\rm singlet}={}&
	(\lambda_1^u)_{ij}
	\left[
	Q_i\cdot H_1\,u_j^c
	+
	L_i\cdot H_1\,\nu_j^c
	\right]
	+
	(\lambda_1^d)_{ij}
	\left[
	H_1^\dagger Q_i\,d_j^c
	+
	H_1^\dagger L_i\,e_j^c
	\right]
	\nonumber\\
	&+
	c_{15}(\lambda_{15}^u)_{ij}
	\left[
	Q_i\cdot H_{15}\,u_j^c
	-
	3L_i\cdot H_{15}\,\nu_j^c
	\right]
	+
	c_{15}(\lambda_{15}^d)_{ij}
	\left[
	H_{15}^\dagger Q_i\,d_j^c
	-
	3H_{15}^\dagger L_i\,e_j^c
	\right]
	+{\rm h.c.}
	\label{eq:app421_singlet_yukawa}
\end{align}
This equation makes the origin of the $1$ and $-3$ Clebsch factors
manifest.

The color-octet interactions are
\begin{equation}
	{\cal L}^{421}_{8}=
	(\lambda_{15}^u)_{ij}(Q_i\cdot\Phi_8)u_j^c
	+
	(\lambda_{15}^d)_{ij}(\Phi_8^\dagger Q_i)d_j^c
	+{\rm h.c.},
	\label{eq:app421_octet}
\end{equation}
with the color-generator contraction implicit.

The scalar-leptoquark interactions are
\begin{align}
	{\cal L}^{421}_{\rm LQ}={}&
	r_{15}(\lambda_{15}^u)_{ij}
	\left[
	(L_i\cdot R_2)u_j^c
	+
	(Q_i \cdot \widetilde R_2^c)\nu_j^c
	\right]
	\nonumber\, +
	r_{15}(\lambda_{15}^d)_{ij}
	\left[
	(R_2^\dagger Q_i)e_j^c
	+
	((\widetilde R_2^c)^\dagger L_i)d_j^c
	\right]
	+{\rm h.c.}.
	\label{eq:app421_LQ}
\end{align}
Thus the same matrices $\lambda_{15}^{u,d}$ that generate the
quark--lepton Clebsch structure also determine the scalar-leptoquark
couplings.

We now use
\begin{equation}
	R_2\sim (3,2,7/6)=
	\begin{pmatrix}
		R_2^{5/3}\\
		R_2^{2/3}
	\end{pmatrix},
	\qquad
	\widetilde R_2 \sim (3,2,1/6)=
	\begin{pmatrix}
		\widetilde R_2^{2/3}\\
		\widetilde R_2^{-1/3}
	\end{pmatrix}.
	\label{eq:app421_LQ_components}
\end{equation}
Then
\begin{align}
	\begin{split}
		{\cal L}^{421}_{\rm LQ}={}&
		r_{15}(\lambda_{15}^u)_{ij}
		\Big[
		(\nu_iR_2^{2/3}-e_iR_2^{5/3})u_j^c
		+
		(\widetilde R_2^{2/3\,*}u_i+
		\widetilde R_2^{-1/3\,*}d_i)\nu_j^c
		\Big]
		\\
		&+
		r_{15}(\lambda_{15}^d)_{ij}
		\Big[
		(R_2^{5/3\,*}u_i+R_2^{2/3\,*}d_i)e_j^c
		-
		(\nu_i\widetilde R_2^{-1/3}-
		e_i\widetilde R_2^{2/3})d_j^c
		\Big]
		+{\rm h.c.}
	\end{split}
	\label{eq:app421_LQ_expanded}
\end{align}
The relative minus signs in $L\cdot R_2$ and
$L\cdot\widetilde R_2$ follow from the antisymmetric $SU(2)_L$
contraction in Eq.~\eqref{eq:appPS_dot}.

\subsubsection{Higgs basis and charged-scalar couplings}

Only the two color-singlet doublets $H_1$ and $H_{15}$ can acquire electroweak vacuum expectation values. For a CP-conserving vacuum we write
\begin{equation}
	\langle H_1^0\rangle=v_1,
	\qquad
	\langle H_{15}^0\rangle=v_{15},
	\qquad
	v^2=v_1^2+v_{15}^2,\qquad 
	c_\beta\equiv\frac{v_1}{v},
	\qquad
	s_\beta\equiv\frac{v_{15}}{v}\,.
	\label{eq:app421_vevs}
\end{equation}
The Higgs basis is
\begin{align}
	H_v&=c_\beta H_1+s_\beta H_{15}\,, \qquad
	H_\perp=-s_\beta H_1+c_\beta H_{15}\,,\qquad \langle H_v^0\rangle=v\,,
	\qquad
	\langle H_\perp^0\rangle=0.
	\label{eq:app421_Higgsbasis}
\end{align}

The Yukawa matrices of the vacuum direction are
\begin{align}
	Y_u&=c_\beta\lambda_1^u+c_{15}s_\beta\lambda_{15}^u,
	&
	Y_\nu&=c_\beta\lambda_1^u-3c_{15}s_\beta\lambda_{15}^u,
	\nonumber\\
	Y_d&=c_\beta\lambda_1^d+c_{15}s_\beta\lambda_{15}^d,
	&
	Y_e&=c_\beta\lambda_1^d-3c_{15}s_\beta\lambda_{15}^d.
	\label{eq:app421_Yf}
\end{align}
and allow to determine
\begin{align}
	\begin{split}
		\lambda_{1}^u
		&=\frac{3\,Y_u+Y_\nu}{4c_\beta}\,,\qquad
		\lambda_{1}^d=\frac{3Y_d+Y_e}{4c_\beta}\,,\\
		\lambda_{15}^u
		&=\frac{Y_u-Y_\nu}{4c_{15}s_\beta}
		=\frac{\sqrt6}{2s_\beta}(Y_u-Y_\nu)\,,\qquad
		\lambda_{15}^d=\frac{Y_d-Y_e}{4c_{15}s_\beta}
		=\frac{\sqrt6}{2s_\beta}(Y_d-Y_e).
	\end{split}
	\label{eq:app421_lambda15_fromY}
\end{align}
Thus a small adjoint vacuum expectation value can correspond to
underlying adjoint Yukawa matrices parametrically larger than the
physical Yukawa couplings.

The orthogonal doublet couples through
\begin{align}
	B^U_\perp&=-s_\beta\lambda_1^u+c_{15}c_\beta\lambda_{15}^u,
	&
	B^N_\perp &=-s_\beta\lambda_1^u-3c_{15}c_\beta\lambda_{15}^u,
	\nonumber\\
	A^D_\perp&=-s_\beta\lambda_1^d+c_{15}c_\beta\lambda_{15}^d,
	&
	A^E_\perp&=-s_\beta\lambda_1^d-3c_{15}c_\beta\lambda_{15}^d.
	\label{eq:app421_rhof}
\end{align}
The color-singlet Yukawa Lagrangian can therefore be written as
\begin{align}
	\begin{split}
		{\cal L}^{421}_{\rm EW}={}&
		Q_i\cdot H_v\,(Y_u)_{ij}u_j^c
		+
		L_i\cdot H_v\,(Y_\nu)_{ij}\nu_j^c +
		H_v^\dagger Q_i\,(Y_d)_{ij}d_j^c
		+
		H_v^\dagger L_i\,(Y_e)_{ij}e_j^c\\
		&+
		Q_i\cdot H_\perp\,(B^U_\perp)_{ij}u_j^c
		+
		L_i\cdot H_\perp\,(B^N_\perp)_{ij}\nu_j^c +
		H_\perp^\dagger Q_i\,(A^D_\perp)_{ij}d_j^c
		+
		H_\perp^\dagger L_i\,(A^E_\perp)_{ij}e_j^c
		+{\rm h.c.}
	\end{split}
	\label{eq:app421_EW_Higgsbasis}
\end{align}
The charged component of $H_v$ is the Goldstone boson, while
$H_\perp^\pm$ is the physical charged scalar. Its interactions are
\begin{align}
	{\cal L}^{421}_{H^\pm}={}&
	H_\perp^+
	\left[
	-d_i(B^U_\perp)_{ij}u_j^c
	-e_i(B^N_\perp)_{ij}\nu_j^c
	\right] +
	H_\perp^-
	\left[
	u_i(A^D_\perp)_{ij}d_j^c
	+\nu_i(A^E_\perp)_{ij}e_j^c
	\right]
	+{\rm h.c.}
	\label{eq:app421_Hpm}
\end{align}
The matrices $B_\perp^{U,N}$ and $A_\perp^{D,E}$ are linear combinations of the PS Yukawa matrices different from the vacuum-aligned $Y_f$, whose singular values are fixed by the measured masses. Charged-scalar
thresholds are therefore controlled by the underlying $\lambda$'s and
are not fixed directly by the small physical Yukawa eigenvalues.

\subsection{$SU(4)_C\times SU(2)_L\times SU(2)_R$}
\label{app:PS422}

We now consider
\begin{equation}
	G_{422}=SU(4)_C\times SU(2)_L\times SU(2)_R,
\end{equation}
with
\begin{equation}
	F_i\sim(4,2,1)\,,
	\qquad
	F_i^c\sim(\bar 4,1,2)\,,\qquad
	\Phi_1\sim(1,2,2)\,,
	\qquad
	\Phi_{15}\sim(15,2,2)\,.
	\label{eq:app422_matter}
\end{equation}
Each complex bidoublet admits two independent renormalizable Yukawa invariants. We write
\begin{align}
	{\cal L}^{422}_Y={}&
	F_i^T
	\left[
	(\lambda_1)_{ij}\Phi_1^*
	-
	(\widetilde\lambda_1)_{ij}\epsilon_L\Phi_1\epsilon_R
	\right]F_j^c +
	F_i^T
	\left[
	(\lambda_{15})_{ij}\Phi_{15}^*
	-
	(\widetilde\lambda_{15})_{ij}\epsilon_L\Phi_{15}\epsilon_R
	\right]F_j^c
	+{\rm h.c.}.
	\label{eq:app422_PS_yukawa}
\end{align}
Here $\epsilon_{L,R}=i\sigma_2$ denote the invariant antisymmetric tensors acting on the $SU(2)_{L,R}$ indices, respectively.
Thus we have four independent flavor matrices $ \lambda_1$, $\widetilde\lambda_1$, $\lambda_{15}$, and $\widetilde\lambda_{15}$. After $SU(2)_R\rightarrow U(1)_R$, each bidoublet gives two
$SU(2)_L$ doublets. After conjugating the doublet with $R=-1/2$, all
four color-singlet electroweak doublets may be written with hypercharge
$Y=+1/2$. We denote them by
\begin{equation}
	H_1^{(1)},\quad H_1^{(2)},\qquad
	H_{15}^{(1)},\quad H_{15}^{(2)}.
	\label{eq:app422_fourdoublets}
\end{equation}
A convenient convention is such that the Yukawa matrices associated
with these four doublets are 
\begin{alignat}{2}
	\kappa_{uI} &=
	\left(
	\lambda_1,
	\widetilde\lambda_1,
	c_{15}\lambda_{15},
	c_{15}\widetilde\lambda_{15}
	\right),\qquad
	&&\kappa_{dI} =
	\left(
	\widetilde\lambda_1,
	\lambda_1,
	c_{15}\widetilde\lambda_{15},
	c_{15}\lambda_{15}
	\right),
	\nonumber\\
	\kappa_{\nu I}
	&=
	\left(
	\lambda_1,
	\widetilde\lambda_1,
	-3c_{15}\lambda_{15},
	-3c_{15}\widetilde\lambda_{15}
	\right),\qquad
	&&\kappa_{eI} =
	\left(
	\widetilde\lambda_1,
	\lambda_1,
	-3c_{15}\widetilde\lambda_{15},
	-3c_{15}\lambda_{15}
	\right),
	\label{eq:app422_kappa}
\end{alignat}
where $I=1,\ldots,4$ labels the doublets in the order displayed in
Eq.~\eqref{eq:app422_fourdoublets}.

For real neutral vacuum expectation values we write
\begin{align}
	\langle H_1^{(1)0}\rangle=u_1,
	\qquad
	\langle H_1^{(2)0}\rangle=u_2,
	\qquad
	\langle H_{15}^{(1)0}\rangle=w_1,
	\qquad
	\langle H_{15}^{(2)0}\rangle=w_2.
	\label{eq:app422_vevs}
\end{align}
The fermion mass matrices are then
\begin{align}
	\begin{split}
		M_u={}&
		\lambda_1u_1
		+
		\widetilde\lambda_1u_2
		+
		c_{15}\lambda_{15}w_1
		+
		c_{15}\widetilde\lambda_{15}w_2\,,\qquad M_d={}
		\widetilde\lambda_1u_1
		+
		\lambda_1u_2
		+
		c_{15}\widetilde\lambda_{15}w_1
		+
		c_{15}\lambda_{15}w_2,\\
		M_\nu^D={}&
		\lambda_1u_1
		+
		\widetilde\lambda_1u_2
		-
		3c_{15}\lambda_{15}w_1
		-
		3c_{15}\widetilde\lambda_{15}w_2\,,\qquad
		M_e={}
		\widetilde\lambda_1u_1
		+
		\lambda_1u_2
		-
		3c_{15}\widetilde\lambda_{15}w_1
		-
		3c_{15}\lambda_{15}w_2.
	\end{split}
	\label{eq:app422_masses}
\end{align}

\subsubsection{Four-doublet Higgs basis}

Collect the four color-singlet doublets into
\begin{equation}
	H_I=
	\left(
	H_1^{(1)},
	H_1^{(2)},
	H_{15}^{(1)},
	H_{15}^{(2)}
	\right),
	\qquad
	I=1,\ldots,4,
	\label{eq:app422_HI}
\end{equation}
with vacuum expectation values
\begin{equation}
	v_I=(u_1,u_2,w_1,w_2),
	\qquad
	v^2=\sum_{I=1}^4|v_I|^2.
	\label{eq:app422_vI}
\end{equation}
The color-singlet electroweak interactions before the Higgs-basis rotation are
\begin{align}
	{\cal L}^{422}_{\rm EW}={}&
	\sum_{I=1}^4
	\Big[
	Q_i\cdot H_I\,(\kappa_{uI})_{ij}u_j^c
	+
	L_i\cdot H_I\,(\kappa_{\nu I})_{ij}\nu_j^c
	+
	H_I^\dagger Q_i\,(\kappa_{dI})_{ij}d_j^c
	+
	H_I^\dagger L_i\,(\kappa_{eI})_{ij}e_j^c
	\Big]
	+{\rm h.c.}.
	\label{eq:app422_EW_prebasis}
\end{align}
The vacuum-aligned Yukawa matrices are
\begin{align}
	Y_{u,\nu}
	&=
	\sum_{I=1}^4\frac{v_I}{v}\,
	\kappa_{u,\nu I}\,,\qquad
	Y_{d,e}
	=
	\sum_{I=1}^4\frac{v_I^*}{v}\,
	\kappa_{d,e I}\,.
	\label{eq:app422_Yf}
\end{align}
These are the matrices entering the corresponding fermion mass
matrices.

In the charged sector, the direction parallel to the vacuum is the
electroweak Goldstone boson. We therefore write
\begin{equation}
	H_I^+
	=
	\frac{v_I}{v}\,G^+
	+
	\sum_{a=1}^3 W_{Ia}H_a^+\,,
	\qquad
	\sum_{I} W^*_{Ia} W_{Ib}=\delta_{a,b}\,,
	\qquad
	\sum_I v_I^*W_{Ia}=0 \,.
	\label{eq:app422_charged_rotation}
\end{equation}
where the three $H_a^+$ span the charged-scalar subspace orthogonal
to the Goldstone direction. Their Yukawa couplings are conveniently
written as
\begin{align}
	B_a^U&=\sum_I W_{Ia}\,\kappa_{uI}\,,
	\qquad
	B_a^N=\sum_I W_{Ia}\,\kappa_{\nu I}\,,
    \qquad
	A_a^D=\sum_I W_{Ia}^*\,\kappa_{dI}\,,
    \qquad
	A_a^E=\sum_I W_{Ia}^*\,\kappa_{eI}\,,\label{eq:app422_charged_coefficients}
\end{align}
such that the charged-scalar interactions are
\begin{align}
	{\cal L}^{422}_{H^\pm}={}&
	\sum_{a=1}^3
	H_a^+
	\left[
	-d_i(B_a^U)_{ij}u_j^c
	-e_i(B_a^N)_{ij}\nu_j^c
	\right]\,
	+\,
	\sum_{a=1}^3
	H_a^-
	\left[
	u_i(A_a^D)_{ij}d_j^c
	+\nu_i(A_a^E)_{ij}e_j^c
	\right]
	+{\rm h.c.}.
	\label{eq:app422_Hpm}
\end{align}

Thus the physical fermion masses determine the vacuum-aligned matrices $Y_f$, but not the couplings of the orthogonal charged-scalar directions. The latter depend on the underlying Pati--Salam Yukawa matrices and on the charged-scalar mixing encoded in the matrix $W$.\\

For real vacuum expectation values, it is useful to parametrize
\begin{align}
	u_1&=u\cos\beta\,,\qquad
	u_2=u\sin\beta\,,\qquad
	w_1=w\cos\gamma\,,\qquad
	w_2=w\sin\gamma,
	\label{eq:app422_simple_vevs}
\end{align}
with $v^2=u^2+w^2$. 
The angles $\beta$ and $\gamma$ are independent.

Within each bidoublet sector, we may separate the charged direction parallel to its vacuum from the orthogonal one:
\begin{alignat}{2}
	G_1^+ &=
	\cos\beta\,H_1^{(1)+} +
	\sin\beta\,H_1^{(2)+}\,,\qquad
	H_{\Phi_1}^+ &&=
	-\sin\beta\,H_1^{(1)+} +
	\cos\beta\,H_1^{(2)+}\,,\\
	G_{15}^+ &=
	\cos\gamma\,H_{15}^{(1)+} +
	\sin\gamma\,H_{15}^{(2)+}\,,\qquad
	H_{\Phi_{15}}^+ &&=
	-\sin\gamma\,H_{15}^{(1)+} +
	\cos\gamma\,H_{15}^{(2)+}\,.
\end{alignat}
The electroweak Goldstone boson and the orthogonal combination are
\begin{align}
	G^+ =	
	\frac{u}{v}G_1^+ 
	+ \frac{w}{v}G_{15}^+\,,\qquad 
	H_R^+ =	
	-\frac{w}{v}G_1^+
	+ \frac{u}{v}G_{15}^+\,.
\end{align}
A convenient orthonormal basis for the three-dimensional charged
subspace orthogonal to the Goldstone is therefore
\begin{equation}
	H_{\Phi_1}^+,\qquad
	H_{\Phi_{15}}^+,\qquad
	H_R^+.
	\label{eq:app422_physical_charged_basis}
\end{equation}
In terms of Eq.~\eqref{eq:app422_charged_rotation}, this corresponds
to the choice
\begin{equation}
	W=
	\begin{pmatrix}
	-\sin\beta & 0 & -\dfrac{w}{v}\cos\beta\\[2mm]
	 \cos\beta & 0 & -\dfrac{w}{v}\sin\beta\\[2mm]
	0 & -\sin\gamma & \dfrac{u}{v}\cos\gamma\\[2mm]
	0 &  \cos\gamma & \dfrac{u}{v}\sin\gamma
	\end{pmatrix},
\end{equation}
where the columns correspond respectively to
$H_{\Phi_1}^+$, $H_{\Phi_{15}}^+$, and $H_R^+$.

In general these states mix through the charged-scalar mass matrix.  In the following we consider the simple limit in which this mixing is negligible, so that the three fields in Eq.~\eqref{eq:app422_physical_charged_basis} can be identified with the physical charged-scalar mass eigenstates.

We now specialize to the flavor structure relevant for the discussion
in the main text. We assume that the singlet bidoublet $\Phi_1$ couples only to the third generation, whereas the adjoint bidoublet $\Phi_{15}$ couples only to the second.  Thus, the only nonvanishing entries are
\begin{align}
	\begin{split}
		(\lambda_1)_{33}&\equiv\lambda_{33}\,,\qquad 
		(\widetilde\lambda_1)_{33}\equiv\widetilde\lambda_{33}\,,\qquad	(\lambda_{15})_{22}\equiv\lambda_{22}\,,\qquad
		(\widetilde\lambda_{15})_{22}\equiv\widetilde\lambda_{22}.
	\end{split}
	\label{eq:app422_simple_texture}
\end{align}

The vacuum-aligned Yukawa couplings are
\begin{alignat}{2}
	(Y_u)_{33}=(Y_{\nu})_{33} &=
	\frac{u}{v}
	\left(
	\lambda_{33}\cos\beta
	+
	\widetilde\lambda_{33}\sin\beta
	\right)\,,\qquad \qquad
	(Y_d)_{33} = (Y_e)_{33} = &&
	\frac{u}{v}
	\left(
	\widetilde\lambda_{33}\cos\beta
	+
	\lambda_{33}\sin\beta
	\right)\,,\\
	(Y_u)_{22} = - \frac{1}{3} (Y_{\nu})_{22}&=
	c_{15}\frac{w}{v}
	\left(
	\lambda_{22}\cos\gamma
	+
	\widetilde\lambda_{22}\sin\gamma
	\right)\,,\qquad
	(Y_d)_{22} = -\frac{1}{3}(Y_e)_{22} = &&
	c_{15}\frac{w}{v}
	\left(
	\widetilde\lambda_{22}\cos\gamma
	+
	\lambda_{22}\sin\gamma
	\right).
\end{alignat}

The charged direction orthogonal to the $\Phi_1$ vacuum couples to the third generation as
\begin{align}
      (B_{\Phi_1}^U)_{33}=(B_{\Phi_1}^N)_{33}
      &=\widetilde\lambda_{33}\cos\beta-
      \lambda_{33}\sin\beta\,,\qquad
      (A_{\Phi_1}^D)_{33}=(A_{\Phi_1}^E)_{33}
      =\lambda_{33}\cos\beta-
      \widetilde\lambda_{33}\sin\beta.
      \label{eq:app422_a3b3}
\end{align}
These are the combinations orthogonal to the vacuum direction. For fixed $u/v$ and
$\beta$, the underlying invariants, and hence
$(A_{\Phi_1})_{33}$ and $(B_{\Phi_1})_{33}$,
are fixed by the corresponding entries of $Y_u$ and $Y_d$ whenever
$\cos2\beta\neq0$.  Explicitly,
\begin{equation}
      \begin{aligned}
              (A_{\Phi_1}^D)_{33}=(A_{\Phi_1}^E)_{33}
              &=\frac{(Y_u)_{33}-\sin2\beta\,(Y_d)_{33}}
              {(u/v)\cos2\beta}\,,\qquad
              (B_{\Phi_1}^U)_{33}=(B_{\Phi_1}^N)_{33}
              =\frac{(Y_d)_{33}-\sin2\beta\,(Y_u)_{33}}
              {(u/v)\cos2\beta}.
      \end{aligned}
      \label{eq:app422_a3b3_Y}
\end{equation}

Similarly, the charged direction orthogonal to the $\Phi_{15}$ vacuum
couples to the second generation as
\begin{align}
      (B_{\Phi_{15}}^U)_{22}
      &=\frac{(B_{\Phi_{15}}^N)_{22}}{-3}=c_{15}\left(
      \widetilde\lambda_{22}\cos\gamma-
      \lambda_{22}\sin\gamma
      \right)\,,\qquad      (A_{\Phi_{15}}^D)_{22}
      =\frac{(A_{\Phi_{15}}^E)_{22}}{-3}=c_{15}\left(
      \lambda_{22}\cos\gamma-
      \widetilde\lambda_{22}\sin\gamma
      \right)\,,
      \label{eq:app422_a2b2}
\end{align}
For fixed $w/v$ and $\gamma$, these couplings are determined by the corresponding entries of $Y_u$ and $Y_d$ whenever $\cos2\gamma\neq0$.
Explicitly,
\begin{equation}
      \begin{aligned}
              (A_{\Phi_{15}}^D)_{22}
              &=\frac{(A_{\Phi_{15}}^E)_{22}}{-3}=\frac{v}{w}\,
              \frac{(Y_u)_{22}-\sin2\gamma\,(Y_d)_{22}}{\cos2\gamma}\,,
              \qquad 
              (B_{\Phi_{15}}^U)_{22}
              =\frac{(B_{\Phi_{15}}^N)_{22}}{-3}=\frac{v}{w}\,
              \frac{(Y_d)_{22}-\sin2\gamma\,(Y_u)_{22}}{\cos2\gamma}\,,
      \end{aligned}
      \label{eq:app422_a2b2_Y}
\end{equation}
The remaining charged scalar $H_R^\pm$ is constructed from the two vacuum-aligned directions. Its couplings in Eq.~\eqref{eq:app422_Hpm} are consequently fixed by the physical Yukawa matrices,
\begin{align}
\begin{split}
(B_R^U)_{33}&=(B_R^N)_{33}=-\frac{w}{u}\,(Y_u)_{33},
\qquad
(A_R^D)_{33}=(A_R^E)_{33}=-\frac{w}{u}\,(Y_d)_{33},
\\
(B_R^U)_{22}&=-\frac13(B_R^N)_{22}=\frac{u}{w}\,(Y_u)_{22},
\qquad
(A_R^D)_{22}=-\frac13(A_R^E)_{22}=\frac{u}{w}\,(Y_d)_{22}.
\end{split}
\label{eq:app422_HR_couplings}
\end{align}

\subsubsection{Scalar-leptoquark couplings}

In $G_{422}$, each adjoint bidoublet gives two $SU(4)_C$ adjoint electroweak doublets after $SU(2)_R$ breaking. Each of these contains a pair of scalar leptoquark doublets. We denote them by
\begin{equation}
	R_{2r}\sim(3,2,7/6),
	\qquad
	\widetilde R_{2r}^c\sim(\bar 3,2,-1/6),
	\qquad
	r=1,2.
	\label{eq:app422_LQfields}
\end{equation}
The two copies are distinct fields below $SU(2)_R$ breaking, but their
Yukawa matrices are correlated by the original bidoublet interaction.
Define
\begin{align}
	\Lambda_u^{(1)}=\lambda_{15}\,,\qquad
	\Lambda_d^{(1)}=\widetilde\lambda_{15}\,,\qquad
	\Lambda_u^{(2)}=\widetilde\lambda_{15}\,,\qquad
	\Lambda_d^{(2)}=\lambda_{15}.
	\label{eq:app422_LQ_Lambda}
\end{align}
The scalar-leptoquark interactions then take the same form as in the
$G_{421}$ theory,
\begin{align}
	{\cal L}^{422}_{\rm LQ}={}&
	r_{15}\sum_{r=1}^2(\Lambda_u^{(r)})_{ij}
	\left[
	(L_i\cdot R_{2r})u_j^c
	+
	(Q_i \cdot \widetilde R_{2r}^c)\nu_j^c
	\right]
	+
	r_{15}\sum_{r=1}^2(\Lambda_d^{(r)})_{ij}
	\left[
	(R_{2r}^\dagger Q_i)e_j^c
	+
	((\widetilde R_{2r}^c)^\dagger L_i)d_j^c
	\right]
	+{\rm h.c.}.
	\label{eq:app422_LQ}
\end{align}
To make contact with the scalar-leptoquark notation commonly used in
the literature, we write $\widetilde R_{2r}^c = -\epsilon \widetilde R_{2r}^*$, with
\begin{equation}
	R_{2r}=
	\begin{pmatrix}
		R_{2r}^{5/3}\\
		R_{2r}^{2/3}
	\end{pmatrix},
	\qquad
	\widetilde R_{2r}=
	\begin{pmatrix}
		\widetilde R_{2r}^{2/3}\\
		\widetilde R_{2r}^{-1/3}
	\end{pmatrix},
\end{equation}
this becomes
\begin{align}
	\begin{split}
		{\cal L}^{422}_{\rm LQ}={}&
		r_{15}\sum_{r=1}^2(\Lambda_u^{(r)})_{ij}
		\Big[
		(\nu_iR_{2r}^{2/3}-e_iR_{2r}^{5/3})u_j^c
		+
		(\widetilde R_{2r}^{2/3\,*}u_i+
		\widetilde R_{2r}^{-1/3\,*}d_i)\nu_j^c
		\Big] \\
		&+
		r_{15}\sum_{r=1}^2(\Lambda_d^{(r)})_{ij}
		\Big[
		(R_{2r}^{5/3\,*}u_i+R_{2r}^{2/3\,*}d_i)e_j^c
		-
		(\nu_i\widetilde R_{2r}^{-1/3}
		-e_i\widetilde R_{2r}^{2/3})d_j^c
		\Big]
		+{\rm h.c.}.
	\end{split}
	\label{eq:app422_LQ_expanded}
\end{align}
The corresponding color-octet interactions, with $\Phi_{8r} \sim (8,2,1/2)$, are
\begin{equation}
	{\cal L}^{422}_{8}=
	\sum_{r=1}^2
	\left[
	(\Lambda_u^{(r)})_{ij}(Q_i\cdot\Phi_{8r})u_j^c
	+
	(\Lambda_d^{(r)})_{ij}(\Phi_{8r}^\dagger Q_i)d_j^c
	\right]
	+{\rm h.c.}.
	\label{eq:app422_octet}
\end{equation}

Equations~\eqref{eq:app422_Hpm} and \eqref{eq:app422_LQ} summarize the
interactions most relevant for threshold corrections to the
quark--lepton Yukawa relations. The measured fermion masses determine
only the combinations $Y_f$ selected by the electroweak vacuum. The
physical charged scalars probe orthogonal combinations of the underlying
Pati--Salam Yukawa matrices, while the scalar leptoquarks couple directly through $\lambda_{15}$ and $\widetilde\lambda_{15}$. Consequently,
neither set of couplings is, in general, fixed by the corresponding small Standard Model Yukawa eigenvalues. This freedom becomes especially important when the adjoint vacuum expectation values are smaller than $v$, in which case the underlying adjoint Yukawa couplings can be parametrically enhanced.

\section{Finite threshold corrections in Pati-Salam Theories}
\label{app:thresholds}

We collect here the finite scalar threshold corrections relevant for the Yukawa ratios. We first diagonalize the tree-level Yukawa matrices,
\begin{equation}
	\widehat Y_f
	\equiv
	U_f^\dagger Y_f V_f
	=
	\operatorname{diag}(y_{f_1},y_{f_2},y_{f_3}),
	\qquad
	y_{f_i}>0,
	\label{eq:app_thr_tree_svd}
\end{equation}
where $U_f$ and $V_f$ denote the left- and right-handed rotations, respectively. All one-loop corrections are then evaluated in this tree-level mass basis. If
\begin{equation}
	\widehat Y_f\longrightarrow \widehat Y_f+\delta\widehat Y_f,
	\qquad
	\delta\widehat Y_f\equiv U_f^\dagger\delta Y_fV_f,
\end{equation}
the strict one-loop shift is simply
\begin{equation}
	\delta y_{f_i}
	=
	\operatorname{Re}
	\left(\delta\widehat Y_f\right)_{ii}.
	\label{eq:app_thr_massshift}
\end{equation}
Indeed, the off-diagonal entries of $\delta\widehat Y_f$ induce additional rotations of the mass eigenstates, but their effect on the singular values starts only at quadratic order in $\delta Y_f$.  Since $\delta Y_f$ is already a one-loop quantity, such effects lie beyond the strict one-loop matching considered here. One should therefore first diagonalize the tree-level matrices and then project the one-loop correction onto the corresponding tree-level mass eigenstates, without rediagonalizing the corrected matrix. For the physical Yukawa ratios we define
\begin{equation}
	\epsilon_i\equiv
	\frac{\delta y_{e_i}}{y_{e_i}}
	-
	\frac{\delta y_{d_i}}{y_{d_i}}\,,
	\label{eq:app_thr_epsilon}
\end{equation}
so that at one loop
\begin{equation}
	R_i
	=
	R_i^{\rm tree}\left(1+\epsilon_i\right)\,.
	\label{eq:app_thr_ratio}
\end{equation}
Writing $R_i^{\rm tree}=1+\Delta_i$, where $\Delta_i$ are the corrections from the flavor diagonalization, this reproduces Eq.~\eqref{eq:matching_condition}, $R_i=(1+\Delta_i)(1+\epsilon_i)$, used in the main text.

\subsection{Massive Gauge Bosons}

The massive gauge bosons associated with Pati--Salam breaking generate one-loop wave-function and vertex corrections.  For a one-step breaking spectrum, the colored vector leptoquarks have masses of order $M_X\sim g_4\Lambda_{\rm PS}$.  Matching at $\mu\simeq M_X$, their
contribution is
\begin{equation}
	\epsilon_i \simeq
	\frac{g_4^2}{4\pi^2}\frac{5}{8}.
	\label{eq:vector_threshold}
\end{equation}
For the values of $g_4$ relevant here this is of order one percent; the massive neutral-vector contribution is smaller. These gauge thresholds are model-independent consequences of embedding color and lepton number into $SU(4)_C$, and are therefore generically present in realistic quark--lepton unification theories, independently of the details of the scalar or flavor sector. In particular, they arise in both the $G_{421}$ and $G_{422}$ realizations considered here. Gauge thresholds thus provide an unavoidable baseline correction to the quark--lepton mass relations, although numerically their effect remains small.

\subsection{Charged-scalar threshold corrections}
\label{app:charged_scalar_thresholds}

We first describe the charged-scalar correction in a form that is independent of the particular Pati--Salam texture.  We use the notation of Appendix~\ref{app:PStheory}, but from now on all charged-scalar coupling matrices are defined in the tree-level fermion mass basis,
\begin{align}
\widehat B_a^U&\equiv U_d^\dagger B_a^U V_u,
&
\widehat A_a^D&\equiv U_u^\dagger A_a^D V_d,
\nonumber\\
\widehat B_a^N&\equiv U_e^\dagger B_a^NV_\nu,
&
\widehat A_a^E&\equiv U_\nu^\dagger A_a^EV_e.
\label{eq:app_charged_massbasis}
\end{align}
Accordingly, the fermion fields appearing below denote tree-level mass
eigenstates, and the charged-scalar interactions take the form
\begin{align}
\mathcal L_{H^\pm}
={}&
\sum_a H_a^+
\left[
-d_i(\widehat B_a^U)_{ij}u_j^c
-e_i(\widehat B_a^N)_{ij}\nu_j^c
\right]
+
\sum_a H_a^-
\left[
 u_i(\widehat A_a^D)_{ij}d_j^c
+\nu_i(\widehat A_a^E)_{ij}e_j^c
\right]
+\mathrm{h.c.}
\label{eq:app_charged_generic}
\end{align}
Here $a$ labels the charged-scalar directions orthogonal to the
electroweak Goldstone. For $G_{421}$ the sum contains a single state,
$a=\perp$, while for $G_{422}$ it runs over the three physical charged
directions.

The down-quark one-loop self-energy contains a chirality-changing contribution proportional to the internal up-type mass. In the
$\overline{\mathrm{MS}}$ scheme, and neglecting corrections of order $m_{d_i}^2/m_{H_a}^2$ in the loop functions, the chirality-enhanced contribution to the fractional Yukawa shift is
\begin{align}
\left.
\frac{\delta y_{d_i}}{y_{d_i}}
\right|_{H^\pm}
=
\frac{1}{16\pi^2}
\sum_{a,j}
\frac{m_{u_j}}{m_{d_i}}
\operatorname{Re}\!\left[
(\widehat B_a^U)_{ij}(\widehat A_a^D)_{ji}
\right]
B_0\!
\left(
0,m_{u_j}^2,m_{H_a}^2 \right),
\label{eq:app_generic_down_threshold}
\end{align}
where
\begin{equation}
B_0(0,m^2,M^2)
\equiv
1-
\frac{
m^2\ln(m^2/\mu^2)
-
M^2\ln(M^2/\mu^2)
}{
m^2-M^2
}
\label{eq:app_B0bar}
\end{equation}
is the finite $\overline{\mathrm{MS}}$ part of the scalar two-point function, and $\mu$ is the $\overline{\mathrm{MS}}$ renormalization scale.

In a Pati--Salam symmetric theory, the charged-lepton
self-energy has the same structure as the down-quark one, with the
internal up-type quark replaced by its neutral-fermion partner. PS symmetry relates the Dirac neutrino mass $m_{D_j}$ to the mass of the corresponding up-type quark, with the relative Clebsch factor determined by the scalar multiplet generating the entry; in the cases of interest here, $m_{D_j}=m_{u_j}$ or $m_{D_j}=-3m_{u_j}$, up to unphysical phases. The same Clebsch structure relates the corresponding charged-lepton and down-quark couplings.  Consequently, in the PS symmetric limit the fractional corrections to the normalized ratios cancel,
\begin{equation}
\epsilon_i^H
\equiv
\frac{\delta y_{e_i}}{y_{e_i}}
-
\frac{\delta y_{d_i}}{y_{d_i}}
=0.
\label{eq:app_ratio_threshold_general}
\end{equation}

A realistic low-scale Pati--Salam theory must, however, substantially
modify the neutral-fermion sector, since Dirac neutrino masses of order the corresponding up-quark masses are incompatible with observation. This breaking of the quark--lepton correspondence can generate a nonvanishing, and potentially sizable, correction to the ratios $R_i$.

To illustrate the mechanism, introduce for each relevant neutral fermion a gauge-singlet left-handed Weyl fermion $N_j$ and a PS-breaking scalar whose vacuum expectation value generates the mass term $M_j N_j\nu_j^c$.  Neglecting the small lepton-number-violating parameters responsible for the observed light-neutrino masses, the neutral-fermion mass matrix for each generation is 
\begin{equation}
\mathcal M_{\nu_j}
=
\begin{pmatrix}
0       & m_{D_j} & 0 \\
m_{D_j} & 0       & M_j \\
0       & M_j     & 0
\end{pmatrix}
\qquad
\text{in the basis }
(\nu_j,\nu_j^c,N_j).
\label{eq:app_neutral_mass_matrix_general}
\end{equation}
Defining
\begin{equation}
m_{S_j}
=
\sqrt{M_j^2+m_{D_j}^2},
\qquad
s_{\theta_j}
=
\frac{m_{D_j}}{m_{S_j}},
\qquad
c_{\theta_j}
=
\frac{M_j}{m_{S_j}},
\label{eq:app_neutral_mixing_general}
\end{equation}
the massive Dirac pair is
\begin{equation}
S_j
=
s_{\theta_j}\nu_j
+
c_{\theta_j}N_j,
\qquad
S_j^c=\nu_j^c,
\qquad
m_{S_j}s_{\theta_j}=m_{D_j},
\label{eq:app_heavy_dirac_general}
\end{equation}
while the orthogonal light state is
\begin{equation}
\nu_{j,\mathrm{light}}
=
c_{\theta_j}\nu_j
-
s_{\theta_j}N_j.
\label{eq:app_light_neutrino_general}
\end{equation}
The mass matrix in Eq.~\eqref{eq:app_neutral_mass_matrix_general} gives an exactly massless state, whose mass can be lifted by a Majorana mass term $(\mu_N)_{ij}N_iN_j$ through an inverse-seesaw scaling $m_\nu\simeq\mu_N(m_D/M_N)^2$~\cite{Mohapatra:1986aw}. 
Equation~\eqref{eq:app_charged_generic} therefore becomes
\begin{align}
\mathcal L_{H^\pm}
\supset{}&
\sum_a H_a^+
\left[
-e_i(\widehat B_a^N)_{ij}S_j^c
\right]
+
\sum_a H_a^-
\left(
s_{\theta_i}S_i
+
c_{\theta_i}\nu_{i,\mathrm{light}}
\right)
(\widehat A_a^E)_{ij}e_j^c
+\mathrm{h.c.}
\label{eq:app_lepton_interactions_mass_basis}
\end{align}
where sums over repeated flavor indices are understood.

The light neutral state does not contribute to the chirality-enhanced
part of the charged-lepton self-energy, since its mass can be neglected.
The heavy state instead gives
\begin{align}
\left.
\frac{\delta y_{e_i}}{y_{e_i}}
\right|_{H^\pm}
&=
\frac{1}{16\pi^2}
\sum_{a,j}
\frac{m_{S_j}s_{\theta_j}}{m_{e_i}}
\operatorname{Re}\!\left[
(\widehat B_a^N)_{ij}(\widehat A_a^E)_{ji}
\right]
B_0
\left(
0,m_{S_j}^2,m_{H_a}^2
\right)
\nonumber\\
&=
\frac{1}{16\pi^2}
\sum_{a,j}
\frac{m_{D_j}}{m_{e_i}}
\operatorname{Re}\!\left[
(\widehat B_a^N)_{ij}(\widehat A_a^E)_{ji}
\right]
B_0
\left(
0,m_{S_j}^2,m_{H_a}^2
\right).
\label{eq:app_generic_lepton_threshold_mixed}
\end{align}
The factor $m_{S_j}$ from the chirality flip on the heavy-fermion
propagator is thus exactly accompanied by the mixing factor
$s_{\theta_j}$, so that the numerator remains proportional to the
Dirac mass $m_{D_j}$.

Combining this result with the corresponding down-quark correction,
the chirality-enhanced charged-scalar contribution to the normalized
Pati--Salam ratio is
\begin{align}
\epsilon_{i}^H
={}&
\frac{1}{16\pi^2}
\sum_{a,j}
\Bigg[
\frac{m_{D_j}}{m_{e_i}}
\operatorname{Re}\!\left[
(\widehat B_a^N)_{ij}(\widehat A_a^E)_{ji}
\right]
B_0
\left(
0,m_{S_j}^2,m_{H_a}^2
\right)
-
\frac{m_{u_j}}{m_{d_i}}
\operatorname{Re}\!\left[
(\widehat B_a^U)_{ij}(\widehat A_a^D)_{ji}
\right]
B_0
\left(
0,m_{u_j}^2,m_{H_a}^2
\right)
\Bigg].
\label{eq:app_generic_ratio_threshold}
\end{align}
In the exact Pati--Salam limit, $M_j\to0$, so that
$m_{S_j}\to |m_{D_j}|$, while the masses and charged-scalar
couplings satisfy their Pati--Salam relations.  The two terms in
Eq.~\eqref{eq:app_generic_ratio_threshold} then coincide, including
the appropriate Clebsch factors entering the definition of $R_i$,
and hence $\epsilon_i^H\to0$.  A nonzero threshold correction arises
only after the Pati--Salam-breaking neutral mass $M_j$ resolves the
neutral-fermion propagator differently from its up-quark counterpart.\\

In the textures considered in this work, the $(3,3)$ entry of the Yukawa matrices is always generated by the color singlet $\Phi_1$ and it is naturally among the largest ones. Therefore, the largest such enhancement is expected to involve the third generation in the loop. With $m_{D_3}\simeq m_t$ and $m_{S_3}^2=M_3^2+m_{D_3}^2$, the correction to the third-generation ratio from a charged scalar $H_a^\pm$ takes the form
\begin{equation}
\epsilon_3^{H_{\Phi_1}^+}
\simeq
-
\frac{y_t}{16\pi^2y_b}
\operatorname{Re}\!\left[
(\widehat B_{\Phi_1})_{33}(\widehat A_{\Phi_1})_{33}
\right]
\left[
F\!\left(\frac{m_{S_3}^2}{m_{H_{\Phi_1}^+}^2}\right)
-
F\!\left(\frac{m_t^2}{m_{H_{\Phi_1}^+}^2}\right)
\right],
\qquad
F(r)\equiv\frac{r\ln r}{r-1}\,,
\qquad F(1)=1\,,
\label{eq:app_R3_top_dominant}
\end{equation}
where $\widehat B_{\Phi_1} =  \widehat B_{\Phi_1}^U = \widehat B_{\Phi_1}^N$ and $\widehat A_{\Phi_1} = \widehat A_{\Phi_1}^D = \widehat A_{\Phi_1}^E$. We note that the common scale-dependent terms cancel between the leptonic and quark contributions.
The charged-scalar couplings entering Eq.~\eqref{eq:app_R3_top_dominant} depend on the realization of the PS symmetry.\\

In the $SU(4)_C\times SU(2)_L\times SU(2)_R$ model, assuming the $(3,3)$ entry is generated by the singlet bidoublet $\Phi_1\sim(1,2,2)$ and is the largest entry of the Yukawa matrices and neglecting the mixing between the charged scalars, we have
\begin{equation}
(\widehat B_{\Phi_1})_{33}\simeq -\frac{\sin2\beta\,y_t-y_b}{(u/v)\cos2\beta},
\qquad
(\widehat A_{\Phi_1})_{33}\simeq \frac{y_t-\sin2\beta\,y_b}
	{(u/v)\cos2\beta},
\label{eq:app_BA_a3b3_422}
\end{equation}
where we used the notation of Appendix~\ref{app:PStheory} In deriving the approximate expressions above and in what follows, we assume that the relevant Yukawa textures are sufficiently hierarchical and approximately aligned with the fermion mass basis, such that their dominant diagonal entries can be identified with the corresponding Yukawa eigenvalues. Therefore, the product $(\widehat B_{\Phi_1})_{33}(\widehat A_{\Phi_1})_{33}$ of the corresponding charged-scalar couplings has no definite sign. The charged-scalar threshold can therefore either increase or decrease $R_3$.  If the physical charged scalars are sizable mixtures of the different charged directions, Eq.~\eqref{eq:app_R3_top_dominant} is applied to each mass eigenstate with the corresponding hatted matrices.

\begin{figure}
    \centering
    \includegraphics[width=0.7\textwidth]{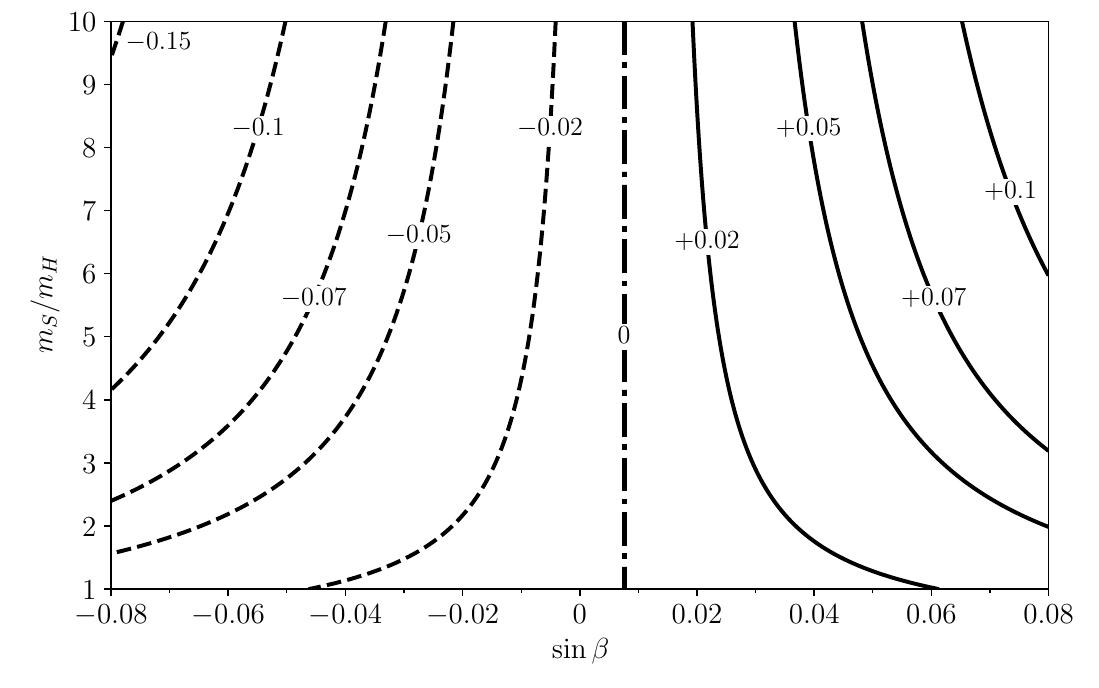}
    \caption{Chirality enhanced charged scalar threshold correction to $y_{\tau}/y_b$, $\epsilon_3^H$, as a function of the heavy neutral-fermion to charged-scalar mass ratio $m_S/m_H$ and $\sin\beta = u_2/(\sqrt{u_1^2+u_2^2})$ in $G_{422}$. For simplicity, $u=v\simeq174$~GeV has been assumed here.}
    \label{fig:3gen_threshold}
\end{figure}

The situation is more restrictive in the
$SU(4)_C\times SU(2)_L\times U(1)_R$, where the physical charged scalar, $H_\perp^\pm$, has couplings
\begin{equation}
(\widehat B_{\Phi_1})_{33}\simeq-y_t\tan\beta,
\qquad
(\widehat A_{\Phi_1})_{33}\simeq-y_b\tan\beta.
\end{equation}
Hence in the $421$ model the sign of the dominant third-family threshold is fixed,
\begin{equation}
\epsilon_3^{H_\perp,\,421}
\simeq
-\frac{y_t^2\tan^2\beta}{16\pi^2}
\left[
F\!\left(\frac{m_{S_3}^2}{m_{H_\perp}^2}\right)
-
F\!\left(\frac{m_t^2}{m_{H_\perp}^2}\right)
\right]
<0,
\label{eq:app_epsilon3_421}
\end{equation}
because $m_{S_3}>m_t$ and
$F(r)=r\ln r/(r-1)$ is monotonically increasing for $r>0$. Thus the dominant charged-scalar correction always lowers $R_3$ in the singlet-supported $421$ realization, whereas its sign is unrestricted in the $422$ theory.

Corrections to $R_2$ and $R_1$ follow from the same general expressions after including the appropriate flavor rotations and charged-scalar projections.

Finally, a comment is in order. The hierarchy of threshold corrections among the three generations need not follow directly the hierarchy of the SM Yukawa couplings. If a given fermion mass is generated by a scalar whose electroweak vacuum expectation value is parametrically smaller than $v$, the corresponding fundamental Yukawa couplings can be much larger than the effective SM Yukawas. This possibility is
particularly relevant for the second generation in the $422$ realization, where the $(15,2,2)$ bidoublet may acquire a small vev. The charged-scalar couplings entering the correction to $R_2$ can
then be $\mathcal O(1)$ even though $y_c$ and $y_s$ are small, so that additional sizable threshold corrections to the lighter-generation
ratios are possible.

\subsection{Scalar-leptoquark threshold corrections}
\label{app:leptoquark_thresholds}

The adjoint scalar multiplets contain scalar leptoquarks whose
interactions were derived in the preceding appendix.  As for the
charged scalars, it is convenient to formulate the threshold directly
in the fermion mass basis.  We denote by
$\widehat\Lambda^{u,d}$ the physical couplings of a scalar
leptoquark $R^{5/3}$,
\begin{equation}
	\mathcal L_{\rm LQ}
	\supset
	-R^{5/3}\,
	e_i\,(\widehat\Lambda^u)_{ij}\,u_j^c
	+
	R^{5/3\,*}\,
	u_j\,(\widehat\Lambda^d)_{ji}\,e_i^c
	+\mathrm{h.c.},
	\label{eq:app_LQ_physical_couplings}
\end{equation}
where the normalization of the adjoint component, the fermion
rotations, and the scalar mixing matrices are included in
$\widehat\Lambda^{u,d}$.  The corresponding chirality-enhanced
charged-lepton correction is
\begin{equation}
	\left.
	\frac{\delta y_{e_i}}{y_{e_i}}
	\right|_{\rm LQ}
	=
	\frac{3}{16\pi^2}
	\sum_j
	\frac{m_{u_j}}{m_{e_i}}
	\operatorname{Re}
	\left[
	(\widehat\Lambda^u)_{ij}
	(\widehat\Lambda^d)_{ji}
	\right]
	B_0\!
	\left(
	0,m_{u_j}^2,m_R^2
	\right).
	\label{eq:app_LQ_lepton_generic}
\end{equation}
The factor of $3$ arises from the color trace of the internal
up-type quark.

The Pati--Salam-related correction to the down-quark Yukawa involves the second leptoquark doublet, $\widetilde R_2$, and the neutral fermion.  Defining the corresponding signed mass-basis couplings by
\begin{align}
	\mathcal L_{\widetilde{\rm LQ}}
	\supset{}&
	-\widetilde R^{-1/3\,*}\,
	d_i\,(\widehat\Lambda^N)_{ij}\,S_j^c
	+
	\widetilde R^{-1/3}
	\left(
	s_{\theta_j}S_j
	+
	c_{\theta_j}\nu_{j,\rm light}
	\right)
	(\widehat\Lambda^E)_{ji}\,d_i^c
	+\mathrm{h.c.},
	\label{eq:app_tildeLQ_physical_couplings}
\end{align}
the light state again does not contribute to the chirality-enhanced
piece. The matrices $\widehat\Lambda^{N,E}$ originate from the same interaction-basis Yukawa couplings as $\widehat\Lambda^{u,d}$, respectively, but involve different fermion rotations to the tree-level mass eigenstate basis. Using $m_{S_j}s_{\theta_j}=m_{D_j}$ gives
\begin{equation}
	\left.
	\frac{\delta y_{d_i}}{y_{d_i}}
	\right|_{\widetilde{\rm LQ}}
	=
	\frac{1}{16\pi^2}
	\sum_j
	\frac{m_{D_j}}{m_{d_i}}
	\operatorname{Re}
	\left[
	(\widehat\Lambda^N)_{ij}
	(\widehat\Lambda^E)_{ji}
	\right]
	B_0\!
	\left(
	0,m_{S_j}^2,m_{\widetilde R}^2
	\right).
	\label{eq:app_LQ_quark_generic}
\end{equation}
The leptoquark contribution to the normalized ratio is therefore
\begin{align}
	\epsilon_i^{\rm LQ}
	={}&
	\frac{1}{16\pi^2}
	\sum_j
	\Bigg[
	3\frac{m_{u_j}}{m_{e_i}}
	\operatorname{Re}
	\left[
	(\widehat\Lambda^u)_{ij}
	(\widehat\Lambda^d)_{ji}
	\right]
	B_0\!
	\left(
	0,m_{u_j}^2,m_R^2
	\right)
	-
	\frac{m_{D_j}}{m_{d_i}}
	\operatorname{Re}
	\left[
	(\widehat\Lambda^N)_{ij}
	(\widehat\Lambda^E)_{ji}
	\right]
	B_0\!
	\left(
	0,m_{S_j}^2,m_{\widetilde R}^2
	\right)
	\Bigg].
	\label{eq:app_LQ_ratio_generic}
\end{align}
The leptoquark pair alone does not cancel as $M_2\to0$, since it does not form a complete $SU(4)_C$ multiplet. Including the color-singlet and color-octet components of the adjoint, the chirality-enhanced correction cancels in the limit $M_2\to0$, $m_{S_2}\to3m_c$, with degenerate adjoint components, up to terms of order $m_c^2/m_R^2$ reflecting the $SU(4)_C$ breaking by the adjoint vacuum expectation value itself.\\

The leptoquark thresholds are particularly relevant for the second generation.  A small electroweak vacuum expectation value of the adjoint scalar allows the underlying adjoint Yukawa couplings to be substantially larger than the corresponding SM Yukawas.  For a predominantly diagonal second-generation coupling, using $m_{D_2}=3m_c$ and $y_\mu\simeq 3y_s$, the chirality-enhanced
contribution to the normalized ratio is
\begin{equation}
	\epsilon_2^{\rm LQ}
	\simeq
	\frac{1}{16\pi^2}
	\frac{y_c}{y_s}
	\operatorname{Re}\!\left[
	(\widehat\Lambda^u)_{22}
	(\widehat\Lambda^d)_{22}
	\right]
	\left[
    3B_0(0,m_c^2,m_R^2) 
    -
    B_0(0,m_{S_2}^2,m_R^2)
    \right]\,,
	\label{eq:app_LQ_R2_charm}
\end{equation}
where $m_R\simeq m_{\tilde{R}}$ denotes the physical leptoquark masses and $m_{S_2}^2=M_2^2+9m_c^2$ is the mass of the heavy neutral partner.

An even larger chirality enhancement is possible when the adjoint Yukawa matrices contain off-diagonal entries.  In particular, mass-basis couplings $(\widehat\Lambda^u)_{23}$ and
$(\widehat\Lambda^d)_{32}$ generate a contribution to the mass ratio $R_2$ with an internal top quark,
\begin{equation}
	\left.
	\epsilon_2^{\rm LQ}
	\right|_{t}
	\simeq
	\frac{1}{16\pi^2}
	\frac{y_t}{y_s}
	\operatorname{Re}\!\left[
	(\widehat\Lambda^u)_{23}
	(\widehat\Lambda^d)_{32}
	\right]
	\left[
    F\!\left(\frac{m_{S_3}^2}{m_R^2}\right)
    -
    F\!\left(\frac{m_t^2}{m_R^2}\right)
    \right]\,,
	\label{eq:app_LQ_R2_top}
\end{equation}
with $m_{S_3}^2=M_3^2+m_t^2$.
The enhancement by $y_t/y_s$ can compensate small flavor
off-diagonal couplings, making this contribution potentially sizable.
Its sign is controlled by the physical coupling product and is not
fixed by Pati--Salam symmetry.  Unlike in Eq.~\eqref{eq:app_LQ_R2_charm}, the correction vanishes continuously when the neutral partner becomes degenerate with the corresponding up-type quark.\\

In general there is a contribution to $R_3$ with an internal charm or top quark quark,
\begin{align}
	\begin{split}
	\left.
	\epsilon_3^{\rm LQ}
	\right|_{c}
	\simeq&-
	\frac{3}{16\pi^2}
	\frac{y_c}{y_\tau}
	\operatorname{Re}\!\left[
	(\widehat\Lambda^u)_{32}
	(\widehat\Lambda^d)_{23}
	\right]
	\left[
    3B_0(0,m_{S_2}^2,m_R^2)
    -
    B_0(0,m_c^2,m_R^2)
    \right]\,,\\
	\left.
	\epsilon_3^{\rm LQ}
	\right|_{t}
	\simeq&
	\frac{3}{16\pi^2}
	\frac{y_t}{y_\tau}
	\operatorname{Re}\!\left[
	(\widehat\Lambda^u)_{33}
	(\widehat\Lambda^d)_{33}
	\right]
	\left[
    F\!\left(\frac{m_{S_3}^2}{m_R^2}\right)
    -
    F\!\left(\frac{m_t^2}{m_R^2}\right)
    \right]\,.
	\end{split}
	\label{eq:app_LQ_R3}
\end{align}
In the textures considered here, however, the adjoint scalar does not
directly generate the $(3,3)$ Yukawa entry.  The corresponding mass-basis leptoquark couplings are therefore induced only by flavor
rotations and are generically suppressed. For example, a purely
$(2,2)$ adjoint coupling generates a $(3,3)$ coupling only at second
order in the $2$--$3$ rotations, while interaction-basis $(2,3)$ or
$(3,2)$ entries induce it at first order.

Consequently, for comparable flavor couplings, scalar-leptoquark
thresholds can affect $R_2$ much more efficiently than $R_3$, owing
to the possibility of the much larger top-quark chirality
enhancement.\\

The same second-generation adjoint coupling also induces a correction
to $R_1$ through the $1$--$2$ flavor rotations.  Using the texture
relation between the first-generation mass and the $1$--$2$ mixings,
the charm contribution is parametrically
\begin{align}
	\begin{split}
	\left.
	\epsilon_1^{\rm LQ}
	\right|_{c}
	&-\simeq
	\frac{1}{16\pi^2}
	\frac{y_c}{y_d}
	\operatorname{Re}\!\left[
	(\widehat\Lambda^u)_{12}
	(\widehat\Lambda^d)_{21}
	\right]
	\left[
    3B_0(0,m_{S_2}^2,m_R^2)
    -
    B_0(0,m_c^2,m_R^2)
    \right]\\
    &\simeq
	\frac{1}{16\pi^2}
	\frac{y_c}{y_s}
	\operatorname{Re}\!\left[
	(\widehat\Lambda^u)_{22}
	(\widehat\Lambda^d)_{22}
	\right]
	\left[
    3B_0(0,m_{S_2}^2,m_R^2)
    -
    B_0(0,m_c^2,m_R^2)
    \right].
	\end{split}
	\label{eq:app_LQ_R1_charm}
\end{align}
Thus, for the contribution generated by the same $(2,2)$ adjoint
interaction, the threshold corrections to $R_1$ and $R_2$ have
opposite signs and comparable parametric size.\\

These effects are present in both Pati--Salam realizations considered
here. In the $421$ theory the relevant couplings arise from the adjoint scalar doublet, while in the $422$ theory they are determined by the $\lambda_{15}$ and $\widetilde\lambda_{15}$ Yukawa matrices and by the projection of the adjoint bidoublet components onto the physical leptoquark states.  In either case the hatted couplings above denote the corresponding physical mass-basis couplings, and their products do not have a fixed sign in general.

\section{Vector-leptoquark contribution to $K_L\to\mu^\pm e^\mp$}
\label{app:Kmue}

Here we summarize the matching and decay rate relevant for the PS vector leptoquark $X_\mu \sim (3,1,2/3)$. In the fermion mass basis,
\begin{equation}
{\cal L}_X=
\frac{g_4}{\sqrt2}X_\mu\,
\bar d_i\gamma^\mu
\left[(V_L)_{i\alpha}P_L+(V_R)_{i\alpha}P_R\right]\ell_\alpha
+\mathrm{h.c.},
\qquad
V_{L,R}\equiv U_{dL,dR}^\dagger U_{eL,eR}.
\end{equation}
where $U_{dL,dR}$ and $U_{eL,eR}$ are the matrices rotating quark and lepton fields from the interaction to the mass basis. We focus on the rare neutral-kaon decays $K_L\to\ell_a^\pm\ell_b^\mp$, which receive contributions from both $s\to d$ and $d\to s$ transitions. In defining the effective Hamiltonian, we explicitly retain only the $s\to d$ operators and sum over all charged-lepton flavour indices, with the Hermitian-conjugate terms accounting for $d\to s$ transition. We define 
\begin{align}
{\cal H}_{\rm eff}^{sd}
=-\sum_{a,b}\big[
&C_9^{ab}(\bar d\gamma_\mu P_Ls)(\bar\ell_a\gamma^\mu\ell_b)
+C_{10}^{ab}(\bar d\gamma_\mu P_Ls)
 (\bar\ell_a\gamma^\mu\gamma_5\ell_b)
\nonumber\\
&+C_S^{ab}(\bar dP_Rs)(\bar\ell_a\ell_b)
+C_P^{ab}(\bar dP_Rs)(\bar\ell_a\gamma_5\ell_b)
+(C_iO_i\to C_i'O_i')
\big]+\mathrm{h.c.},
\end{align}
where the primed operators are obtained by interchanging $P_L\leftrightarrow P_R$ in the quark bilinears.
Tree-level exchange of $X_\mu$ gives, at $\mu=M_X$,
\begin{align}
C_9^{ab}=-C_{10}^{ab}
&=-\frac{g_4^2}{4M_X^2}(V_L)_{db}(V_L)_{sa}^*,
&
C_9^{\prime ab}=C_{10}^{\prime ab}
&=-\frac{g_4^2}{4M_X^2}(V_R)_{db}(V_R)_{sa}^*,
\nonumber\\
C_S^{ab}=-C_P^{ab}
&=\frac{g_4^2}{2M_X^2}(V_L)_{db}(V_R)_{sa}^*,
&
C_S^{\prime ab}=C_P^{\prime ab}
&=\frac{g_4^2}{2M_X^2}(V_R)_{db}(V_L)_{sa}^* .
\label{eq:KL_matching}
\end{align}
At leading order in QCD, the vector coefficients do not run, while the scalar and pseudoscalar coefficients evolve in the same way as the quark masses. Consequently, the combinations $C_{S,P}^{(\prime)}\mu)/[m_s(\mu)+m_d(\mu)]$ are RG invariant under QCD evolution. We therefore evaluate both the Wilson coefficients and the running quark masses at the matching scale $\mu=M_X$.\\
We use the phase convention
\begin{equation}
K_L=\frac{K^0+\bar K^0}{\sqrt2},\qquad
\langle0|\bar s\gamma^\mu\gamma_5d|K^0\rangle
=\langle0|\bar d\gamma^\mu\gamma_5s|\bar K^0\rangle
=if_Kp^\mu ,\qquad \langle0|\bar s\gamma_5d|K^0\rangle
=\langle0|\bar d\gamma_5s|\bar K^0\rangle
=-if_K \mathcal{B}_0\,,
\end{equation}
with $\mathcal{B}_0=m_K^2/(m_s+m_d)$. 
For the ordered final state $\ell_a^-\ell_b^+$, the amplitude can then be
written as
\begin{equation}
{\cal A}(K_L\to\ell_a^-\ell_b^+)
=\frac{if_K}{2}
\left[
{\cal S}_L^{ab}\,\bar u_av_b+
{\cal P}_L^{ab}\,\bar u_a\gamma_5v_b
\right],
\end{equation}
where
\begin{align}
\begin{split}
{\cal S}_L^{ab}
={1\over\sqrt2}\Big\{&
(m_a-m_b)
\big[C_9^{ab}-C_9^{\prime ab}
+C_9^{ba*}-C_9^{\prime ba*}\big]
 +\mathcal{B}_0
\big[C_S^{ab}-C_S^{\prime ab}
-C_S^{ba*}+C_S^{\prime ba*}\big]\Big\},
\\
{\cal P}_L^{ab}
={1\over\sqrt2}\Big\{&
(m_a+m_b)
\big[C_{10}^{ab}-C_{10}^{\prime ab}
+C_{10}^{ba*}-C_{10}^{\prime ba*}\big]
 + \mathcal{B}_0
\big[C_P^{ab}-C_P^{\prime ab}
+C_P^{ba*}-C_P^{\prime ba*}\big]\Big\}.
\end{split}
\label{eq:KL_SP}
\end{align}
The corresponding rate is
\begin{align}
\begin{split}
\Gamma(K_L\to\ell_a^-\ell_b^+)
={f_K^2\lambda^{1/2}(m_K^2,m_a^2,m_b^2)\over32\pi m_K^3}
\Big\{&
\big[m_K^2-(m_a+m_b)^2\big]|{\cal S}_L^{ab}|^2 
+
\big[m_K^2-(m_a-m_b)^2\big]|{\cal P}_L^{ab}|^2
\Big\}\,,
\end{split}
\label{eq:KL_width}
\end{align}
where $\lambda(x,y,z)=x^2+y^2+z^2-2xy-2yz-2zx$ is the K\"all\'en function. The charge-inclusive branching ratio relevant for the experimental constraint is obtained by summing over both final states $\Gamma(K_L \to \mu^+ e^-) + \Gamma(K_L \to \mu^- e^+)$. For illustration, taking a diagonal flavor structure, $V_L=V_R= \mathds{1}$, gives, for
$M_X=5\times10^5\,{\rm GeV}$,
\begin{align}
\alpha_s(M_X)=0.055,\qquad
m_s(M_X)+m_d(M_X)=35.1\,{\rm MeV},\qquad
{\rm Br}(K_L\to\mu^\pm e^\mp)\simeq3.1\times10^{-9}.
\end{align}
Using the experimental constraint
${\rm Br}(K_L\to\mu^\pm e^\mp)<4.7\times10^{-12}$~\cite{BNL:1998apv} and evolving the coupling and quark masses consistently with $M_X$ gives
\begin{align}
M_X\gtrsim2.5\times10^6~{\rm GeV}\left(\frac{4.7\times 10^{-12}}{\text{Br}(K_L \to \mu^\pm e^\mp)}\right)^{1/4}
\qquad
(V_L=V_R=\mathbf 1).
\label{eq:KL_bound}
\end{align}

We now consider the zeroth-order Yukawa textures studied in the main text, in which the third generation decouples and mixing occurs only between the first two generations. The corresponding down-quark and charged-lepton Yukawa matrices are symmetric and, up to common field rephasings, their light blocks take the form
\begin{equation}
Y_d=
\begin{pmatrix}0&a\\a&b\end{pmatrix},
\qquad
Y_e=
\begin{pmatrix}0&a\\a&-3b\end{pmatrix},
\end{equation}
with $a,b>0$. Their orthogonal diagonalization gives one negative eigenvalue in each sector. Requiring positive physical fermion masses therefore introduces relative signs between the left- and right-handed rotations. In the bi-orthogonal convention, these can be chosen as
\begin{equation}
\begin{aligned}
U_{dL}&=R(\theta_d),
&
U_{dR}&=R(\theta_d)\operatorname{diag}(-1,1),\\
U_{eL}&=R(\theta_e),
&
U_{eR}&=R(\theta_e)\operatorname{diag}(1,-1),
\end{aligned}
\end{equation}
where
\begin{equation}
R(\theta)=
\begin{pmatrix}
\cos\theta&\sin\theta\\
-\sin\theta&\cos\theta
\end{pmatrix}.
\end{equation}
Defining $\alpha=\theta_e-\theta_d$, the vector-leptoquark coupling matrices become
\begin{equation}
V_L=
\begin{pmatrix}
\cos\alpha&\sin\alpha&0\\
-\sin\alpha&\cos\alpha&0\\
0&0&1
\end{pmatrix},
\qquad
V_R=
\begin{pmatrix}
-\cos\alpha&\sin\alpha&0\\
-\sin\alpha&-\cos\alpha&0\\
0&0&1
\end{pmatrix}.
\end{equation}

Neglecting the electron mass and retaining only the dominant chirality-enhanced contribution to Eq.~\eqref{eq:KL_width}, the charge-inclusive decay width becomes proportional to 
\begin{equation}
    C_P^{e\mu}-C_P^{\prime e\mu}
+C_P^{\mu e*}-C_P^{\prime\mu e*}
=\frac{g_4^2}{M_X^2} (\cos^2 \alpha + \sin^2 \alpha) = \frac{g_4^2}{M_X^2} .
\end{equation}
Interestingly, the dependence on the $1$--$2$ mixing angle cancels at this order.  Consequently,
\begin{equation}
\frac{\Gamma(K_L\to\mu^\pm e^\mp)}
{\Gamma(K_L\to\mu^\pm e^\mp)|_{V_L=V_R=\mathbf1}}
\simeq1.
\end{equation}
Thus, although the symmetric light-generation texture predicts $\theta_e\simeq-\theta_d/3$, with $\theta_d\simeq\sqrt{m_d/m_s}$, these rotations do not suppress the leading vector-leptoquark contribution to $K_L\to\mu^\pm e^\mp$. The flavour-diagonal mass bound therefore remains essentially unchanged.

\section{Chiral suppression of $K_L\to\mu^\pm e^\mp$ from vector-like fermion mixing}
\label{app:chiralKmue}

The dominant constraint on a PS vector leptoquark originates from the scalar operators in Eq.~\eqref{eq:KL_matching}, which arise from the simultaneous presence of left- and right-handed couplings,
\begin{align}
C_{S,P} \propto (V_L)_{db}(V_R)_{sa}^{*}
\qquad {\rm and}\qquad
C_{S,P}^{\prime} \propto (V_R)_{db}(V_L)_{sa}^{*}.
\label{eq:scalar_op}
\end{align}
These contributions are not helicity suppressed: the relevant hadronic matrix element is proportional to $B_0=m_K^2/(m_s+m_d)$, whereas the vector operators contribute only through terms proportional to the charged-lepton mass. Therefore, suppressing one chirality of the vector-leptoquark coupling to the light fermions can eliminate the mixed-chirality operators and leave only the helicity-suppressed vector contribution.\\

Such a situation can occur in PS models with additional vector-like fermions~\cite{Dolan:2020doe}. The mechanism can be understood from a single-generation example. Consider a charged lepton $e$ contained in a PS multiplet and a vector-like partner $E$ whose mass matrix, after PS breaking, is
\begin{equation}
{\cal L}_m\supset
(\bar e_L,\bar E_L)
\begin{pmatrix}
m_e&m_{eE}\\
m_{Ee}&M_E
\end{pmatrix}
\begin{pmatrix}
e_R\\
E_R
\end{pmatrix}
+\mathrm{h.c.},
\label{eq:vectorlike_mass}
\end{equation}
and consider the limit $|m_{eE}|\gg |m_e|,|m_{Ee}|,|M_E|$.
The large off-diagonal entry pairs $e_L$ and $E_R$ into a heavy Dirac
state. To leading order in the small ratios of mass parameters, the relation between interaction and mass eigenstates is
\begin{align}
e_L &\simeq e_L^{\rm heavy}
+\epsilon_L e_L^{\rm light},
&
E_L&\simeq e_L^{\rm light}
-\epsilon_L e_L^{\rm heavy},
\nonumber\\
e_R &\simeq e_R^{\rm light}
+\epsilon_R e_R^{\rm heavy},
&
E_R&\simeq e_R^{\rm heavy}
-\epsilon_R e_R^{\rm light},
\end{align}
where the small mixing parameters are
\begin{equation}
\epsilon_L\simeq
\frac{m_e m_{Ee}^*+m_{eE}M_E^*}
{|m_{eE}|^2},
\qquad
\epsilon_R\simeq
\frac{m_e^*m_{eE}+m_{Ee}^*M_E}
{|m_{eE}|^2}.
\end{equation}
Thus, although the vector leptoquark couples with full gauge strength to the PS fields $e_{L,R}$, the light lepton has a suppressed overlap with
the left-handed PS component,
\begin{equation}
e_L\rightarrow \epsilon_L e_L^{\rm light},
\end{equation}
while the right-handed overlap can remain unsuppressed. The opposite
hierarchy, $|m_{Ee}|\gg |m_e|,|m_{eE}|,|M_E|$, gives the analogous suppression of the right-handed coupling.\\

This mechanism is particularly interesting if the Pati--Salam breaking
scale is to be identified with the scale $\mu_3$ at which
\begin{equation}
    \frac{y_\tau}{y_b}
    =
    \frac{y_\mu}{3y_s}
    =1 .
\end{equation}
We therefore do not want to appreciably mix the muon, since such a mixing
would in general spoil the successful second-generation Yukawa relation.
Instead, we modify only the first-generation charged-lepton sector.
Equation~\eqref{eq:scalar_op} then shows that both chiral couplings of the
light electron to the Pati--Salam vector leptoquark must be suppressed.
Equivalently, the light electron must have a small overlap with both the
$e\subset F_1$ and $e^c\subset F_1^c$ interaction eigenstates.

Such a modification necessarily affects the first-generation
quark--lepton mass relation. A viable construction should therefore not
only suppress $K_L\to\mu e$, but also accommodate
\begin{equation}
    \frac{y_e}{y_d}\simeq 0.28
\end{equation}
at $\mu\simeq\mu_3$, as shown in Fig.~\ref{fig:running}. We now present a
simple realization in which these two requirements can be simultaneously
satisfied.\\

We consider $SU(4)_c\times SU(2)_L\times SU(2)_R$, broken to the Standard Model gauge group by $\chi_R\sim(4,1,2)$, with $\langle\chi_R\rangle=v_R$.
The electroweak scalar sector contains a bidoublet
$\Phi_1\sim(1,2,2)$ and an adjoint bidoublet
$\Phi_2\sim(15,2,2)$. The former gives the usual quark--lepton equality,
whereas the latter generates the relative Clebsch factor of $3$ between
charged leptons and down-type quarks. These interactions account for the
second- and third-generation Yukawa relations at the scale $\mu_3$.

The three chiral families transform as
\begin{equation}
    F_i\sim(4,2,1),
    \qquad
    F_i^c\sim(\bar4,1,2),
\end{equation}
and we supplement the first-generation charged-lepton sector with
\begin{equation}
    \Psi_1,\Psi_1^c\sim(1,1,3),
    \qquad
    \Psi_2,\Psi_2^c\sim(1,2,2).
\end{equation}
The interactions relevant for the charged-lepton mass matrix are
\begin{align}
\begin{split}
\mathcal L\supset {}&
    \mu_1\,\Psi_1\Psi_1^c
    +\mu_2\,\Psi_2\Psi_2^c
    +y_1\,\Psi_1\chi_R F_1^c
    +y_2\,F_1\chi_R^\dagger\Psi_2^c +
    y_{12}\,\Psi_1\Phi_1\Psi_2^c
    +y_{21}\,\Psi_2\Phi_1\Psi_1^c
    +{\rm h.c.}\, .
\end{split}
\label{eq:mixing_lagrangian}
\end{align}

After Pati--Salam and electroweak symmetry breaking,
$\langle\chi_R\rangle=v_R$ and $\langle\Phi_1\rangle=v_1\sim v$
(up to the appropriate $\beta$ dependence), the charged-lepton mass
terms are
\begin{align}
\mathcal L_m \supset {}&
    m\,e e^c
    +M_2\,eE_2^c
    +M_1\,E_1e^c
    +\mu_1\,E_1E_1^c
    +\mu_2\,E_2E_2^c 
    + m_{12}\,E_1E_2^c
    +m_{21}\,E_2E_1^c
    +{\rm h.c.},
\end{align}
where
\begin{equation}
    M_1=y_1v_R,
    \qquad
    M_2=y_2v_R,
    \qquad
    m_{12}=y_{12}v_1,
    \qquad
    m_{21}=y_{21}v_1 .
\end{equation}
Equivalently,
\begin{align}
    \begin{pmatrix}
        e & E_1 & E_2
    \end{pmatrix}
    M
    \begin{pmatrix}
        e^c\\
        E_1^c\\
        E_2^c
    \end{pmatrix}
    +{\rm h.c.},
    \qquad
    M=
    \begin{pmatrix}
        m & 0 & M_2\\
        M_1 & \mu_1 & m_{12}\\
        0 & m_{21} & \mu_2
    \end{pmatrix}.
\label{eq:massmatrix_mixing}
\end{align}

The physical content of Eq.~\eqref{eq:massmatrix_mixing} is particularly
transparent in the limit
\begin{equation}
    M_1,M_2\gg
    m,\mu_{1,2},m_{12},m_{21}.
\end{equation}
The large entries $M_2$ and $M_1$ form two heavy Dirac pairs,
approximately $(e,E_2^c)$ and $(E_1,e^c)$, respectively. The remaining
light chiral states are therefore approximately $E_2$ and $E_1^c$.
Consequently, the entry $m_{21}$ directly couples the two light states,
whereas $m_{12}$ couples predominantly heavy states.

This is also manifest from
\begin{equation}
    \det (MM^\dagger)
    =
    |M_1M_2m_{21}
    -m\,m_{12}m_{21}
    +m\,\mu_1\mu_2|^2 .
\end{equation}
Assuming all parameters real for simplicity, since the two heavy singular values are $M_1$ and $M_2$ at leading order,
the light charged-lepton mass is
\begin{align}
    m_e^{\rm phys}
    &\simeq
    \sqrt{\frac{\det (MM^\dagger)}{M_1^2M_2^2}}
    \simeq
    m_{21} -\frac{m\,m_{12}m_{21}}{M_1M_2} +\frac{m\,\mu_1\mu_2}{M_1M_2}
    \simeq m_{21}.
\label{eq:electron_light_mass}
\end{align}

At the same time, the overlap of the light eigenstate with the original
Pati--Salam charged-lepton fields is suppressed. To leading order,
\begin{equation}
    \epsilon_L\simeq\frac{\mu_2}{M_2},
    \qquad
    \epsilon_R\simeq\frac{\mu_1}{M_1}.
\label{eq:electron_overlaps}
\end{equation}
These are precisely the factors that suppress the left- and right-handed
vector-leptoquark couplings.

It is instructive to compare this with the limit $m_{21}=0$. In that
case,
\begin{equation}
    m_e^{\rm phys}
    \simeq
    m\,\frac{\mu_1}{M_1}\frac{\mu_2}{M_2}
    =
    m\,\epsilon_L\epsilon_R ,
\end{equation}
so suppressing both leptoquark couplings would simultaneously suppress
the electron mass by two powers of the small mixing angles. The
$m_{21}$ interaction avoids this correlation.

Indeed, the scalar operators relevant for $K_L\to\mu e$ contain only one
suppressed electron overlap,
\begin{equation}
    C_{S,P}\propto\frac{\epsilon_L}{M_X^2},
    \qquad
    C^{\prime}_{S,P}\propto\frac{\epsilon_R}{M_X^2},
\end{equation}
up to flavor factors of order unity in the muon sector. Lowering the vector-leptoquark bound from $\simeq 2.5 \times10^6 \,\mathrm{GeV}$ to
${\cal O}(5\times10^5)\,\mathrm{GeV}$ therefore requires
$\epsilon_{L,R}$ at the level of a few times $10^{-2}$.

The role of $m_{21}$ is therefore crucial: it allows the electron mass to remain of electroweak origin while its Pati--Salam gauge couplings are strongly suppressed. The first-generation relation between $y_e$ and $y_d$ is consequently no longer fixed by the original Pati--Salam Yukawa structure. Instead, $y_e/y_d$ is controlled by the additional mixing sector and can naturally differ from the second- and third-generation quark--lepton relations. Thus the price paid for lowering the Pati--Salam scale is the loss of a first-generation Yukawa prediction, while preserving the successful relations involving the muon and tau.

As written, this minimal field content should be regarded as a proof of principle for the charged-lepton suppression mechanism rather than as a complete neutrino model. The neutral components of the added multiplets participate in the same symmetry structure and lead to an accidental relation in the neutral sector. A realistic completion must lift this relation through additional neutrino-sector interactions or states while preserving the small charged-lepton overlaps in Eq.~\eqref{eq:electron_overlaps}. This issue does not modify the parametric suppression of the vector-leptoquark contribution derived above.

\section{Future experimental constraints}
\label{app:future_constraints}

The current strongest constraint on the Pati--Salam scale is provided by $K_L\to\mu^\pm e^\mp$. The current E871 branching-fraction limit of $4.7\times10^{-12}$ at $90\%$ C.L.~\cite{BNL:1998apv} corresponds to $M_X\gtrsim2.5\times10^6\,\mathrm{GeV}$ for unsuppressed light-family currents, as discussed in Appendix~\ref{app:Kmue} . What are the prospects to improve on this constraint? To our knowledge, no approved dedicated experimental program is currently planned to improve this specific limit. Given the dependence of the branching ratio to $M_X^{-4}$, testing masses near $10^7\,\mathrm{GeV}$ through this channel would require branching-fraction sensitivity near $10^{-14}$. Other kaon measurements are not expected to improve this constraint either: LHCb Upgrade~II studies envisage sensitivity to $K_S\to\mu^+\mu^-$ near $5\times10^{-12}$, corresponding to a leptoquark reach of a few times $10^5\,\mathrm{GeV}$, with a different dependence on flavor and CP phases~\cite{LHCb:2018roe}.

Several other low-energy observables are sensitive to different combinations of the fermion rotations and need not constrain the same flavor structure. To compare their sensitivities, we quote indicative vector-leptoquark mass reaches for unsuppressed relevant flavor products, and no cancellations.

Lepton-flavor-violating $B$ decays probe lower masses. LHCb limits on $B^0\to e^\pm\mu^\mp$ and $B_s^0\to e^\pm\mu^\mp$, at about the $10^{-9}$ levels, correspond to characteristic scales of order $(1$--$2)\times10^5\,\mathrm{GeV}$~\cite{LHCb:2013tno,LHCb:2017hag}. Even a hundredfold improvement in branching-fraction sensitivity would
increase these mass reaches by only a factor of about three, leaving them below the kaon reference scale. Similarly, Belle~II projections for lepton-flavor-violating $\tau$ decays reach branching fractions of $10^{-10}$--$10^{-9}$ with $50\,\mathrm{ab}^{-1}$; representative tree-level semileptonic channels then probe leptoquark masses of order $10^4$--$10^5\,\mathrm{GeV}$~\cite{Banerjee:2022vdd}. These measurements remain valuable because they access bottom-quark and tau-lepton couplings that need not enter the leading light-family constraints.

Pion universality provides a more sensitive complementary test. PIONEER aims to measure the ratio of electronic to muonic pion decay rates with relative precision $10^{-4}$~\cite{PIONEER:2025idw}. The interference between the SM and the leptoquark mediated amplitudes allows to reach scales of approximately $(1$--$2)\times10^6\,\mathrm{GeV}$, compared with the present sensitivity of a few $10^5\,\mathrm{GeV}$. PIONEER can therefore approach the kaon reference scale while testing different coupling products~\cite{Valencia:1994cj}.

Purely leptonic flavor-violating searches also offer significant improvements. MEG~II targets $\mu\to e\gamma$ branching-fraction sensitivity near $6\times10^{-14}$, while Mu3e aims at single-event sensitivities of approximately $2\times10^{-15}$ and $10^{-16}$ in its two phases~\cite{MEGII:2025gzr,Mu3e:2020gyw}. For unsuppressed mixed-chirality couplings to an internal bottom quark, a leading dipole estimate gives mass reaches of several $10^5\,\mathrm{GeV}$ for MEG~II and order $10^6\,\mathrm{GeV}$ for Mu3e Phase~II.

Magnetic moments and the muon EDM are less sensitive to this heavy gauge sector. Present electron and muon magnetic-moment sensitivities correspond to masses of order $10^3$--$10^4\,\mathrm{GeV}$, even allowing an internal bottom-quark chirality enhancement. The PSI frozen-spin muon-EDM program targets
$6\times10^{-23}\ e\cdot\mathrm{cm}$, corresponding to only a few $10^3\,\mathrm{GeV}$ for an unsuppressed CP-odd bottom-quark coupling product~\cite{Altmannshofer:2020ywf,Adelmann:2025nev}. Neutron-EDM searches such as n2EDM, targeting the $10^{-27}\ e\cdot\mathrm{cm}$ level, can instead probe characteristic scales around $10^5$--$10^6\,\mathrm{GeV}$ in estimates dominated by tau-enhanced light-quark dipoles~\cite{n2EDM:2021yah,Altmannshofer:2020ywf}.

Molecular EDM measurements and coherent $\mu\to e$ conversion provide the principal prospective gains in sensitivity to the vector-leptoquark sector. The full Mu2e and COMET programs can probe mixed-chirality mass scales of order $10^7\,\mathrm{GeV}$, with the proposed Mu2e-II upgrade extending this reach to a few  times $10^7\,\mathrm{GeV}$. Longer-term proposals aim to detect coherent $\mu\to e$ conversion at rates as small as $10^{-19}$ of the ordinary nuclear muon-capture
rate~\cite{CGroup:2022tli}. For an aluminum target and unsuppressed mixed-chirality flavor products, such sensitivity would probe vector-leptoquark masses of several $10^7\,\mathrm{GeV}$. Electron-EDM sensitivities approaching $10^{-31}\ e\cdot\mathrm{cm}$ can probe scales in the $10^7$--$10^8\,\mathrm{GeV}$ range for suitable CP-odd products, while the accompanying tree-level electron--nucleon interactions provide additional sensitivity~\cite{Roussy:2022cmp,Diver2026ACME}. These opportunities are complementary: conversion requires lepton-flavor violation, whereas EDM measurements require CP violation. We therefore discuss their matching, present bounds, and future reach in detail below, keeping the flavor dependence explicit.

\subsection{Electric dipole moments}

The charged-lepton EDM operator is
\begin{equation}
	\mathcal L_{\rm EDM}
	= i d_\ell\,(\ell^c\sigma^{\mu\nu}\ell)F_{\mu\nu}
	+\mathrm{h.c.}
	\label{eq:edm_definition}
\end{equation}
Vector-leptoquark exchange generates this operator at one loop and we get~\cite{Altmannshofer:2020ywf}
\begin{equation}
	\frac{d_\ell(M_X)}{e_{\rm em}}
	=-\frac{3g_4^2}{16\pi^2M_X^2}
	\sum_{q=d,s,b}\overline m_q(M_X)
	\operatorname{Im}\!\left[(V_L)_{q\ell}(V_R)_{q\ell}^*\right].
	\label{eq:edm_matching}
\end{equation}
The internal down-type quark supplies the chirality flip. The contribution vanishes for real mass matrices or $V_R=V_L$.
The HfF$^+$ bound $|d_e|<4.1\times10^{-30}\ e\cdot\mathrm{cm}$ at $90\%$ C.L. assumes that other CP-odd interactions are absent~\cite{Roussy:2022cmp}.
Taking representative matching-scale masses
$(\overline m_d,\overline m_s,\overline m_b) =(0.002,0.04,2)\,\mathrm{GeV}$, it implies
\begin{equation}
	M_X\gtrsim1.1\times10^7\,\mathrm{GeV}
	\left(\frac{g_4}{0.8}\right)
	\left|
	\sum_{q=d,s,b}
	\frac{\overline m_q(M_X)}{2\,\mathrm{GeV}}\,
	\operatorname{Im}\!\left[(V_L)_{qe}(V_R)_{qe}^*\right]
	\right|^{1/2}.
	\label{eq:edm_electron_bound}
\end{equation}
For a single contributing quark, the corresponding coefficients
multiplying $(g_4/0.8)\sqrt{|\operatorname{Im}[(V_L)_{qe}(V_R)_{qe}^*]|}$ are approximately $3.4\times10^5$, $1.5\times10^6$, and $1.1\times10^7\,\mathrm{GeV}$ for $q=d,s,b$, respectively. A future sensitivity of $10^{-31}\ e\cdot\mathrm{cm}$ would increase these mass reaches by a factor of about $6.4$, reaching approximately $7\times10^7\,\mathrm{GeV}$ for an unsuppressed bottom-quark CP-odd product.\\

\paragraph{Electron--nucleon interactions.}
Molecular EDM measurements are also sensitive to CP-odd electron--nucleon interactions. Vector-leptoquark exchange generates these already at tree level through
\begin{equation}
	\mathcal L_{eq}^{\rm CP}
	=\frac{i g_4^2}{M_X^2}\sum_{q=d,s,b}
	\operatorname{Im}\!\left[(V_L)_{qe}(V_R)_{qe}^*\right]
	\left[(ee^c)(qq^c)^\dagger-(ee^c)^\dagger(qq^c)\right].
	\label{eq:edm_eqCP}
\end{equation}
Matching the quark scalar densities onto nucleons gives the spin-independent electron--nucleon coupling conventionally denoted by $C_S$~\cite{Kitano:2002mt,Roussy:2022cmp}. Assuming a negligible electron EDM, the HfF$^+$ measurement translates into the approximate upper bound
\begin{equation}
	|C_S|\lesssim4.4\times10^{-10}
	\qquad (90\%~\mathrm{C.L.}),
\end{equation}
using the confidence prescription of Ref.~\cite{Roussy:2022cmp}. For a dominant down-quark contribution, standard nucleon matrix elements and QCD evolution give the indicative constraint
\begin{equation}
	M_X\gtrsim3\times10^7\,\mathrm{GeV}\,
	\left(\frac{g_4}{0.8}\right)
	\left|
	\operatorname{Im}\!\left[(V_L)_{de}(V_R)_{de}^*\right]
	\right|^{1/2}.
	\label{eq:edm_semileptonic_bound}
\end{equation}
The tree-level interaction can therefore be substantially more constraining than the corresponding down-quark dipole contribution. In general, molecular measurements constrain a linear combination of $d_e$ and $C_S$, so both contributions must be included when comparable. Measurements using different molecules help separate
them~\cite{Roussy:2022cmp}. For fixed coupling products, an improvement in the sensitivity to $C_S$ increases the mass reach as the square root of that improvement.

\subsection{$\mu\to e$ conversion}

Vector-leptoquark exchange generates $\mu\to e$ conversion in nuclei at tree level through
\begin{align}
	\mathcal L_{\mu e}
	=-\frac{g_4^2}{2M_X^2}\sum_{q=d,s,b}\Big\{&
	(V_L)_{qe}^*(V_L)_{q\mu}
	(e^\dagger\bar\sigma^\rho\mu)
	(q^\dagger\bar\sigma_\rho q)
	+(V_R)_{qe}^*(V_R)_{q\mu}
	(\mu^{c\dagger}\bar\sigma^\rho e^c)
	(q^{c\dagger}\bar\sigma_\rho q^c)
	\nonumber\\
	&-2(V_L)_{qe}^*(V_R)_{q\mu}
	(e\mu^c)^\dagger(qq^c)
	-2(V_R)_{qe}^*(V_L)_{q\mu}
	(e^c\mu)(qq^c)^\dagger
	\Big\}+\mathrm{h.c.}
	\label{eq:conv_matching}
\end{align}
The same-chirality terms generate nucleon vector currents, whereas the mixed-chirality terms generate scalar densities. The coherent
rate follows by evolving these operators to hadronic scales and matching onto nucleons, including the nuclear overlap integrals~\cite{Kitano:2002mt,Crivellin:2014cta}.
Unlike EDMs, conversion requires flavor-changing coupling products but no CP-violating phase.

The experimental observable is the conversion rate normalized to ordinary nuclear muon capture,
\begin{equation}
	\mathrm{CR}_{\mathcal N}
	=\frac{\Gamma(\mu^-\mathcal N\to e^-\mathcal N)}
	{\Gamma_{\rm capt}(\mathcal N)}.
	\label{eq:conv_rate}
\end{equation}
The SINDRUM~II limit $\mathrm{CR}_{\rm Au}<7\times10^{-13}$ at $90\%$ C.L. implies the indicative constraint~\cite{SINDRUMII:2006dvw,Kitano:2002mt}
\begin{equation}
	\begin{split}
		M_X\gtrsim{}&
		4.3\times10^5\,\mathrm{GeV}
		\left(\frac{g_4}{0.8}\right)
		\left(\frac{7 \times 10^{-13}}{\text{CR}_{\rm Au}}\right)^{1/4}\\
        &\qquad \times \Big\{
		\left|(V_L)_{de}^*
		\left[(V_L)_{d\mu}-\rho_{Au}(M_X)(V_R)_{d\mu}\right]\right|^2
		+\left|(V_R)_{de}^*
		\left[(V_R)_{d\mu}-\rho_{Au}(M_X)(V_L)_{d\mu}\right]\right|^2
		\Big\}^{1/4} .
\label{eq:conv_current_bounds}
	\end{split}
\end{equation}
This expression retains all four chirality combinations and their scalar--vector interference. The relative scalar-to-vector coefficient $\rho_{\rm Au}$ incorporates the scalar nucleon matrix elements, QCD evolution, and gold nuclear overlaps. The vector-leptoquark coupling is evaluated at the matching scale, where $g_4(M_X)=g_3(M_X)$. Since the scalar Wilson coefficients evolve in the same way as the running quark masses, their products with the corresponding scalar nucleon matrix elements are invariant under QCD running. The numerical value of $\rho_{\rm Au}$ depends on the adopted hadronic inputs; using the original Kitano scalar form factors~\cite{Kitano:2002mt} gives a coefficient $\rho_{\rm Au}\simeq 11$ at PeV-scale matching.

For the real symmetric $1$--$2$ texture discussed in Appendix~\ref{app:Kmue}, the mixing matrices are specified by a single angle $\alpha\simeq-\frac{4}{3}\sqrt{m_d/m_s}$. The two chiral contributions have equal magnitudes, and the above bound simplifies to
\begin{equation}
\begin{split}
M_X\gtrsim{}&
\, 4.3\times10^5\text{ GeV}
\left(\frac{g_4(M_X)}{0.8}\right)
\left(\frac{7\times10^{-13}}{\text{CR}_{\rm Au}^{\rm lim}}\right)^{1/4}\
\bigg|
\frac{\sin(2\alpha)}{\sqrt{2}}
(1-\rho_{\rm Au}(M_X))
\bigg|^{1/2}\\
\simeq&
\, 8.6\times10^5\text{ GeV}
\left(\frac{g_4(M_X)}{0.8}\right)
\left(\frac{7\times10^{-13}}{\text{CR}_{\rm Au}^{\rm lim}}\right)^{1/4}
\left(\frac{|1-\rho_{\rm Au}(M_X)|}{10}\right)^{1/2}.
\label{eq:mu2e_Au}
\end{split}
\end{equation}

This estimate neglects the strange-quark scalar contribution, which adds coherently in the symmetric light-generation texture but gives a smaller contribution because of the inverse quark-mass dependence of the scalar nucleon matrix elements.

Mu2e and COMET will substantially improve this sensitivity. The full Mu2e program has a projected $90\%$ C.L. sensitivity near $\mathrm{CR}_{\rm Al}=6\times10^{-17}$, corresponding to mixed-chirality mass scales of order $10^7\,\mathrm{GeV}$ for unsuppressed coupling products; COMET Phase~II aims at a comparable reach~\cite{COMET:2018auw,Mu2e:2022ggl,COMET:2025sdw}. Turning to coherent $\mu\to e$ conversion in aluminum, the corresponding projected bound takes the form,
\begin{equation}
    M_X \gtrsim{} 4.9 \times 10^6 \text{ GeV} \left(\frac{g_4(M_X)}{0.8}\right) \left(\frac{6.2 \times 10^{-16}}{\text{CR}_{\rm Al}^{\rm lim}}\right)^{1/4} \left(\frac{|1-\rho_{\rm Al}(M_X)|}{17}\right)^{1/2}
    \label{eq:mu2e_Al}
\end{equation}
In Eqs.~\eqref{eq:mu2e_Au} and~\eqref{eq:mu2e_Al}, the last factors are normalized to unity for the values obtained with the scalar form factors and overlap integrals of Ref.~\cite{Kitano:2002mt}, $\rho_{\rm Au}\simeq11$ and $\rho_{\rm Al}\simeq18$; scalar nucleon matrix elements based on the pion--nucleon sigma term~\cite{Crivellin:2014cta} give values of $\rho$ larger by a factor $1.4$--$1.8$, which would raise these reaches by about $20$--$40\%$, so the quoted constraints are conservative.
The proposed Mu2e-II upgrade projects a $90\%$ C.L. sensitivity
near $6\times10^{-18}$~\cite{Mu2e-II:2022blh}, extending the corresponding mass reach to approximately $1.5\times10^7\,\mathrm{GeV}$.
Longer-term concepts motivate conversion sensitivities near $10^{-19}$~\cite{CGroup:2022tli}.

\end{document}